\documentclass[aps,prb,twocolumn,showpacs,eqsecnum]{revtex4}
\usepackage{bm}
\usepackage{amsmath}
\usepackage{amssymb}
\usepackage{graphicx}

\newcommand{\beq}{\begin{equation}}
\newcommand{\eeq}{\end{equation}}
\newcommand{\beqar}{\begin{eqnarray*}}
\newcommand{\eeqar}{\end{eqnarray*}}

\newcommand{\ETh}{E_{\text{Th}}}

\newcommand{\sgn}{\text{sgn}\,}
\newcommand{\vsc}{v_\text{sc}}

\newcommand{\Db}{\bar{D}}

\newcommand{\sea}{\searrow}
\newcommand{\nea}{\nearrow}
\newcommand{\swa}{\swarrow}
\newcommand{\nwa}{\nwarrow}

\newcommand{\Hh}{\hat{H}}

\newcommand{\Gc}{\mathcal{G}}
\newcommand{\Dc}{\mathcal{D}}

\newcommand{\Vc}{\mathcal{V}}
\newcommand{\Pc}{\mathcal{P}}

\newcommand{\Vt}{\widetilde{V}}

\newcommand{\rtarr}{\rightarrow}

\newcommand{\Hb}{{\bf H}}
\newcommand{\nb}{{\bf n}}
\newcommand{\tb}{{\bf t}}

\newcommand{\kb}{{\bf k}}
\newcommand{\lb}{{\bf l}}
\newcommand{\ib}{{\bf i}}
\newcommand{\jb}{{\bf j}}
\newcommand{\ab}{{\bf a}}
\newcommand{\bb}{{\bf b}}
\newcommand{\eb}{{\bf e}}

\newcommand{\s}{{\bf s}}

\newcommand{\Rb}{{\bf R}}

\newcommand{\xb}{{\bf x}}
\newcommand{\qb}{{\bf q}}
\newcommand{\pb}{{\bf p}}

\newcommand{\Ab}{{\bf A}}
\newcommand{\rb}{{\bf r}}
\newcommand{\sbf}{{\bf s}}

\newcommand{\dg}{\dagger}

\newcommand{\lan}{\langle}
\newcommand{\ran}{\rangle}
\newcommand{\om}{\omega}
\newcommand{\Om}{\Omega}
\newcommand{\al}{\alpha}
\newcommand{\be}{\beta}
\newcommand{\ga}{\gamma}
\newcommand{\Ga}{\Gamma}
\newcommand{\de}{\delta}

\newcommand{\la}{\lambda}
\newcommand{\sig}{\sigma}

\newcommand{\eps}{\varepsilon}

\newcommand{\lt}{\left}
\newcommand{\rt}{\right}
\begin{document}

\title{Hall Transport in Granular Metals and Effects of Coulomb Interactions}
\author{Maxim Yu. Kharitonov$^{1}$ and Konstantin B. Efetov$^{1,2}$}
\affiliation{$^{1}$ Theoretische Physik III, Ruhr-Universit\"{a}t Bochum, Germany,\\
$^{2}$L.D. Landau Institute for Theoretical Physics, Moscow, Russia.}
\date{\today}

\begin{abstract}
We present a theory of Hall effect in granular systems at large
tunneling conductance $g_{T}\gg 1$. Hall transport is essentially
determined by  the intragrain electron dynamics, which, as we find
using the Kubo formula and diagrammatic technique, can be
described by nonzero diffusion modes inside the grains. We show
that in the absence of Coulomb interaction the Hall resistivity
$\rho_{xy}$ depends neither on the tunneling conductance nor on
the intragrain disorder and is given by the classical formula
$\rho_{xy}=H/(n^* e c)$, where $n^*$ differs from the carrier
density $n$ inside the grains by a numerical coefficient
determined by the shape of the grains and type of granular
lattice. Further, we study the effects of Coulomb interactions by
calculating first-order in $1/g_T$ corrections and find that (i)
in a wide range of temperatures $T \gtrsim \Ga$ exceeding the
tunneling escape rate $\Ga$, the Hall resistivity $\rho_{xy}$ and
conductivity $\sig_{xy}$ acquire logarithmic in $T$ corrections,
which are of local origin and absent in homogeneously disordered
metals; (ii) large-scale ``Altshuler-Aronov'' correction to
$\sig_{xy}$, relevant at  $T\ll\Ga$, vanishes in agreement with
the theory of homogeneously disordered metals.

\end{abstract}
\maketitle


\section{Introduction}
Hall transport in different systems has always been a subject of extensive research.
Already the classical
Drude-Boltzmann theory provides us with an interesting result.
It is well-known that the Hall resistivity (HR)
\begin{equation}
\rho _{xy}=\frac{H}{nec}
\label{eq:rxy}
\end{equation}
of a disordered metal does not depend on the mean free path
and is determined solely by the carrier concentration $n$
allowing one to extract it experimentally.
%
At  low enough temperatures quantum effects
of Coulomb interaction and weak localization
(see, e.g., Refs.~\onlinecite{AA,LR})
influence the Hall transport, giving corrections to Eq.~(\ref{eq:rxy}).

Dense-packed arrays of metallic or semiconducting nanoparticles imbedded into
an insulating matrix, usually called {\em granular systems}
or {\em nanocrystals}, have recently received much attention
from both experimental and theoretical sides
(see a Review~\onlinecite{BELVreview} and references therein).
The longitudinal transport
in such systems is theoretically well understood now,
both in the metallic and insulating regimes.
At the same time, Hall transport in such granular materials has not been
addressed theoretically before, neither in the insulating nor in the metallic regimes. 
The absence of a theoretical description is apparently one of the
reasons, why measurements of the Hall resistivity have not become a standard
tool for characterization of granular metals, although they do not seem
to be very difficult.

Trying to apply the conventional theory of disordered metals to granular systems,
the following questions can be asked:
To what extent is the formula (\ref{eq:rxy}) applicable to granular metals?
How is the carrier concentration extracted from
Eq.~(\ref{eq:rxy}) related to the actual carrier concentration inside the
grains?
%
%
%
%
What impact do quantum effects have on Hall transport of a granular system?

In this paper we present a theory of Hall effect
in a granular system in the metallic regime
and answer these questions.

\begin{figure}
\includegraphics[width=.48\textwidth]{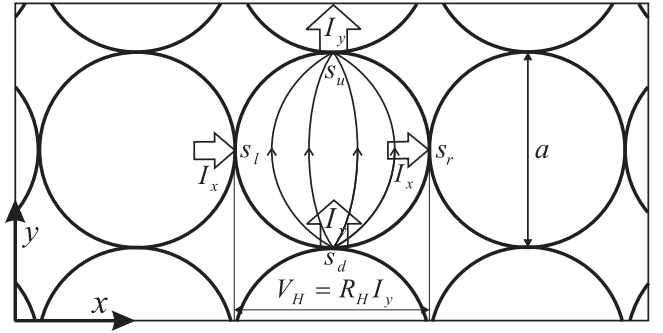}
\caption{ Granular system and classical picture of Hall conductivity.
The external Ohmic voltage $V_y$ is applied to the contacts in the $y$ direction.
The resulting Ohmic current $I_y=G_T V_y$ running through
the grain in the $y$ direction causes the Hall voltage drop $V_H=R_H
I_y$ between its opposite banks in the $x$ direction.
Since for calculating Hall conductivity $\sig_{xy}$
the total voltage drop per lattice period
in the $x$ direction is assumed 0, the Hall voltage $V_H$
is also applied (with an opposite sign)
to the contacts in the $x$ direction,
causing the Hall current $I_x=G_T V_H = G_T^2 R_H V_y$ [Eq.~(\ref{eq:sigxy0})]. }
\label{fig:system}
\end{figure}

In the metallic regime (``{\em granular metal}''), when the intergrain tunneling conductance
$G_T=(2e^2/\hbar) g_T$ is large, $g_T \gg 1$ (further we set $\hbar=1$),
the granular system as a whole is, roughly speaking, a good conductor and its properties
are quite similar to those of ordinary homogeneously disordered metals (HDMs).
At the same time the granularity of
the system brings a new physical aspect:
confinement of electrons  inside the grains.
In a system with ``well-pronounced'' granularity
electron traverses each grain many times
before it escapes from it to some neighboring grain
due to tunneling. This is ensured by the condition that
the tunneling escape rate $\Ga$ is much smaller than
the Thouless energy $\ETh$ of the grain:
\beq
    \Ga \ll \ETh,
\label{eq:granular1}
\eeq
or, equivalently, the tunneling conductance $G_T$
is much smaller than the longitudinal conductance $G_0=(2e^2/\hbar)g_0$
of the grain:
\beq
    g_T \ll g_0,
\label{eq:granular2}
\eeq
since $\Ga =g_T \de$ and $\ETh \propto g_0 \de$
($\de$ is the mean level spacing of the grain).


The conditions (\ref{eq:granular1}) and (\ref{eq:granular2}),
leading to new physics\cite{BELVreview} absent in HDMs, simplify
calculations at the same time.
Consider, for example,  the
classical
(in the absence of quantum effects, such as Coulomb interaction and weak
localization)
longitudinal conductivity (LC) $\sig_{xx}^{(0)}$
of a regular quadratic or cubic granular lattice (Fig.~\ref{fig:system})
with all contacts having equal conductances $G_T$.
In the limit $g_T\ll g_0$ the main contribution to
the longitudinal resistivity (LR)
$\rho_{xx}^{(0)}=1/\sig_{xx}^{(0)}$ comes from the tunnel barriers
between the grains rather than from scattering on impurities
inside the grains and LC equals \beq
    \sig^{(0)}_{xx}= G_T a^{2-d} ,
\label{eq:sigxx0}
\eeq
where $a$ is the size of the grains and $d=2,3$
is the dimensionality of the array.
The longitudinal conductance $G_0$ of the grain itself, which,
in principle, should be obtained
from a solution of a classical electrodynamics problem
for the distribution of the electric potential inside the grain
and is, therefore, determined by the properties of the intragrain electron dynamics,
does not enter Eq.~(\ref{eq:sigxx0}).

Thus, when studying
longitudinal transport one may neglect the details of electron dynamics inside
the grains, which is a significant simplification.
Technically,
this is equivalent to considering only the {\em zero}
({\em coordinate-independent inside the grains})
{\em spatial modes} of the diffusons or phases in the phase functional\cite{ET,BELVreview}.
%
%
Owing to the conditions (\ref{eq:granular1}) and (\ref{eq:granular2}),
the {\em zero-mode approximation suffices for studying the longitudinal transport}.

For Hall transport, however, the situation appears to be more complicated.
The Hall current
originates from the transversal drift
in the crossed magnetic and electric fields {\em inside} the grains.
The classically prohibited regions of tunnel contacts are neglegibly
small for dense-packed arrays and cannot contribute
to the Hall transport.
%
From simple classical
considerations (see Fig.~\ref{fig:system}) one obtains,
that Hall conductivity (HC) $\sig_{xy}^{(0)}$
in the leading in $g_T/g_0 \ll 1$ order is
\beq
    \sig^{(0)}_{xy}= G_T^2 R_H a^{2-d},
\label{eq:sigxy0}
\eeq
where $R_H$ is the Hall resistance of the grain.
Just like $G_0$, the Hall resistance $R_H$ should be obtained
from a solution of a classical electrodynamics problem
for the distribution of the electric potential inside the grain.
We come to the situation when
one is forced to take
the intragrain electron dynamics into account,
no matter how well the conditions (\ref{eq:granular1}) and (\ref{eq:granular2}) are satisfied.
%
In other words, {\em the zero-mode
approximation is not sufficient for the description of the Hall
transport} of a granular system.


However, a purely classical approach to the problem, 
giving a quick answer (\ref{eq:sigxy0}),
does not allow one to include quantum effects (such as Coulomb interaction
and weak localization) into considerations,
which come into play at sufficiently low temperatures
and can significantly affect transport properties.

In this work we develop a method of calculating conductivity of a granular system
in the metallic regime,
which allows one to take the  intragrain electron dynamics
into account.
Using the Kubo formula and diagrammatic technique,
we show that this can be done
%
by considering {\em nonzero} ({\em coordinate-dependent}) 
{\em modes} of standard two-particle
propagators ({\em ``diffusons''}) inside the grains.
%
This
procedures accounts for the finiteness of the ratio $g_T/g_0$
and reproduces the
solution of the classical electrodynamics problem
for the conductivity of a granular medium.
The generality of our approach allows one, in principle, to study
both LC and HC of the granular system for arbitrary ratio $g_T/g_0$
and for arbitrary type of the intragrain electron dynamics,
either ballistic or diffusive.
Nonzero modes of the diffusons are
eventually related to the longitudinal $G_0^{-1}$
and Hall $R_H$ resistances of the grain.


We apply our method to the problem of Hall transport,
for which considering intragrain dynamics is inevitable.
Neglecting quantum effects, we do recover the classical formula (\ref{eq:sigxy0})
for the Hall conductivity and obtain quite a universal result for  the Hall resistivity.
%
%
Diagrammatic approach allows us to include quantum effects
of Coulomb interaction and weak localization straightforwardly
into the developed scheme.
%
%
We study the influence of Coulomb interactions on HC and HR
by calculating first-order corrections
and find that the major temperature dependence of both HC and HR
of a granular metal comes from the contributions which are absent in HDMs.
We also announce our results for weak localization corrections,
detailed calculations of which will be presented elsewhere \cite{KEprepWL}.
%
Part of the results of our work (for temperatures $T\gtrsim \Ga$) was presented
in a brief form in Ref.~\onlinecite{KEletter}.

The paper is organized as follows.
In Sec.~\ref{sec:results} we present our results
for Hall conductivity and resistivity and Coulomb interaction corrections to them.
In Sec.~\ref{sec:model} the model for the
granular system is formulated and discussed.
In Sec.~\ref{sec:tech} the main features of the diagrammatic technique are explained,
and important building blocks, namely, the
intragrain diffuson in the presence of magnetic field
and the screened Coulomb interaction, are obtained.
In Sec.~\ref{sec:Hall} we calculate Hall conductivity
neglecting quantum effects of Coulomb interaction
and obtain the correspondence with the classical result.
Quantum effects of Coulomb interaction are studied in Sec.~\ref{sec:QCoulomb}.
Concluding remarks are presented in Sec.~\ref{sec:conclusion}.
In Appendix~\ref{app:bc} the boundary condition for the intragrain
diffuson in the presence of magnetic field is derived.

\section{\label{sec:results}Results}
In this section we list the main results of this work.
We perform calculations for magnetic fields $H$ such that
$\om_H \tau_0 \ll 1$,
where $\om_H = e H/(mc)$ is the cyclotron frequency and $\tau_0$ is the  electron scattering time
inside the grain. Since the (effective) electron mean free path $l=v_F \tau_0 \lesssim a$ does not
exceed the grain size $a$, and typically  $a \approx 1-100 \text{nm}$,
the condition $\om_H \tau_0 \ll 1$ is well fulfilled even
for experimentally 
high fields $H$.
We also assume that the granularity of the system is ``well-pronounced'',
i.e., the conditions (\ref{eq:granular1}) and (\ref{eq:granular2}) are satisfied.
Other assumptions and approximations are formulated in Sec.~\ref{sec:model}.

{\em Classical Hall conductivity and resistivity.}
First, we neglect quantum effects of
Coulomb interaction and
obtain  Eq.~(\ref{eq:sigxy0}) for HC $\sig^{(0)}_{xy}$
in the lowest nonvanishing order in $g_T/g_0 \ll 1 $.
This result obtained by diagrammatic methods is of completely
classical origin provided the tunneling contacts are viewed as  surface resistors with conductance $G_T$.
The HR of the system
\beq
    \rho^{(0)}_{xy} =
    \frac{\sig^{(0)}_{xy}}{(\sig^{(0)}_{xx})^2}=
    R_H  a^{d-2}
\label{eq:rhoxy00}
\eeq
following from Eqs.~(\ref{eq:sigxx0}) and (\ref{eq:sigxy0}),
thus, {\em does not depend on the tunneling conductance $G_T$}
and is expressed solely through the Hall resistance $R_H$ of a single
grain.
%
%
%
%
Further, the Hall resistance $R_H$ of the grain
{\em does not depend on the intragrain disorder},
but only on {\em the geometry of the grain} and carrier density $n$
of the grain material.
For grains of a simple geometry
(e.g., having reflectional symmetry
in all three dimensions)
$R_H= \rho_{xy}^\text{gr}a/S$
,
 where
$\rho_{xy}^\text{gr}=H/(n e c)$ is the specific Hall resistivity of the grain material
and $S$ is the area of the largest cross section of  the grain.


Therefore, akin to the universal result (\ref{eq:rxy}) for ordinary disordered metals,
for the classical Hall resistivity of a granular metal we obtain
\beq
    \rho^{(0)}_{xy} = \frac{H}{n^* e c}
\label{eq:rhoxy0}
\eeq
in  the case of a three-dimensional sample (3D, $d=3$, many grain monolayers).
Here,
\[      n^* = A n, \mbox{ }   A=\frac{S}{a^2} \leq 1,
\]
is the effective carrier density of the system,
which differs from the actual carrier density $n$
inside the grains only by a numerical factor $A$ determined by the shape of the grains
($A=\pi/4$ for spherical and $A=1$ for cubic grains).
For a two-dimensional sample
(2D, $d=2$, one or a few grain monolayers) the expression (\ref{eq:rhoxy0}) must
divided by the thickness $d_z$ of the sample or, equivalently,
$n^*=d_z A n$ in this case\footnote{ Note that
although the granular array may be two- ($d=2$) or
three-dimensional ($d=3$), the grains themselves are
three-dimensional, and $n$ is a three-dimensional density.}.

The result (\ref{eq:rhoxy0}) for the Hall resistivity $\rho^{(0)}_{xy}$
is quite {\em universal}. It is valid even if
(i) the tunneling conductances $G_T$ fluctuate from contact to contact
and
(ii) the mean free path $l$ fluctuates from grain to grain:
HR is simply independent of the distributions of $G_T$ and $l$;
%
Therefore, Eq.~(\ref{eq:rhoxy0}) is applicable
to real granular arrays
in which such irregularities are always present
(provided such system is still in the metallic regime).
We also note that although Eq.~(\ref{eq:rhoxy00}) was obtained for a regular
quadratic/cubic granular lattice,
the result (\ref{eq:rhoxy0})
with a different factor $A \leq 1$
remains valid for other regular lattices
(e.g., more common for real experimental samples triangular lattice).
We also expect Eq.~(\ref{eq:rhoxy0}) to hold for arrays
with moderate structural disorder,
i.e., in which the positions of the grains deviate
from regular 
and their sizes and shapes are not identical.


{\em Coulomb interaction corrections.}
Next, we calculate the first-order in $1/g_T$ corrections to HC
$\sig^{(0)}_{xy}$ [Eq.~(\ref{eq:sigxy0})] due to
Coulomb interaction.
%
%
We find significant
contributions for temperatures $T \lesssim g_T E_c$
not exceeding the inverse $RC$ time $g_T E_c$ of the system ``grain+contact''
[$E_c=e^2/(\kappa a)$ is the charging energy and
$\kappa$ is the dielectric constant of the array],
whereas for
$T\gtrsim g_T E_c$ the relative corrections are of the order of
$1/g_T$ or smaller.
%
Three types of corrections to HC $\sig^{(0)}_{xy}$ [Eq.~(\ref{eq:sigxy0})]
can be identified:
\beq
    \sig_{xy}=\sig^{(0)}_{xy}+\de\sig_{xy}^{TA}+\de\sig_{xy}^{VD}+\de\sig_{xy}^{AA}
    \label{eq:sigxy}.
\eeq
The first correction $\de\sig_{xy}^{TA}$ can be attributed to
the renormalization of the individual tunneling conductances $G_T$
[{\em tunneling anomaly} (TA) \cite{AA,TA1,TA2}] in the granular medium
and has the form
\beq
    \frac{\de\sig^{TA}_{xy}(T)}{\sig^{(0)}_{xy}} =-\frac{1}{\pi g_T d}
        \ln \lt[ \frac{g_T E_c}{\max(T,\Ga)} \rt] \mbox{ for } T \lesssim g_T E_c.
\label{eq:dsigxyTA}
\eeq
This correction renormalizes the tunneling conductances $G_T$
in Eq.~(\ref{eq:sigxy0}), but does not affect the Hall resistance $R_H$ of the grain.

The second correction $\de\sig_{xy}^{VD}$ corresponds
to the process of {\em virtual diffusion} (VD) of electrons through the grain
and equals
\beq
    \frac{\de\sig^{VD}_{xy}(T)}{\sig^{(0)}_{xy}} =\frac{c_d}{4 \pi g_T }
        \ln \lt[\frac{\min(g_T E_c,\ETh)}{\max(T,\Ga)}\rt]
\label{eq:dsigxyVD}
\eeq
for $T \lesssim \min(g_T E_c,\ETh)$,
where $c_d \sim 1$ is a numerical lattice structure factor~(\ref{eq:cd}).
Contrary to $\de\sig_{xy}^{TA}$, the correction $\de\sig_{xy}^{VD}$
is suppressed at temperatures greater than the Thouless energy of the grain $\ETh$,
which emphasizes its diffusion character.
%

For $T\gtrsim \Ga$
both corrections
$\de\sig_{xy}^{TA}$ and  $\de\sig_{xy}^{VD}$
are $\ln T$-dependent.
This dependence saturates at temperatures $T \sim \Ga$,
so that
$\de\sig_{xy}^{TA}$ and  $\de\sig_{xy}^{VD}$
remain logarithmically large constants at $T \lesssim \Ga$.
These two corrections are specific for granular systems
and, in essence,
due to the strong discrepancy of timescales of
the intra- ($\ETh^{-1}$) and intergrain ($\Ga^{-1}$) electron dynamics described by Eq.~(\ref{eq:granular1}).
They
arise from spatial scales of the order of the grain size $a$ and
are absent in HDMs. The logarithmic behavior of the corrections
(\ref{eq:dsigxyTA}) and (\ref{eq:dsigxyVD}) is due to the form of
the screened Coulomb interaction in granular systems\cite{BEAH}.
They have the same logarithmic form in 2D and 3D, but the
coefficients are not universal and lattice-dependent: $1/d$ and
$c_d$ [Eq.~(\ref{eq:cd})] are the  results for the cubic (3D) or
quadratic (2D) lattice, which we assumed in our calculations.

The third correction $\de\sig_{xy}^{AA}$ in Eq.~(\ref{eq:sigxy}) is analogous
to the one present in homogeneously disordered metals\cite{AA}
[``Altshuler-Aronov'' (AA) corrections].
It can be significant at low temperatures $T\ll\Ga$ only,
when the thermal length $L_T= \sqrt{D_0^*/T} \gg a$
for the intergrain motion exceeds the grain size $a$
($D_0^*=\Ga a^2$ is the effective diffusion coefficient for the intergrain electron motion
at scales greater than $a$).
However, we find that this correction
{\em vanishes identically}
both in 2D and 3D:
\beq
    \de\sig_{xy}^{AA} =0.
\label{eq:dsigxyAA}
\eeq
It is always instructive to compare the
results for a granular metal with those for a HDM.
The quantities arising from large spatial scales
(exceeding the grain size $a$ for a granular metal
and the mean free path $l$ for a HDM)
are expected to behave universally,
because at such scales the microscopic structure of the system becomes irrelevant.
Indeed, the result (\ref{eq:dsigxyAA}) for $\de\sig_{xy}^{AA}$
agrees with the one for HDMs first obtained in Ref.~\onlinecite{AKLL}.
Being an exact cancellation, however, Eq.~(\ref{eq:dsigxyAA})
is valid not only for low $T \ll \Ga$, but for arbitrary relevant temperatures.

The quantity directly measured in experiments is
the Hall resistivity
\beq
    \rho_{xy}=\frac{\sig_{xy}}{\sig_{xx}^2} =\rho_{xy}^{(0)}+\de\rho_{xy},
\label{eq:rhoxy}
\eeq
where $\rho_{xy}^{(0)}$ is the ``bare'' HR [Eq.~(\ref{eq:rhoxy0})],
$\de\rho_{xy}$ is the total Coulomb interaction correction to HR
and HC $\sig_{xy}$ is given by Eq.~(\ref{eq:sigxy}).
The interaction corrections to LC
were studied in Refs.~\onlinecite{ET,BELV}
and the following result was obtained:
\beq
    \sig_{xx}=\sig^{(0)}_{xx}+\de\sig_{xx}^{TA}+\de\sig_{xx}^{AA}
\label{eq:sigxx}
\eeq
[$\de\sig_{xx}^{TA}$ and $\de\sig_{xx}^{AA}$
correspond to $\de\sig_1$ [Eq.~(2b)] and $\de \sig_2$ [Eq.(2c)]
in Ref.~\onlinecite{BELV}, respectively].
The correction $\de\sig_{xx}^{TA}$
is due to the tunneling anomaly in granular metal. It renormalizes the
tunneling conductance $G_T$ in Eq.~(\ref{eq:sigxx0}) and
equals
\beq
    \frac{\de\sig^{TA}_{xx}(T)}{\sig^{(0)}_{xx}} =-\frac{1}{2\pi g_T d}
        \ln \lt[ \frac{g_T E_c}{\max(T,\Ga)} \rt] \mbox{ for } T \lesssim g_T E_c.
\label{eq:dsigxxTA}
\eeq
This correction is of {\em local} origin and governs the temperature dependence
of LC $\sig_{xx}(T)$ in a wide temperature range.
The Hall counterpart of $\de\sig_{xx}^{TA}$  is $\de\sig_{xy}^{TA}$ [Eq.~(\ref{eq:dsigxyTA})].

The correction $\de\sig_{xx}^{AA}$ is analogous to that in a HDM,
first obtained by Altshuler and Aronov (AA) in
Ref.~\onlinecite{AAjetp} and its Hall counterpart is
$\de\sig_{xy}^{AA}$ [Eq.~(\ref{eq:dsigxyAA})]. The AA correction
$\de\sig_{xx}^{AA}$ does not diverge at large spatial scales in 3D
case, being smaller than the logarithmic contributions
(\ref{eq:dsigxyTA}), (\ref{eq:dsigxyVD}), (\ref{eq:dsigxxTA}) for
all relevant temperatures down to very low ones\cite{BELV}:
$\de\sig_{xx}^{AA}/\sig_{xx}^{(0)} \lesssim 1/g_T$.

In 2D case\footnote{
Dealing with the Hall transport, we do not discuss one-dimensional case of granular ``wires''
in this paper,
for which the ``Altshuler-Aronov'' correction $\de\sig_{xx}^{AA}$ is also divergent.},
i.e. for granular films of thickness $d_z$ consisting of one or a few grain monolayers
($d_z/a$ is the number of monolayers),
the correction $\de\sig_{xx}^{AA}$ is diverging at large spatial scales.
This divergence is relevant for low temperatures $T \ll \Ga (a/d_z)^2$ (when $L_T^* \gg d_z$),
for which $\de\sig_{xx}^{AA}$ acquires a logarithmic dependence\cite{BELV}:
\beq
    \frac{\de\sig^{AA}_{xx}(T)}{\sig^{(0)}_{xx}}=
    -\frac{1}{4 \pi^2 g_T} \ln  \lt[ \frac{\Ga}{T} \lt(\frac{a}{d_z}\rt)^2 \rt],
\label{eq:dsigxxAA}
\eeq


The total Coulomb interaction  correction
\beq
    \de\rho_{xy}=
\de\rho_{xy}^{TA}+\de\rho_{xy}^{VD}+\de\rho_{xy}^{AA}
\label{eq:drhoxy}
\eeq
to Hall resistivity (\ref{eq:rhoxy}) is given by the sum of the contributions arising from
the corresponding corrections to Hall (\ref{eq:sigxy}) and longitudinal  (\ref{eq:sigxx})  conductivities
as
\[
    \frac{\de\rho_{xy}^{(i)}}{\rho_{xy}^{(0)}} =\frac{\de\sig_{xy}^{(i)}}{\sig^{(0)}_{xy}}-
    2 \frac{\de\sig_{xx}^{(i)}}{\sig^{(0)}_{xx}}, \mbox{ } (i)=TA,VD,AA.
\]

Since the TA effects leads to the renormalization of the tunneling
conductance $G_T$ only,
it cannot affect the HR
$\rho^{(0)}_{xy}$ [Eq.~(\ref{eq:rhoxy0})], which does
not contain $G_T$.
Indeed, it follows from Eqs.~(\ref{eq:dsigxyTA}) and  (\ref{eq:dsigxxTA})
that
\[
    \frac{\de\sig^{TA}_{xy}}{\sig^{(0)}_{xy}} =  2 \frac{\de\sig^{TA}_{xx}}{\sig^{(0)}_{xx}}
\]
and, therefore, the correction to Hall resistivity from the ``tunneling anomaly''
effect vanishes:
\beq
    \de \rho_{xy}^{TA}=0.
\label{eq:drhoxyTA}
\eeq
Further, since the ``virtual diffusion'' correction is absent for longitudinal conductivity
[Eq.~(\ref{eq:sigxx})]\footnote{Roughly, the reason is that diagram for the bare
LC $\sig_{xx}^{(0)}$ [Eq.~(\ref{eq:sigxx0}), Fig.~\ref{fig:eloops}(d)]
does not contain diffusons, contrary
to the diagrams for HC $\sig_{xy}^{(0)}$ (Figs.~\ref{fig:Hall1},\ref{fig:Hall}), and consequently, the diagrams
for the Coulomb interaction corrections
describing the ``virtual diffusion'' process do not arise.
See also the footnote on page \pageref{VDpage}.}
we obtain
\beq
    \frac{\de\rho_{xy}^{VD}(T)}{\rho_{xy}^{(0)}} =\frac{\de\sig^{VD}_{xy}(T)}{\sig^{(0)}_{xy}},
\label{eq:drhoxyVD}
\eeq
where $\de\sig^{VD}_{xy}(T)$ is given by Eq.~(\ref{eq:dsigxyVD}).
Finally, since the ``Altshuler-Aronov''-type correction $\de\sig_{xy}^{AA}$
to Hall conductivity vanishes [Eq.~(\ref{eq:dsigxyAA})], we get
\beq
    \frac{\de\rho_{xy}^{AA}(T)}{\rho_{xy}^{(0)}}
    =-2\frac{\de\sig^{AA}_{xx}(T)}{\sig^{(0)}_{xx}}.
\label{eq:drhoxyAA}
\eeq
%
Therefore, according to Eqs.~(\ref{eq:drhoxy}), (\ref{eq:drhoxyTA}), (\ref{eq:drhoxyVD}),
and (\ref{eq:drhoxyAA}),
the total Coulomb interaction correction $\de\rho_{xy}$ to Hall resistivity
is
\beq
    \frac{\de\rho_{xy}(T)}{\rho_{xy}^{(0)}} =\frac{\de\sig^{VD}_{xy}(T)}{\sig^{(0)}_{xy}}-
    2 \frac{\de\sig^{AA}_{xx}(T)}{\sig^{(0)}_{xx}},
\label{eq:drhoxy2}
\eeq
where $\de\sig^{VD}_{xy}$ is given by Eq.~(\ref{eq:dsigxyVD})
and $\de\sig^{AA}_{xx}$ [Eq.~(\ref{eq:dsigxxAA})] can be significant
in 2D at $T\ll \Ga (a/d_z)^2$ only.

Equations~(\ref{eq:dsigxyTA})-(\ref{eq:dsigxyAA}),
(\ref{eq:drhoxy})-(\ref{eq:drhoxy2})
constitute our main result for Coulomb interaction corrections
to Hall conductivity and resistivity.
Another effect occurring at similar temperatures is weak
localization (WL). The WL corrections to LC of a granular metal
were studied in Refs.~\onlinecite{BelUD,BVG,BCTV}.
Significant (logarithmic) contributions may arise in 2D samples only,
from spatial scales greater than the grain size $a$, when the inverse dephasing time
$1/\tau_\phi \lesssim \Ga$ (if $1/\tau_\phi \propto T/g_T$
\cite{BelUD,BVG}, this corresponds to $T \lesssim g_T \Ga$).
However, we find \cite{KEprepWL} that the first-order in $1/g_T$
WL correction to Hall resistivity {\em vanishes identically} both in 2D and 3D
in correspondence with the result for HDMs \cite{Fukuyama, AKLL,
Khodas}:
\beq
    \de\rho_{xy}^{WL} = 0.
\label{eq:drhoxyWL}
\eeq
Therefore weak localization does not affect Hall resistivity
at least in the first order in $1/g_T$, in which we obtain significant
contributions from the Coulomb interaction.

\begin{figure}
\includegraphics[width=0.47\textwidth]{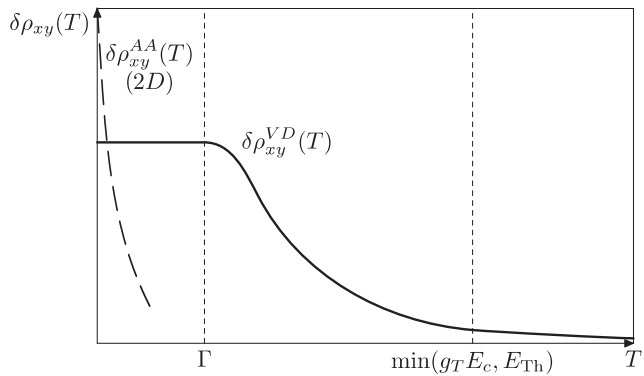}
\caption{ Temperature dependence of the  Coulomb interaction
corrections $\de\rho_{xy}^{VD}(T)$ [Eq.~(\ref{eq:drhoxyVD})]
and $\de\rho_{xy}^{AA}(T)$ [Eq.~(\ref{eq:drhoxyAA})]
to the Hall resistivity [second and third terms in Eq.~(\ref{eq:rhoxyfinal}), respectively].
The most significant contribution in a wide range of
temperatures [$T\lesssim \min(g_T E_c, \ETh)$] both for two- and
three-dimensional samples comes from the correction
$\de\rho_{xy}^{VD}(T)$ (thick solid line), which is
due to the process of ``virtual diffusion'' of electrons through a single grain.
The contribution $\de\rho_{xy}^{VD}(T)$ is of {\em local} origin and absent in homogeneously disordered metals.
It depends
logarithmically on temperature $T$ in the range $\Ga \lesssim
T\lesssim \min(g_T E_c, \ETh)$, saturating at $T\sim \Ga$ and
remaining constant for $T \lesssim \Ga$. The correction
$\de\rho_{xy}^{AA}(T)$ is analogous to the ``Altshuler-Aronov''
correction in ordinary disordered metals, it can be significant
[Eq.~(\ref{eq:dsigxxAA})] for sufficiently thin granular films and
low enough temperatures [$T\ll \Ga (a/d_z)^2$] only, the latter 2D
case shown in figure (dashed line).
%
}
\label{fig:drhoxy}
\end{figure}

{\em Summarizing} our findings for the classical result (\ref{eq:rhoxy0}),
Coulomb interaction
[Eqs.~(\ref{eq:dsigxyTA})-(\ref{eq:dsigxyAA})
(\ref{eq:drhoxy})-(\ref{eq:drhoxy2})]
and weak localization (\ref{eq:drhoxyWL}) corrections,
we predict the following behavior of the Hall resistivity
of a granular metal:
%
\[
    \rho_{xy}(T)= \rho_{xy}^{(0)}+\de\rho_{xy}^{TA}+\de\rho_{xy}^{VD}
        +\de\rho_{xy}^{AA}+\de\rho_{xy}^{WL}=
\]
\beq
= \frac{H}{n^* e c}
    \lt(1+\frac{c_d}{4 \pi g_T} \ln \lt[\frac{\min(g_T E_c,\ETh)}{\max(T,\Ga)}\rt]
        -2 \frac{\de \sig_{xx}^{AA}(T)}{\sig_{xx}^{(0)}} \rt).
\label{eq:rhoxyfinal}
\eeq

(i) At high enough temperatures $T\gtrsim \min(g_T E_c,\ETh)$
the Hall resistivity
$ \rho_{xy}(T)=\rho_{xy}^{(0)}$
is given by the Drude-like expression (\ref{eq:rhoxy0}) [first term in Eq.~(\ref{eq:rhoxyfinal})]
and is independent  of both the intragrain and
tunnel contact disorder.
Measuring $\rho_{xy}$  at such $T$
and using Eq.~(\ref{eq:rhoxy0}) one can extract an important characteristic
of the granular system: its effective carrier density $n^*=A n$, which is related
to the actual carrier density $n$ of the grain material through a geometry-dependent factor $A \leq 1$.

(ii) In a wide temperature range $\Ga \lesssim T \lesssim \min(g_T E_c,\ETh)$
both for 2D and 3D samples,
{\em local} effects of Coulomb interaction lead to the logarithmic
in $T$ correction
to the Hall resistivity
[$\de\rho_{xy}^{VD}(T)$, second term in Eq.~(\ref{eq:rhoxyfinal}),
see Eqs.~(\ref{eq:dsigxyVD}),(\ref{eq:drhoxyVD})].
This $\ln T$-dependence saturates at  $T \sim \Ga$ and
$\de\rho_{xy}^{VD}(T)$ remains constant for  $T \lesssim \Ga$.
We emphasize that this correction is absent in homogeneously disordered metals, but
it appears to be the major quantum correction to the Hall resistivity of a granular metal
that governs the $T$-dependence of $\rho_{xy}(T)$
in a wide temperature range both for 2D and 3D samples.

(iii) An additional $T$-dependence  of $ \rho_{xy}(T)$
may arise
due to the ``Altshuler-Aronov'' correction $\de\sig_{xx}^{AA}(T)$ to the
longitudinal conductivity
[$\de\rho_{xy}^{AA}(T)$, third term in Eq.~(\ref{eq:rhoxyfinal}),
see Eqs.~
(\ref{eq:dsigxyAA}),(\ref{eq:dsigxxAA}),(\ref{eq:drhoxyAA})]
at much lower temperatures $T \ll \Ga$,
the most significant logarithmic contribution
expected for sufficiently thin granular films
[one or a few grain monolayers, Eq.~(\ref{eq:dsigxxAA})].

The temperature behavior of the contributions
$\de\rho_{xy}^{VD}(T)$
and $\de\rho_{xy}^{AA}(T)$
is shown in Fig.~\ref{fig:drhoxy}.

We expect our result Eq.~(\ref{eq:rhoxyfinal}) to hold for
realistic arrays with moderate structural disorder
and, most importantly, with randomly distributed
tunneling conductances,
which is inevitable in real systems:
(i) $n^*$ simply does not depend on the  distribution of $G_T$;
(ii) the logarithmic form of the major quantum correction
[$\de\rho_{xy}^{VD}(T)$, second term in Eq.~(\ref{eq:rhoxyfinal})] persists in this case,
although the structure factor $c_d \sim 1$   does depend on
the distribution of conductances and $g_T$ should be substituted by some
averaged quantity.

%

Comparison of our findings
with experimental data may serve as a good check of the theory developed here.
The experimental situation related to our theory is mentioned in the Conclusion.

%
%

\section{\label{sec:model}Model and Hamiltonian}
%
We consider a quadratic ($d=2$, 2D) or cubic ($d=3$, 3D) lattice of
metallic grains
coupled to each other by tunnel contacts (Fig.~\ref{fig:system}).


Aiming to concentrate on the Hall effect,
we assume the
simplest
case of translationally invariant lattice,
i.e., equal tunneling conductances $G_T$ of all contacts,
translationally invariant capacitance matrix,
and identical properties of all grains
(the same form and size, mean free path, electron density, density of states, etc.).
After the main properties of the Hall effect in such system have been established,
we  argue that
they also hold for realistic arrays.
%
In real systems in the metallic regime,
the major  type of irregularities
that (could) affect electron transport
even for structurally quite regular arrays
seems to be the randomness of tunneling conductances $G_T$,
while other assumptions can be well met or are inessential.

To provide more explicit analysis
we further simplify the calculations
assuming the intragrain electron dynamics diffusive, i.e.,
that the bulk mean free path $l$ in the grain is much smaller than
the size $a$ of the grain, $l \ll a$.
In this case details of electron scattering off
the grain boundary are irrelevant.
However, our approach
is also entirely applicable to the case of ballistic ($l \gtrsim a$)
intragrain disorder, when surface scattering becomes important.
The main results, listed in Sec.~\ref{sec:results},
are valid for both diffusive and ballistic grains.

In the metallic regime ($g_T \gg 1$)
quantum effects of Coulomb interaction
can be considered perturbatively
with an expansion parameter $1/g_T$ as long as the relative corrections remain small.

We write the Hamiltonian describing the system as
\beq
    \Hh = \Hh_0 + \Hh_t + \Hh_c.
\label{eq:H}
\eeq

In Eq.~(\ref{eq:H}) the first term
\beq
    \Hh_0= \sum_\ib \int d\rb_\ib \psi^\dg(\rb_\ib)
    \lt[ \xi\lt(\pb_\ib-\frac{e}{c} \Ab(\rb_\ib)\rt)+ U(\rb_\ib) \rt] \psi(\rb_\ib)
\label{eq:H0}
\eeq
is the Hamiltonian of isolated grains,
$\xi(\pb)=\pb^2/(2m)-\epsilon_F$,
$\Ab(\rb_\ib)$ is the vector potential describing the uniform magnetic field $\Hb=H\eb_z$
directed along the $z$ axis, $U(\rb_\ib)$ is the random disorder potential of the grains,
$\ib=(i_1, \ldots , i_d)~\in~\mathbb{Z}^d$ is an integer vector numerating the grains.
The integration with respect to $\rb_\ib$ is done over the volume of the grain $\ib$.
Since we do not deal with spin-related phenomena in this paper,
we omit the spin indices of the operators $\psi(\rb_\ib)
$.
Accounting for spin degeneracy in the course of calculations
is simple: each electron loop comes with the factor 2.
We consider  white-noise disorder and perform averaging
using the Gaussian distribution with the variance
\beq
    \lan U(\rb_\ib) U(\rb'_\ib) \ran_U = \frac{1}{2 \pi \nu \tau_0} \de(\rb_\ib-\rb'_\ib),
\label{eq:UU}
\eeq
where $\nu$ is the density of states in the grain
at the Fermi level per one spin projection and
$\tau_0$ is the scattering time.

The tunneling Hamiltonian $\Hh_t$ in Eq.~(\ref{eq:H}) is given by
\beq
    \Hh_t= \sum_{\lan \ib,\jb \ran} (X_{\ib,\jb}+X_{\jb,\ib}),
\label{eq:Ht}
\eeq
where the operator $X_{\ib,\jb}$ describes  tunneling
from the grain $\jb$ to the grain $\ib$,
the summation is taken over the neighboring grains connected by a
tunnel contact, such that each contact is counted only once.
%
For studying Hall effect
the geometry of the grains and contacts is essential,
therefore we write the tunneling operators $X_{\ib,\jb}$
in the coordinate representation:
\beq
    X_{\ib,\jb}=\int
        d\s_\ib d\s_\jb \,
        t(\s_\ib, \s_\jb) \psi^\dg(\s_\ib) \psi(\s_\jb),
\label{eq:X}
\eeq
where the integration is carried out over two surfaces of
the contact: one of them ($\s_\ib$) belonging to the $\ib$-th grain,
whereas the other ($\s_\jb$) to the $\jb$-th grain.
Such form implies that tunneling occurs
from
a close  vicinity of the contact of atomic size, but not from the bulk of the grain.
This is a natural assumption,
since we consider a good metallic limit for the grains, i.e.,
the size $a$ of the grains is much greater than the Fermi length,
$p_F a \gg 1$ ($p_F$ is the Fermi momentum).
Fast oscillations
of the wave functions in the grains result
in a fast decay of the overlap of the wave functions
in different grains.
Since $\Hh_t^\dg=\Hh_t$, we have $X_{\ib,\jb}^\dg=X_{\jb,\ib}$ and
$t^*(\sbf_\ib, \sbf_\jb) =t(\sbf_\jb, \sbf_\ib)$.

Without further assumptions about the tunneling amplitudes
$t(\s_\ib, \s_\jb)$ in Eq.~(\ref{eq:X}),
electrons  can tunnel from a given point $\s_\jb$ to an arbitrary
point $\s_\ib$ on the other side of the contact.
However, it is  physically clear
that
(i) electrons effectively
tunnel from the point
$\s_\jb$ to the points $\s_\ib$ in the vicinity of $\s_\jb$ of atomic size only,
therefore $t(\s_\ib, \s_\jb)$  should decay rapidly on atomic scale
as a function of 
$\s_\ib-\s_\jb$; (ii) the amplitude $t(\s_\ib, \s_\jb)$ can also fluctuate
as a function of $\s_\ib$ for fixed $\s_\ib-\s_\jb$
due to irregularities of the contact
on atomic scale.

To effectively model this behavior of the tunneling amplitudes
we consider $
t(\sbf_\ib, \sbf_\jb)$ as Gaussian random variables
and average over them with the variance
\beq
    \lan t(\sbf_\ib, \sbf_\jb) t(\sbf_\jb, \sbf_\ib) \ran_t = t_0^2 \de(\sbf_\ib-\sbf_\jb),
\label{eq:tt}
\eeq
where $\de(\sbf_\ib-\sbf_\jb)$ is an atomic scale
$\de$-function on the contact surface and
$t_0^2$ has a meaning of tunneling probability
per unit area of the contact.
As we will see, the assumption $p_F a \gg 1$
will enable us to neglect the contributions containing the regular parts
$ \lan t(\s_\ib,\s_\jb) \ran_{t}$ of the tunneling amplitudes.
%

The third term in Eq.~(\ref{eq:H}) stands
for the Coulomb interaction between the electrons.
In principle, one has to start with its the bare form
\beq
    \Hh_c= \frac{1}{2} \sum_{ \ib,\jb } \int d\rb_\ib d\rb_\jb\,
    \psi^\dg(\rb_\ib)\psi^\dg(\rb_\jb)
        \frac{e^2}{|\rb_\ib-\rb_\jb|}
    \psi(\rb_\jb)\psi(\rb_\ib).
\label{eq:Hc}
\eeq
Proceeding with the calculations we will have to take the screening
of Coulomb potential by electron motion into account.
One should distinguish between the {\em intragrain}
and {\em intergrain} electron motion.
%
In the static limit (classical electrostatics) for the intragrain motion
the Coulomb interaction is reduced to the effective charging energy $E_{\ib\jb}$ interaction
between the total excess charges of the grains.
Accounting for tunneling yields
the screened form of the charging energy interaction\cite{BEAH},
which is sufficient for  studying the {\em intergrain} transport.
We will see, however, that
coordinate-dependent interaction modes inside each grain
arising from the {\em intragrain} motion
will be necessary to get a correct frequency dependence
of the  classical  Hall resistance $R_H$ of a single grain.

\subsection{Kubo formula}
The conductivity in a homogeneous external electric field
is calculated using the Kubo formula\cite{ET} in Matsubara technique\cite{AGD}:
\beq
    \sig_{\ab\bb}(\om)= 2e^2 a^{2-d}\frac{1}{|\om|}
        \lt[ \Pi_{\ab\bb}(\om)-\Pi_{\ab\bb}(0) \rt]
\label{eq:sig}
\eeq
\beq
    \Pi_{\ab\bb}(\om)= \sum_\jb \Pi_{\ab\bb}(\om, \ib-\jb),
\label{eq:Pi}
\eeq
where
\beq
    \Pi_{\ab\bb}(\om,\ib-\jb)=\int_0^{1/T} d\tau\,e^{i \om \tau}
    \lan T_\tau I_{\ib,\ab}(\tau) I_{\jb,\bb}(0) \ran
\label{eq:Pi2}
\eeq
is the current-current correlation function,
\beq
    I_{\ib,\ab}(\tau) =X_{\ib+\ab,\ib}(\tau)-X_{\ib,\ib+\ab}(\tau).
\label{eq:I}
\eeq
Here $\om \in 2\pi T \mathbb{Z}  $
is an external bosonic Matsubara frequency ($\mathbb{Z}$ is the set of integers),
$\ab$ and $\bb$ are the lattice unit vectors.
The factor $2$ in Eq.~(\ref{eq:sig})
stands for the spin degeneracy coming from one electron loop.
The vector $\ab$ denotes the direction of the current component
and $\bb$ points along the external electric field that causes the current.
For example, if $\bb=\eb_y$, then $\ab=\eb_x$ for Hall conductivity
$\sig_{xy}=\sig_{\eb_x\eb_y}$ and $\ab=\eb_y$  for longitudinal conductivity
$\sig_{yy}=\sig_{\eb_y\eb_y}$.
Further,
$A(\tau)=e^{\Hh\tau}A e^{-\Hh \tau}$ is the Heisenberg operator in Matsubara technique.
The operator of the tunneling current through the contact
connecting the grains $\ib$ and $\ib+\ab$ actually equals $-ie
I_{\ib,\ab}(\tau)$,
we extracted  $(-ie)^2$ from $\Pi_{\ab\bb}(\om,\ib-\jb)$
for further convenience.
The average $\lan\ldots\ran$  in Eq.~(\ref{eq:Pi2}) implies
both the quantum mechanical thermodynamic averaging with respect to $\Hh$
and averaging over the intragrain and contact disorder
according to Eqs.~(\ref{eq:UU}) and (\ref{eq:tt}).
The contact between the neighboring grains $\ib+\ab$ and $\ib$
will be further identified by the pair $(\ib+\ab,\ib)$.

The correlation function $\Pi_{\ab\bb}(\om, \ib-\jb)$ represents the current
running through the contact $(\ib+\ab,\ib)$ in response to the voltage applied
to the contact $(\jb+\bb,\jb)$ only. The sum over $\jb$ in
Eq.~(\ref{eq:Pi}) means that the contributions from all contacts
have to be considered.


\section{\label{sec:tech}Technique}

The current-current correlation function $\Pi_{\ab\bb}(\om, \ib-\jb)$ is calculated
using diagrammatic technique.
Let us first discuss its details neglecting the Coulomb interaction
$\Hh_c$ [Eq.~(\ref{eq:Hc})] in $\Hh$ [Eq.~(\ref{eq:H})] completely.
Technically,
for a given pair $(\ib+\ab,\ib)$ and $(\jb+\bb,\jb)$ of contacts
one expands Eq.~(\ref{eq:Pi2})
both in the disorder potential $U( \rb_\ib)$ [Eq.~(\ref{eq:H0})]
and tunnelling Hamiltonian $\Hh_t$ [Eqs.~(\ref{eq:Ht}) and (\ref{eq:X})].
Each diagrammatic contribution to $\Pi_{\ab\bb}(\om, \ib-\jb)$
is a loop of two Green functions
%
connecting the contacts $(\ib+\ab,\ib)$ and $(\jb+\bb,\jb)$.
Then one averages this loop over the intragrain and contact disorder
according to Eqs.~(\ref{eq:UU}) and (\ref{eq:tt}).

\begin{figure}
\includegraphics{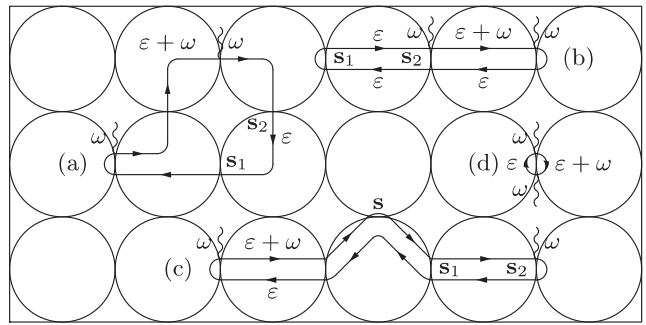}
\caption{Different types of diagrams for the current-current correlation function
$\Pi_{\ab\bb}(\om, \ib-\jb)$ [Eq.~(\ref{eq:Pi2})] neglecting
Coulomb interaction [Eq.~(\ref{eq:Hc})].
Diagrams of types (a) and (b)
that contain oscillating at Fermi wavelength $\la_F=p_F^{-1}$
functions in coordinate representation vanish after the integration of the contacts surfaces
give 0.
In diagram (a) two different contacts are connected by a single Green function
$\Gc(\eps, \s_1, \s_2)$;
in diagram (b) two different contacts are connected by two Green functions
$\Gc(\eps, \s_1, \s_2)$ and $\Gc(\eps, \s_2, \s_1)$
having the same (sign of) energies.
(c) The only type of ``allowed'' diagrams that do not contain oscillating functions
and give nonvanishing contributions:
the two contacts $(\ib+\ab,\ib)$ and $(\jb+\bb,\jb)$
with external tunneling vertices (wavy lines) are
``capped'' by the Green functions $G(\eps,\s,\s)$
from  one of their sides and
connected by two Green functions
$\Gc(\eps+\om,\s_{\ib+\ab},\s_\jb)$ and
$\Gc(\eps,\s_\jb,\s_{\ib+\ab})$, the ``paths'' of which through
other contacts {\em coincide}. For energies, such that
$(\eps+\om)\eps <0$, the diffuson $D$ [Eq.~(\ref{eq:Ddef})] in
each grain along this path arises.
(d) Diagram for the longitudinal conductivity $\sig_{xx}^{(0)}=a^{2-d} G_T$ [Eq.~(\ref{eq:sigxx0})]
in the leading order in $g_T/g_0 \ll 1$.}
\label{fig:eloops}
\end{figure}

Of course, many different possibilities of drawing such loop can be considered
(see Fig.~\ref{fig:eloops}).
However due to the general properties of the Green functions
in the coordinate representation and the assumption employed
in Eq.~(\ref{eq:X}) that tunneling
occurs from the vicinity of the contacts, but not from the bulk of the grain,
a lot of them can be ruled out even before averaging over $U(\rb)$.

Consider a Matsubara Green function $\Gc(\eps,\rb,\rb')$ of an arbitrary grain
for a given realization of the disorder potential $U(\rb)$.
The Green function
 $\Gc(\eps,\rb,\rb')\propto e^{i p_F |\rb-\rb'| \sgn \eps}$
oscillates
 at the Fermi wavelength $\la_F=p_F^{-1}$ as a function
of the difference $\rb-\rb'$.
Since we assume the grain size $a$ and the size of the area of the contact
much greater than $\la_F$, this fact excludes the following possibilities.

(i) If two different contacts  $\s_1$ and $\s_2$
are connected by a single Green function $\Gc(\eps,\s_1,\s_2)$ in a given
grain (Fig.~\ref{fig:eloops}a),
then integration over the contacts surfaces
$\int d\s_1 d\s_2 \Gc(\eps,\s_1,\s_2)$ gives zero
due to the rapid oscillations of the integrand.

(ii)
If two different contacts  $\s_1$ and $\s_2$
are connected by two Green functions $\Gc(\eps,\s_1,\s_2)$
and $ \Gc(\eps',\s_2,\s_1)$ (or $\Gc(\eps',\s_1,\s_2)$)
in a given grain having the same signs of energies, $\eps\eps'>0$,
(Fig.~\ref{fig:eloops}b)
then, again, their product is an oscillating function
and
$\int d\s_1 d\s_2 \Gc(\eps,\s_1,\s_2) \Gc(\eps',\s_2,\s_1)$ also gives zero.

So,
the  only objects of the diagrammatic technique that ``survive'' inside the grains are those
that do not contain oscillations at the Fermi wavelength $\la_F$
in their coordinate dependence (Fig.\ref{fig:eloops}c).
These are:
(1) the single Green  function
$\Gc(\eps,\s,\s)$
with coinciding coordinates $\s$ on the contact surface;
(2) the product of two Green functions
with pairwise coinciding coordinates and opposite signs of energies:
$\Gc(\eps,\s_1,\s_2) \Gc(\eps',\s_2,\s_1)$ or
$\Gc(\eps,\s_1,\s_2) \Gc(\eps',\s_1,\s_2)$ with $\eps\eps'<0$.
After disorder averaging such products of two
Green function give well-known electron propagators
for a single isolated grain:
the {\em ``diffuson''}
\footnote{
Although we term the propagator $D(\om,\rb,\rb')$ defined by Eq.~(\ref{eq:Ddef})
as ``diffuson'', there is yet no need to assume the diffusive limit ($l \ll a$)
for these general considerations and they are valid in the ballistic case ($l \gtrsim a$)
as well.
}
\beq
    D(\om,\rb,\rb') \equiv \frac{1}{2 \pi \nu} \lan \Gc(\eps+\om,\rb,\rb')
            \Gc(\eps,\rb',\rb)\ran_U,
            \mbox{ } (\eps+\om)\eps < 0.
\label{eq:Ddef}
\eeq
and the {\em ``Cooperon''}
\[
    C(\om,\rb,\rb') \equiv \frac{1}{2 \pi \nu} \lan \Gc(\eps+\om,\rb,\rb') \Gc(\eps,\rb,\rb')\ran_U,
            \mbox{ } (\eps+\om)\eps < 0.
\]
Cooperons will be important for weak localization effects \cite{KEprepWL},
while in this paper we only consider the diffuson~(\ref{eq:Ddef})
in detail.

We are left with the following general type of
diagram
(in the absence of Coulomb interaction and weak localization effects)
for $\Pi_{\ab\bb}(\om, \ib-\jb)$ shown in Fig.\ref{fig:eloops}c:
(1)
each contact $(\ib+\ab,\ib)$ and $(\jb+\bb,\jb)$
with external tunneling vertices must be ``capped''
by the Green function $G(\eps,\s,\s)$  from  one of its sides
(one cannot ``construct'' a diffuson from $G(\eps,\s,\s)$, since only one energy $\eps$
is available, see Fig.~\ref{fig:eloops}b);
(2)
two Green functions
$\Gc(\eps+\om,\s_{\ib+\ab},\s_\jb)$ and $\Gc(\eps,\s_\jb,\s_{\ib+\ab})$
connect the contacts $(\ib+\ab,\ib)$ and $(\jb+\bb,\jb)$
from the opposite sides and their
``paths''
through different contacts {\em must coincide}.
Therefore, in each grain along this path
the diffuson $D$ [Eq.~(\ref{eq:Ddef})] of this particular
 grain arises.
The arising product of two Green functions with pairwise
coinciding coordinates defines
the diffuson of the whole granular system
\beq
    \Dc(\om,\rb_\ib,\rb'_\jb) \equiv
    \frac{1}{2 \pi \nu} \lan \Gc(\eps+\om,\rb_\ib,\rb'_\jb)
        \Gc(\eps,\rb'_\jb,\rb_\ib)\ran_{U,t},
            \mbox{ } (\eps+\om)\eps < 0.
\label{eq:Dcdef}
\eeq
Contrary to Eq.~(\ref{eq:Ddef}),
the points $\rb_\ib$ $\rb'_\jb$ may belong
to different distant grains $\ib$ and $\jb$ now.
Each 
diagrammatic contribution to $\Dc$ [Eq.~(\ref{eq:Dcdef})]
is factorized into the product of intragrain diffusons $D$ [Eq.~(\ref{eq:Ddef})]
connecting different contacts inside the grain
and tunneling probabilities expressed via the tunneling escape rate $\Ga$.

%


%

To get the conductivity $\sig_{\ab\bb}(\om)$
one should sum $\Pi_{\ab\bb}(\om,\ib-\jb)$
over all $\jb$ according to Eq.~(\ref{eq:Pi}).
An important observation is that due to this summation
the intragrain diffusons
always enter the expression for $\sig_{\ab\bb}(\om)$
as a {\em difference}
$D(\om, \sbf_1,\sbf_2)-D(\om, \sbf_3,\sbf_4)$ of the diffusons connecting different contacts
Therefore the {\em zero mode} $1/(|\om| \Vc)$
(see Sec. \ref{sec:diffuson} below)
{\em drops out} and the contribution to $\sig_{\ab\bb}(\om)$
comes from nonzero modes with ``excitation energies'' of the order
of the Thouless energy $\ETh$\footnote{
To avoid misunderstanding, we emphasize
that  zero modes drop out
from the expression for the classical conductivity
only for the system
with identical tunneling conductances $G_T$ of all contacts.
If  conductances are not equal, then, for example
for LC in the limit $G_T/G_0=0$, one needs to consider
not only the contribution from a single given contact
[a simple diagram (d) in Fig.~\ref{fig:eloops}], but also
those from all other contacts,
which have to be connected to a given contact by zero-mode diffusons.
Briefly,
zero modes take care of the fact that conductances of different contacts are not equal,
while nonzero modes take care of the finite resistances of the grains themselves
(that $G_T/G_0 \neq0$ or $G_T R_H \neq 0$).
Our aim is to discuss the latter point and to show that this is crucial
for the Hall effect.}.
Each pair ``grain + contact'' brings a factor
$\Ga/\ETh \propto g_T/g_0$,
given by the ratio of the tunneling conductance $g_T$
to the conductance of the grain $g_0$.

What does the above procedure amount to?
It appears, that
this procedure reproduces exactly the solution
of the classical electrodynamics problem
for the conductivity of a granular medium, provided each
tunnel contact is viewed as a surface resistor with
conductance $G_T$
\footnote{
In fact, the Coulomb interaction has to be also taken into account
in a certain way to get a correct classical expression
for $\sig_{\ab\bb}(\om)$ at nonzero $\om$.
This point will be  discussed in detail in the case of Hall conductivity in Sec.\ref{sec:Hall}.}.
In principle, this approach
allows one to study both LC and HC of the
granular system for arbitrary ratio $g_T/g_0$.
For example, the classical formula
\[
    \sig^{(0)}_{xx}= a^{2-d} \frac{G_T G_0}{G_T+G_0}
\]
for LC can be obtained
(the contact $G_T^{-1}$ and grain $G_0^{-1}$ resistances connected in series).
Its expansion
\beq
    \sig^{(0)}_{xx}= a^{2-d} (G_T -G_T^2/G_0 +G_T^3/G_0^2-\ldots)
\label{eq:sigxxexp}
\eeq
in $g_T/g_0$ corresponds to the  expansion of the
diffuson $\Dc$ in the intragrain diffusons $D$.
Each subsequent term in Eq.~(\ref{eq:sigxxexp})
corresponds to including contacts $(\jb+\eb_x,\jb)$ more and more
remote from $(\ib+\eb_x,\ib)$ in Eq.~(\ref{eq:Pi}).

However, for the system with well-pronounced granularity
[$g_T \ll g_0$, Eqs.~(\ref{eq:granular1}) and (\ref{eq:granular2})]
one does not need to sum the contributions from
all distant contacts $(\jb+\bb,\jb)$ in Eq.~(\ref{eq:Pi}).

It is sufficient to consider the lowest nonvanishing order
in $g_T /g_0 \ll 1$, given by the closest contacts.
In fact, for LC $\sig_{xx}^{(0)}$ ($\ab=\bb=\eb_x$)
considering non-zero-mode intragrain diffusons
is not necessary at all,
since the first term $G_T$ [Eq.~(\ref{eq:sigxx0})]
of the expansion (\ref{eq:sigxxexp}) is obtained
from a {\em single contact} ($\jb=\ib$)
without expanding Eq.~(\ref{eq:Pi2})
in $\Hh_t$ (see Fig.~\ref{fig:eloops}d).
Including the closest contacts ($\jb=\ib,\ib \pm \eb_x$ in Eq.~(\ref{eq:Pi}))
via the intragrain diffusons $D$ will give the
next term $-G_T^2/G_0$ in Eq.~(\ref{eq:sigxxexp}),
which is a small correction to $G_T$.

On the contrary, for the Hall conductivity $\sig_{xy}$ ($\ab=\eb_x, \bb=\eb_y$)
the expansion in $g_T/g_0$ starts
from the term $G_T^2 R_H$
[Eq.~(\ref{eq:sigxy0})] analogous to $-G_T^2/G_0$ in Eq.~(\ref{eq:sigxxexp}).
To get this term one must connect the contacts $(\jb+\eb_y,\jb)$
in the $y$ direction
closest to the contact $(\ib+\eb_x,\ib)$ in the $x$ direction
via the intagrain diffusons $D$
[i.e., take into account the terms with $\jb= \ib-\eb_y, \ib,
\ib+\eb_x, \ib+\eb_x-\eb_y$ in Eq.~(\ref{eq:Pi})].
Thus, considering nonzero diffusion modes for Hall transport is inevitable.




%

The above considerations also explain
why expanding in the tunneling Hamitonian $\Hh_t$
is a ``legal'' procedure in the metallic regime,
even though the dimensionless tunneling coupling
constant $g_T\gg 1$ is large.
The answer is that the actual expansion parameter is
the ratio $g_T/g_0$.

Before we proceed with 
the Hall conductivity,
we consider important building blocks
of our diagrammatic technique: the intragrain diffuson in the presence
of magnetic field and the screened Coulomb interaction.

\subsection{\label{sec:diffuson}Intragrain diffuson}


\begin{figure}
\includegraphics{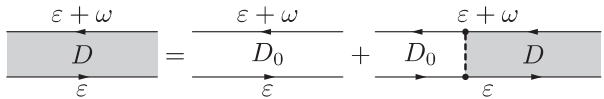}
\caption{Diagrammatic representation of the integral equation (\ref{eq:Dint})
for the diffuson $D(\om,\rb,\rb')$ (gray block), defined by Eq.~(\ref{eq:Ddef}).
Fermionic lines stand for the disorder-averaged Green function $G(\eps,\rb,\rb')$,
dashed line denotes the correlation function (\ref{eq:UU}) of the random potential.
}
\label{fig:D}
\end{figure}
%
The diffuson $D(\om,\rb,\rb')$ of a single isolated grain
is defined by Eq.~(\ref{eq:Ddef}).
The averaging over the disorder potential $U(\rb)$ in Eq.~(\ref{eq:Ddef})
is done using the conventional diagrammatic technique \cite{AGD}.
In the ``noncrossing'' approximation valid for weak disorder
($p_F l \gg 1$, where $l=v_F \tau_0$ is the electron mean free path
and $v_F$ is the Fermi velocity)
the diffuson is given by the sum of ladder-type diagrams.
This standard series can be expressed in terms of the solution of the
integral equation (Fig.~\ref{fig:D})
\beq
       D(\om, \rb,\rb')= D_0(\om,\rb,\rb')+ \frac{1}{\tau_0} \int d \xb
        D_0(\om,\rb,\xb)    D(\om,\xb,\rb'),
\label{eq:Dint}
\eeq
where
\[
    D_0(\om,\rb,\rb')=\frac{1}{2 \pi \nu} G(\eps+\om,\rb,\rb') G(\eps,\rb',\rb)
\]
is the ``ladder step'' and $G(\eps+\om,\rb,\rb') =\lan \Gc(\eps+\om,\rb,\rb') \ran_U$
is the disorder-averaged Green function of the grain.

In the diffusive limit ($l \ll a$, $\om \tau_0 \ll 1$)
the integral equation (\ref{eq:Dint})
can be reduced to the differential diffusion equation (we assume $\om \geq 0$ from now on)
\beq
    (\om- D_0 \nabla_\rb^2) D(\om,\rb,\rb')=\de(\rb-\rb')
\label{eq:D} \eeq where $D_0=v_F l /3$ is the classical diffusion
coefficient in the grain ($D_0$ is not affected by magnetic field,
such that $\om_H\tau_0 \ll 1$). Equation~(\ref{eq:D}) must be
supplied by a proper boundary condition at the grain surface,
which we derive in Appendix \ref{app:bc} in the presence of
magnetic field:
\beq
    (\nb, \nabla_\rb D)|_{\rb \in S}= \om_H  \tau_0 (\tb, \nabla_\rb D)|_{\rb \in S}.
\label{eq:Dbc}
\eeq
Here $\nb$ is the normal unit vector pointing outside the grain,
$\tb = [\nb,\Hb]/H$ is the tangent vector
pointing in the direction opposite to the edge drift.
Equation~(\ref{eq:Dbc}) is due to the fact that the current component
normal to the grain surface vanishes,
its RHS describes the edge drift caused by the magnetic part of the Lorentz force.
Note that only due to the boundary condition~(\ref{eq:Dbc})
the diffuson $D$ ``knows'' about the magnetic field.
{\em All} information about the magnetic field in the system
is now contained in this boundary condition and
the nonzero HC we will obtain is due to the nonzero
RHS of Eq.~(\ref{eq:Dbc}) only.
The main consequence of Eq.~(\ref{eq:Dbc}),
{\em crucial} for the Hall effect,
is the {\em directional asymmetry}
\[
     D(\om, \rb,\rb') \neq D(\om, \rb',\rb)
\]
of the diffuson  for $H \neq 0$.
For $H=0$, Eq.~(\ref{eq:Dbc}) reduces to the well-known Neumann
boundary condition.

The solution to Eqs.~(\ref{eq:D}) and (\ref{eq:Dbc}) can be presented in the form
\beq
      D(\om, \rb,\rb')=
        \frac{1}{\om \Vc}+ \sum_{n>0}
    \frac{ \phi_n(\rb) \phi^*_n(\rb')}{\om+\ga_n},
\label{eq:Dexpr}
\eeq
where $\phi_n$ are the eigenfunctions of the problem
\[ -  \nabla_\rb^2 \phi_n= q^2_n \phi_n, \mbox{ }
    (\nb, \nabla_\rb \phi_n)|_S= \om_H  \tau_0 (\tb, \nabla_\rb \phi_n)|_S,
\]
$\ga_n=D_0 q_n^2$ is the ``diffusion spectrum'', and $\Vc$ is the grain volume.
The functions $\phi_n$ satisfy the orthonormality condition
\beq
 \int d\rb \phi_n^*(\rb) \phi_{n'}(\rb)=\de_{n n'}.
\label{eq:ortho}
\eeq
There always exists a uniform solution $\phi_0(\rb)=1/\sqrt{\Vc}$
with the zero eigenvalue $\ga_0=0$
giving the zero mode $1/(\om \Vc)$ in Eq.~(\ref{eq:Dexpr}).
The lowest excited mode $\ga_1 \sim \ETh \equiv D_0/a^2$
defines the Thouless energy scale $\ETh$.
%
The zero mode $1/(\om \Vc)$ describes the fact that at
time scales much greater than the traversal time $1/\ETh$
the probability density to find an electron
is distributed uniformly over the grain volume.
Information about nontrivial intragrain dynamics
is contained in nonzero modes:
\beq
      \Db(\om, \rb,\rb')= \sum_{n>0}
    \frac{ \phi_n(\rb) \phi^*_n(\rb')}{\om+\ga_n}.
\label{eq:Dbexpr}
\eeq
We will see that for Hall effect, for
which the intragrain dynamics is essential,
only the non-zero mode part $\Db(\om, \rb,\rb')$
of the diffuson $D(\om,\rb,\rb')$
enters the expressions for HC and HR,
whereas the zero mode
$1/(\om \Vc)$
simply drops out.

\subsection{\label{sec:Coulomb} Screened Coulomb interaction}
\begin{figure}
\includegraphics{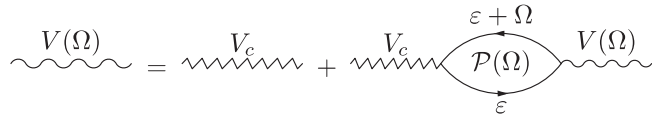}
\caption{
Diagrammatic representation of the integral equation (\ref{eq:Vint})
for the screened Coulomb interaction $V(\Om,\rb_\ib,\rb'_\jb)$
[wavy line, see Eqs.~(\ref{eq:V}),(\ref{eq:Vq}),(\ref{eq:v})  below].
Zigzag line represents the bare Coulomb potential
$V_c(\rb-\rb')=e^2/|\rb-\rb'|$ and electron loop the
polarization operator   $\Pc(\Om,\rb_\ib,\rb'_\jb)$ [Eq.~(\ref{eq:Pcdef})].}
\label{fig:V}
\end{figure}

Within the random phase approximation
the screened Coulomb interaction is given
by a diagrammatic series, that can be obtained
as a solution of the integral equation (Fig.~\ref{fig:V})
\begin{eqnarray}
    V(\Om,\rb_\ib,\rb'_\jb)=V_c(\rb_\ib-\rb'_\jb)-
\sum_{\kb,\lb}\int d\xb_\kb d\xb'_\lb \times
\nonumber \\
     \times V_c(\rb_\ib-\xb_\kb)
                \, \nu \Pc(\Om,\xb_\kb,\xb'_\lb)\,
    V(\Om,\xb'_\lb,\rb'_\jb),
\label{eq:Vint}
\end{eqnarray}
where $\Om \in 2\pi T \mathbb{Z}$ is a bosonic Matsubara frequency
and $V_c(\rb-\rb')=e^2/|\rb-\rb'|$ is the bare Coulomb interaction Eq.~(\ref{eq:Hc}).
%
Just like for ordinary disordered metals
the polarization operator of the granular system is
defined as an electron-hole loop

\beq
    \Pc(\Om,\rb_\ib,\rb'_\jb)=
        - \frac{2}{\nu}
    T \sum_\eps \lan \Gc(\eps+\Om,\rb_\ib,\rb'_\jb) \Gc(\eps,\rb'_\jb,\rb_\ib)\ran_{U,t}
\label{eq:Pcdef}
\eeq
(2 comes from the spin degeneracy) and can be expressed
in terms of the diffuson (\ref{eq:Dcdef}) of the system (we assume $\Om \geq 0$):
\beq
    \Pc(\Om,\rb_\ib,\rb'_\jb)=
        2 \lt[ \de_{\ib\jb}\de(\rb_\ib-\rb'_\jb)-\Om \Dc(\Om,\rb_\ib,\rb'_\jb)
 \rt]
\label{eq:Pc}
\eeq
Since the Coulomb potential $V_c(\rb-\rb')$ satisfies
 the Poisson equation
\[
    -\nabla_\rb^2 V_c(\rb-\rb')= 4\pi e^2 \de(\rb-\rb')
\]
Eq.~(\ref{eq:Vint})  can be rewritten in a differential form
\begin{eqnarray}
    -r_D^2 \nabla_\rb^2 V(\Om,\rb_\ib,\rb'_\jb)=
        \frac{1}{\nu} \de_{\ib\jb}\de(\rb_\ib-\rb'_\jb)- \nonumber \\
    -\sum_{\kb}\int d\xb_\kb \Pc(\Om,\rb_\ib,\xb_\kb)
    V(\Om,\xb_\kb,\rb'_\jb),
\label{eq:Vdiff}
\end{eqnarray}
where $r_D$ is the Debye screening radius, $1/r_D^2=4\pi e^2 \nu$.


Depending on the approximations used for $\Dc$, one obtains
different forms of the screened potential $V(\Om,\rb_\ib,\rb'_\jb)$.

\subsubsection{Coulomb interaction $V(\Om,\rb_\ib,\rb'_\jb)$ for isolated grains}

First we obtain the screened potential $V(\Om,\rb_\ib,\rb'_\jb)$
neglecting tunneling between the grains.
In this case the polarization operator~(\ref{eq:Pc}) takes the form
\[
    \Pc(\Om,\rb_\ib,\rb'_\jb)=
        \de_{\ib\jb} P(\Om,\rb_\ib,\rb'_\jb),
\]
where
\[
    P(\Om,\rb,\rb')= 2 \lt[ \de(\rb-\rb')-\Om D(\Om,\rb,\rb') \rt]
\]
is the polarization operator of a single isolated grain:
\beq
      P(\Om, \rb,\rb')=
        \sum_{n>0}
    P_n(\Om) \phi_n(\rb) \phi^*_n(\rb'), \mbox{ } P_n(\Om)=2 \frac{\ga_n}{\Om+\ga_n},
\label{eq:P}
\eeq
[see Eq.~(\ref{eq:Dexpr})].
Considering the limit when (i) the spatial scales $q_n^{-1} \sim a\gg r_D$ are much greater than
the Debye screening radius $r_D$,
(ii) the frequencies $\Om \ll \sig_{xx}^\text{gr}$ are
much smaller than the grain conductivity $\sig_{xx}^\text{gr}$
($\sig_{xx}^\text{gr} \propto D_0/r_D^2$),
we can neglect the LHS of Eq.~(\ref{eq:Vdiff}) altogether.
Following the lines of Ref.~\onlinecite{ABG}, we get
\beq
    V(\Om,\rb_\ib,\rb'_\jb)=
        E_{\ib\jb}
        +\de_{\ib\jb}\, v(\Om,\rb_\ib,\rb'_\jb).
\label{eq:Vnotunnel}
\eeq
Here $E_{\ib\jb}=e^2 (C^{-1})_{\ib\jb}$ is the charging
energy matrix of the granular array
($C_{\ib\jb}$ is the capacitance matrix, see e.g. Ref. \onlinecite{LLECM}).
The characteristic scale of $E_{\ib\jb}$ is $E_c=e^2/(\kappa a)$,
where $\kappa$ is the dielectric constant of the array.
The charging energy $E_{\ib\jb}$  appears from the zero mode
$1/(\Om \Vc)$ of the diffuson $D(\Om,\rb,\rb')$.
On the contrary, the coordinate-dependent part of the interaction
inside the grain is due to the nonzero diffusion modes of $D(\Om,\rb,\rb')$
and equals
\beq
    v(\Om,\rb,\rb')= \frac{1}{\nu} \sum_{n>0} v_n(\Om) \phi_n(\rb) \phi^*_n(\rb'), \mbox{ }
    v_n(\Om)=\frac{1}{2}\frac{\Om +\ga_n}{\ga_n}.
\label{eq:v}
\eeq
For $q_n r_D \ll 1$ and $\Om \ll \sig_{xx}^\text{gr}$
this part is completely screened and equal to
the inverse intragrain polarization operator~(\ref{eq:P}), $v_n(\Om)=1/[P_n(\Om)]$.


\subsubsection{Coulomb interaction $V(\Om,\rb_\ib,\rb'_\jb)$ taking tunneling into account}
Now we take tunneling into account.
This modifies the expression for the diffuson
$\Dc(\Om,\rb_\ib,\rb'_\jb)$ in Eq.~(\ref{eq:Pcdef}),
which now becomes nondiagonal in  the grain indices $\ib,\jb$.
Let us rewrite the diffuson $\Dc$ in the following form:
\[
    \Dc(\Om,\rb_\ib,\rb'_\jb)=\de_{\ib\jb} D(\Om,\rb_\ib,\rb'_\jb)+
    \de D(\Om,\rb_\ib,\rb'_\jb).
\]
The part $\de D(\Om,\rb_\ib,\rb'_\jb)$
is responsible for tunneling and vanishes, if tunneling is absent.
If we leave only the zero intragrain modes (0D limit) in
$\de D(\Om,\rb_\ib,\rb'_\jb)$, the diffuson equals
\beq
\Dc(\Om,\rb_\ib,\rb'_\jb)=\de_{\ib\jb} \Db(\Om,\rb_\ib,\rb'_\jb)
        +\frac{1}{\Vc} \Dc_0(\Om,\ib,\jb),
\label{eq:Dcapprox}
\eeq
where $\Db(\Om,\rb,\rb')$ [Eq.~(\ref{eq:Dbexpr})] is the non-zero-mode part of the intragrain diffuson
%
and
\begin{eqnarray}
      \Dc_0(\Om,\ib,\jb)=\sum_\qb
        e^{i a \qb (\ib-\jb)} \Dc_0(\Om,\qb),
\nonumber \\
    \Dc_0(\Om,\qb)=1/(\Om+\Ga_\qb),
\label{eq:Dc0}
\end{eqnarray}
is the diffuson for the whole granular system with 0D limit in each grain.
The ``kinetic term'' $\Ga_\qb$ in Eq.~(\ref{eq:Dc0}) equals
\[
    \Ga_\qb=2\Ga \sum_\be (1-\cos q_\be a),
\]
where $\Ga=2 \pi \nu t_0^2 S_0/\Vc$ is the tunneling escape rate
($S_0$ is the area of the contact),
$\be=x,y$ for $d=2$ and
$\be=x,y,z$ for $d=3$,
$\qb\in[-\pi/a,\pi/a]^d$ is the quasimomentum of the granular lattice,
and the sum $\sum_\qb \ldots = \int \frac{a^d d^d\qb}{(2\pi)^d}  \ldots$
denotes the integration over the first Brillouin zone $[-\pi/a,\pi/a]^d$ .

According to Eq.~(\ref{eq:Dcapprox})
the polarization operator (\ref{eq:Pc}) takes  the form
\[
    \Pc(\Om,\rb_\ib,\rb'_\jb)=
         \de_{\ib\jb} P(\Om,\rb_\ib,\rb'_\jb)+\frac{1}{\Vc} \Pc_0(\Om,\ib,\jb),
\]
where
\[
    \Pc_0(\Om,\ib,\jb)=2\lt[ \de_{\ib\jb}- \Om\Dc_0(\Om,\ib,\jb) \rt], \mbox{ }
    \Pc_0(\Om,\qb)=2 \frac{\Ga_\qb}{\Om+\Ga_\qb}
\]
is the zero-mode polarization operator of a granular system\cite{BEAH}.

Accounting for tunneling according to Eq.~(\ref{eq:Dcapprox})
results in the screening of the charging energy $E_{\ib\jb}$ in
Eq.~(\ref{eq:Vnotunnel}) only, whereas $v(\Om,\rb_\ib,\rb'_\jb)$
remains unchanged. As a result, we get for the screened Coulomb
interaction of the granular system:
\beq
    V(\Om,\rb_\ib,\rb'_\jb)=
      V(\Om,\ib,\jb)+\de_{\ib\jb}\, v(\Om,\rb_\ib,\rb'_\jb),
\label{eq:V}
\eeq
where
\[
      V(\Om,\ib,\jb)=\sum_\qb
        e^{i a \qb (\ib-\jb)} V(\Om,\qb)
\] is the screened form of the zero-mode interaction \cite{BEAH},
\begin{eqnarray}
    V(\Om,\qb)=   \frac{E_c(\qb)}{1+ [E_c(\qb) /\de] \Pc_0(\Om,\qb) },
\label{eq:Vq}
\\
    E_c(\qb)=\sum_\ib e^{-ia \qb (\ib-\jb)} E_{\ib-\jb},
\nonumber
\end{eqnarray}
and $v(\Om,\rb,\rb')$ is given by Eq.~(\ref{eq:v}).

The form~({\ref{eq:V}) of the screened interaction
will be sufficient for us. We will see that
the nonzero interaction modes $v(\Om,\rb,\rb')$
inside the grain will be necessary to get a correct
classical expression for the Hall resistance $R_H$ of the grain
and the screened zero-mode interaction $V(\Om,\ib,\jb)$
will be sufficient for calculating quantum corrections
to the classical result.
Significant quantum corrections to HC and HR
arise from the frequency range $\Om \ll g_T E_c$
($g_T E_c$ is the inverse $RC$-time of the pair ``contact+grain''),
when $V(\Om,\ib,\jb)$ is completely screened by the intergrain motion:
\[
        V(\Om,\qb)=  \frac{\de}{\Pc_0(\Om, \qb)}=
    \frac{ \de }{2} \frac{\Om+\Ga_\qb}{ \Ga_\qb }, \mbox{ } \Om \ll g_T E_c.
\]

\section{\label{sec:Hall}Hall conductivity}
Having obtained important building blocks of the diagrammatic
technique, the intragrain diffuson and screened Coulomb interaction,
we can proceed with our main goal: calculating Hall conductivity.

%

\begin{figure}
\includegraphics{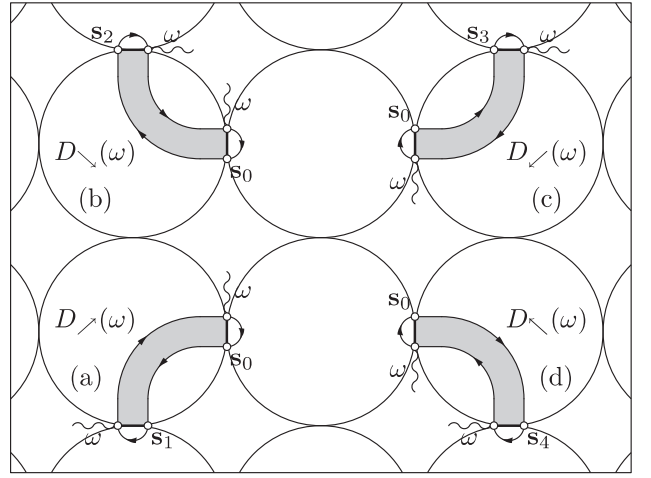}
\caption{Diagrams giving the contribution
$\sig_{xy}^{(0,1)}(\om)$ [Eq.~(\ref{eq:sigxy01D})]
to the bare (without quantum effects)
Hall conductivity $\sig_{xy}^{(0)}(\om)$ [Eq.~(\ref{eq:sigxy0D})]
of the granular metal.
The contacts $\s_a$, $a=1,2,3,4$, must be connected
to the contact $\s_0$ by the intragrain diffusons, as shown
in diagrams (a), (b), (c), and (d).
The diagrams are offset for clarity, the contact $\s_0$ in each
diagrams denotes the same contact.
For each diagram  four possibilities of attaching external tunneling vertices (wavy lines)
must be considered, as shown in Fig.~\ref{fig:Hall11}, only one choice is shown here.}
%
%
\label{fig:Hall1}
\end{figure}

\begin{figure}
\includegraphics{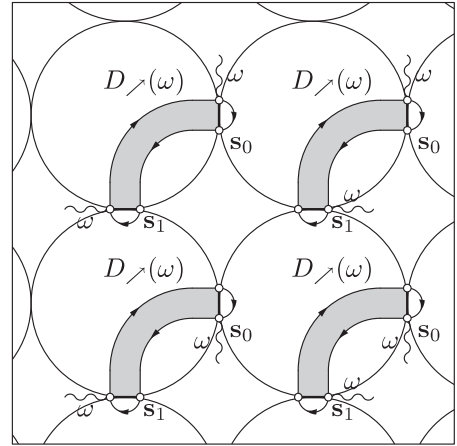}
\caption{
For each diagram in Fig.~\ref{fig:Hall1}
four possibilities [two for each contact according to Eq.~(\ref{eq:I})]
of attaching external tunneling vertices (wavy lines)  must be considered.
}
\label{fig:Hall11}
\end{figure}

We start by neglecting Coulomb interaction
$\Hh_c$ [Eq.~(\ref{eq:Hc})] in $\Hh$ [Eq.~(\ref{eq:H})] completely.
As explained in Sec.~\ref{sec:tech},
in order to compute Hall conductivity $\sig_{xy}$ ($\ab=\eb_x, \bb=\eb_y$)
in the lowest nonvanishing order in $g_T/g_0 \ll 1$
one has to consider the contacts
$(\jb+\eb_y,\jb)$ in the $y$ direction
closest to the contact $(\ib+\eb_x,\ib)$ in the $x$ direction.
Calculating the current through the contact
$(\ib+\eb_x,\ib)$ (denoted further $\s_0$)
one has to connect the contacts $\s_1, \s_2,\s_3,\s_4$  to $\s_0$,
corresponding to $\jb= \ib-\eb_y, \ib, \ib+\eb_x, \ib+\eb_x-\eb_y$
in Eqs.~(\ref{eq:Pi}) and (\ref{eq:Pi2}),  respectively,
by the diffusons $D(\om, \s_0, \s_a)$, $a=1,2,3,4$, of a single grain,
as shown in Fig.~\ref{fig:Hall1}.

Let us consider the contribution
$\Pi_\nea^{(0,1)}(\om)$ to the correlation function
$\Pi_{xy}[\om,\ib-(\ib-\eb_y)]$  [Eq.~(\ref{eq:Pi2})]
from the contact $\s_1$.
From now on we assume $\om \geq 0$,
the arrow subscript $\nea$ denotes the direction of
the diffuson according to Fig.~\ref{fig:Hall1},
the superscript ``0'' stands for the ``bare value''
without quantum effects of Coulomb interaction,
the superscript ``1'' is introduced,
since there will be another contribution ``2'' to HC, see Sec.~\ref{sec:Hall2}.
According to the diagram in Fig.~\ref{fig:Hall1}, we get:
\beqar
    \Pi_\nea^{(0,1)}(\om) &=&-2\pi \nu t_0^4 T \sum_{-\om<\eps<0} \int d\s_0 d\s_1
                D_\ib(\om,\s_0,\s_1) \\\
                & \times &[G_{\ib+\eb_x}(\eps+\om, \s_0,\s_0)-G_{\ib+\eb_x}(\eps, \s_0,\s_0)] \\
                & \times &[G_{\ib-\eb_y}(\eps+\om, \s_1,\s_1)
-G_{\ib-\eb_y}(\eps, \s_1,\s_1)]. \\
\eeqar
Each end of the diffuson $D_\ib$ of the grain $\ib$ is ``capped'' by the Green functions
$G_{\ib+\eb_x}$ and $G_{\ib-\eb_y}$ of the adjacent grains $\ib+\eb_x$ and $\ib-\eb_y$.
We do not write the grain subscripts 
further.
The difference $G(\eps+\om)-G(\eps)$  for each contact
arises due to two possibilities
of choosing the external tunneling vertex
 in $I_{\ib,\ab}$:
$X_{\ib+\ab,\ib}$ or $-X_{\ib,\ib+\ab}$, see Eq.~(\ref{eq:I}) and Fig.~\ref{fig:Hall11}.
For the Green functions at coinciding points one can use their bulk expression (with $H=0$)
$G^{-1}(\eps, \pb)=i\eps-\xi(\pb)+\frac{i}{2\tau_0} \sgn \eps$:
\beqar
    G(\eps+\om, \s_0,\s_0)-G(\eps, \s_0,\s_0)= \\
    = \nu \int d\xi [G(\eps+\om, \pb)-G(\eps, \pb)]=
            \lt\{ \begin{array}{ll} -2\pi i \nu, &\mbox{  } (\eps+\om)\eps<0,\\
                        0,     &\mbox{  } (\eps+\om)\eps>0.\\
                \end{array} \rt.
\eeqar
Therefore, we get
\beq
    \Pi_\nea^{(0,1)}(\om) =\om \frac{g_T^2}{\nu} \frac{1}{S_0^2} \int d\s_0 d\s_1
                D(\om,\s_0,\s_1),
\label{eq:Pine01D}
\eeq
where $g_T=2 \pi (\nu t_0)^2 S_0$ is the conductance of a tunnel contact,
$S_0$ is the area of the contact,
and $\om$ arises as  $ 2\pi T \sum_{-\om<\eps<0}
 1=\om$.

Carrying out the same procedure for the remaining contacts
$\s_2,\s_3,\s_4$  and paying special attention to the signs of the contributions,
for the total contribution
\beq
        \Pi_{xy}^{(0,1)}(\om)=  \Pi_\nea^{(0,1)}(\om)
        +\Pi_\sea^{(0,1)}(\om) +\Pi_\swa^{(0,1)}(\om) +\Pi_\nwa^{(0,1)}(\om)
\eeq
to $\Pi_{xy}(\om)$ [Eq.~(\ref{eq:Pi})] from the diagrams in Fig.~\ref{fig:Hall1},
we obtain
\beq
    \Pi^{(0,1)}_{xy}(\om)=  \om \frac{g_T^2}{\nu}
        \lt[ D_\nea(\om) -D_\sea(\om) + D_\swa(\om) - D_\nwa(\om)\rt].
\label{eq:Pixy01D}
\eeq
Here
\beq
    D_\al(\om)=\frac{1}{S_0^2} \int d\s_0 d\s_a D(\om,\s_0,\s_a)
\label{eq:Daldef}
\eeq
with $a=1,2,3,4$ for $\al=\nea,\sea,\swa,\nwa$, respectively (Fig.~\ref{fig:Hall1}).

Using the expansion~(\ref{eq:Dexpr}) for the diffuson,
we see that
due to the sign structure of Eq.~(\ref{eq:Pixy01D})
the zero mode $1/(\om \Vc)$ drops out of it.
Therefore, retaining only the zero mode in Eq.~(\ref{eq:Dexpr})
would give just $0$ in Eq.~(\ref{eq:Pixy01D}), and we are forced
to take all nonzero modes into account.

According to the structure of Eq.~(\ref{eq:Pixy01D}) we introduce
the following auxiliary quantity:
\beq
    f_n=f_{n,\nea} -f_{n,\sea} + f_{n,\swa} - f_{n,\nwa}
\label{eq:fn}
\eeq
where
\beq
    f_{n,\al}=\frac{1}{S_0^2} \int d\s_0 d\s_a \phi_n(\s_0)\phi_n^*(\s_a)
\label{eq:fnal}
\eeq
with $a=1,2,3,4$ for $\al=\nea,\sea,\swa,\nwa$, respectively (Fig.~\ref{fig:Hall1}).
The factor $f_n$ takes care about the geometry
and gives a  convenient compact form of the contributions.
It will be especially helpful for studying interaction corrections to HC.
We can rewrite Eq.~(\ref{eq:Pixy01D}) as
\beq
    \Pi^{(0,1)}_{xy}(\om)=  \om \frac{g_T^2}{\nu}
        \sum_{n>0}  \frac{f_n}{\om+\ga_n}
\label{eq:Pixy01}
\eeq
and the contribution to Hall conductivity~(\ref{eq:sig})
corresponding to $\Pi^{(0,1)}_{xy}(\om)$ takes the form
\beq
    \sig^{(0,1)}_{xy}(\om)=  2 e^2 a^{2-d} \frac{g_T^2}{\nu}
        \sum_{n>0} \frac{f_n}{\om+\ga_n}.
\label{eq:sigxy01D}
\eeq

\subsection{Correct $\om$-dependence \label{sec:Hall2}}
As we show further in Sec.~\ref{sec:classics},
the expression~(\ref{eq:sigxy01D})
for $\sig^{(0,1)}_{xy}(\om)$
at zero frequency $\om=0$
reproduces exactly the result~(\ref{eq:sigxy0}) for HC  of a  granular medium
obtained from the solution of the classical electrodynamics problem.
Therefore, it would be natural
to expect such correspondence with classics for all $\om$.

However, the obtained result~(\ref{eq:sigxy01D})
does not agree with the classical formula (\ref{eq:sigxy0}) at finite frequency $\om>0$.
Indeed, in classical electrodynamics  the resistance of a
metallic sample {\em itself } is frequency-independent
 up to very high frequencies
$\om \sim \sig^{\text{gr}}_{xx}$ of the order of the grain conductivity
$\sig^{\text{gr}}_{xx}$
(
$\sig^{\text{gr}}_{xx} \propto g_0 E_c$ is the inverse  $RC$-time of the grain)\footnote{
We neglect the dispersion $\sig^{\text{gr}}_{xx}(\om)$ at $\om \sim 1/\tau_0$
of the grain conductivity itself in these considerations.}.
So the HC we are looking for should be given by the zero-frequency expression
$\sig^{(0,1)}_{xy}(\om=0)$ for all $\om \ll \sig^{\text{gr}}_{xx}$.
We see, however, from Eq.~(\ref{eq:sigxy01D}) that
$\sig^{(0,1)}_{xy}(\om)$ has a diffusion-like dispersion at
Thouless energy (when $\om \sim \ETh$)
which contradicts
 the classical picture.

\begin{figure*}
\includegraphics{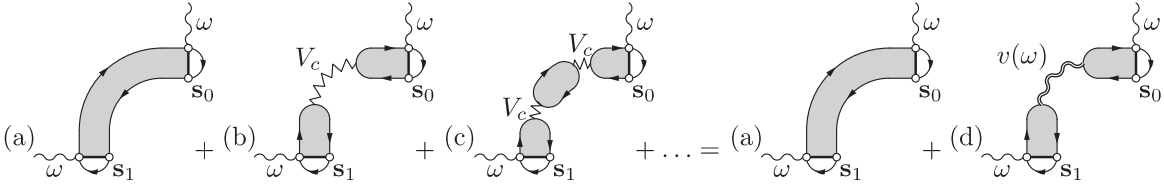}
\caption{Complete set of diagrams
for the bare (without quantum effects) Hall conductivity
$\sig_{xy}^{(0)}(\om)$ [Eqs.~(\ref{eq:sigxy0D})] of the granular system.
(a) One starts by connecting the relevant contacts by the intragrain diffusons
(see Fig.~\ref{fig:Hall1}).
(b) To get a correct $\om$-dependence one should take the Coulomb interaction
into account by inserting the interaction line into the diffuson.
(c) Coulomb interaction, in its turn, is screened by the intragrain motion
and one should consider insertions of the polarization bubbles into the interaction line.
(d) The summation of the resulting series yields an additional
to $\Pi_{xy}^{(0,1)}(\om)$ [(a), Eq.~(\ref{eq:Pixy01})] contribution
$\Pi_{xy}^{(0,2)}(\om)$ [(d), Eq.~(\ref{eq:Pixy02})]
to the current-current correlation function.
The sum $\Pi_{xy}^{(0)}(\om)=\Pi_{xy}^{(0,1)}(\om)+\Pi_{xy}^{(0,2)}(\om)$
[Eq.~(\ref{eq:Pixy0})] 
gives a correct classical expression (\ref{eq:sigxy0D})
for the Hall conductivity $\sig_{xy}^{(0)}(\om)$. 
}
\label{fig:Hall}
\end{figure*}

What is yet missing in our approach?
Apparently, one must consider the
Coulomb interaction inside the grain.
Indeed, the diffuson  $D(\om,\rb,\rb')$,
appearing in Eqs.~(\ref{eq:Pixy01D}),(\ref{eq:Pixy01}),(\ref{eq:sigxy01D}),
describes the propagation of
electron density, but it does not take into account that electrons
are charged and can interact. The correct way to include the {\em classical effect}
of Coulomb interaction
is to insert the interaction line into the diffuson
as shown in Fig.~\ref{fig:Hall}b.
In its turn, this interaction is screened by electron motion,
and this can be accounted for by inserting polarization operators
into the interaction lines, as shown in Fig.~\ref{fig:Hall}c.
The summation of the resulting series in Fig.~\ref{fig:Hall}
yields the screened form of the Coulomb interaction in a grain
$V(\om,\rb,\rb')= V(\om,\ib,\ib)+v(\om,\rb,\rb')$ [Eq.~(\ref{eq:V})],
and we obtain an additional to $\Pi_\nea^{(0,1)}(\om)$ [Eq.~(\ref{eq:Pine01D})]
contribution
\beqar
    \Pi_\nea^{(0,2)}(\om) =2 g_T^2 \frac{1}{S_0^2} \int d\s_0 d\s_1
                \int d\rb d\rb' \times
\\
    \om D(\om,\s_0,\rb) V(\om,\rb,\rb') \om D(\om,\rb',\s_1).
\eeqar
Here the integration with respect to $\rb$ and $\rb'$ is done over the grain volume,
the spin degeneracy factor 2 comes from an additional electron loop.
Due to the orthogonality of the eigenfunctions $\phi_n$ [Eq.~(\ref{eq:ortho})]
the  zero modes of $D$ and $V$ drop out
of the needed combination
\beq
        \Pi_{xy}^{(0,2)}(\om)=  \Pi_\nea^{(0,2)}(\om)
        +\Pi_\sea^{(0,2)}(\om) +\Pi_\swa^{(0,2)}(\om) +\Pi_\nwa^{(0,2)}(\om),
\label{eq:Pi02}
\eeq
and we get an additional to $\Pi_{xy}^{(0,1)}(\om)$
[Eq.~(\ref{eq:Pixy01})] contribution (Fig.~\ref{fig:Hall}d):
\beq
    \Pi_{xy}^{(0,2)}(\om) =2 \frac{g_T^2}{\nu} \sum_{n>0} f_n
            \frac{\om}{\om+\ga_n} v_n(\om) \frac{\om}{\om+\ga_n}
\label{eq:Pixy02}
\eeq
Summing the contributions (\ref{eq:Pixy01}) 
and (\ref{eq:Pixy02}) 
we obtain
\beq
    \Pi_{xy}^{(0)}(\om) \equiv \Pi_{xy}^{(0,1)}(\om)+\Pi_{xy}^{(0,2)}(\om)=
        \frac{g_T^2}{\nu} \sum_{n>0} \om f_n r_n,
\label{eq:Pixy0}
\eeq
where
\beq
    r_n \equiv \frac{1}{\om+\ga_n} \lt( 1+ v_n(\om) \frac{2\om}{\om+\ga_n} \rt) =\frac{1}{\ga_n}
\label{eq:rn}
\eeq
for $\om \ll \sig^{\text{gr}}_{xx}$ [see Eq.~(\ref{eq:v})].
Equations~(\ref{eq:Pixy0}) and (\ref{eq:rn}) lead to the final classical expression
for the Hall conductivity [Eq.~(\ref{eq:sig})]:
\beq
    \sig^{(0)}_{xy}(\om)=  2 e^2 a^{2-d} \frac{g_T^2}{\nu}
        \sum_{n>0} \frac{f_n}{\ga_n},
\eeq
or, going back to diffusion propagators,
\beq
    \sig^{(0)}_{xy}(\om)=  2 e^2 a^{2-d} \frac{g_T^2}{\nu}
        (\Db_\nea -\Db_\sea + \Db_\swa - \Db_\nwa),
\label{eq:sigxy0D}
\eeq
where
$\bar{D}_\al=\frac{1}{S_0^2} \int d\s_0 d\s_a \bar{D}(\s_0,\s_a)$,
with $a=1,2,3,4$ for $\al=\nea,\sea,\swa,\nwa$, respectively,
and
\[  \Db(\rb,\rb')=
        \sum_{n>0}
    \frac{\phi_n(\rb) \phi^*_n(\rb')}{\ga_n}
\]
is the diffuson without the zero mode at $\om=0$
satisfying Eqs.~(\ref{eq:D}) and (\ref{eq:Dbc}) with $\om=0$:
\[
   - D_0 \nabla_\rb^2 \Db(\rb,\rb')=\de(\rb-\rb'),
\]
\beq
    (\nb, \nabla_\rb \Db)|_S= \om_H  \tau_0 (\tb, \nabla_\rb \Db)|_{S}.
\label{eq:Db}
\eeq
%
It follows from Eqs.~(\ref{eq:Db}) that $\Db(\rb,\rb')$ is a Green function for the Poisson equation.
Actually the propagator $\Db(\rb,\rb')$ should not
be termed ``diffuson'' anymore, since it describes
the propagation of electron density with Coulomb interaction taken into account,
i.e. the real conduction process.

Equation~(\ref{eq:sigxy0D}) constitutes our main result
for HC in the absence of quantum effects.
We stress that the diagrammatic series in Fig.~\ref{fig:Hall}
leading to Eq.~(\ref{eq:sigxy0D}) describes the {\em classical} effect:
propagation of electron density in a disordered metallic sample.
The temperature-independent result (\ref{eq:sigxy0D})
is valid for arbitrary temperature $T$ and arbitrary size $a$ of the grains
(not necessarily small grains and $T \ll \ETh$).
Temperature will be relevant for {\em quantum} effects of Coulomb
interaction, which we consider in Sec. \ref{sec:QCoulomb}.

\subsubsection{Properties of Eq.~(\ref{eq:sigxy0D}) for Hall conductivity $\sig^{(0)}_{xy}$.}
Let us discuss the basic  properties of Eq.~(\ref{eq:sigxy0D}).
For simplicity, we assume that grains have reflectional symmetry
in all three dimensions.
Then $\Db_\nea = \Db_\swa$ and  $\Db_\sea= \Db_\nwa$
 due to this symmetry (for $H\neq 0$, too).
At zero magnetic field ($H=0$) we have $\Db_\nea = \Db_\sea$ and  $\Db_\swa  = \Db_\nwa $
due to the time-reversal symmetry $D(\om,\rb,\rb')=D(\om,\rb',\rb)$,
and therefore $\sig^{(0)}_{xy}(\om, H=0)=0$.
The nonzero differences $\Db_\nea - \Db_\sea=\Db_\swa - \Db_\nwa$
arise only due to nonzero RHS of the boundary condition Eq.~(\ref{eq:Dbc}),
which represents the edge drift.
To understand the sign of $\Db_\nea - \Db_\sea$ and $\Db_\swa - \Db_\nwa$
we recall that the diffuson $D(\om,\rb,\rb')
$ describes the probability
of getting from point $\rb'$ to point $\rb$.
In nonzero field ($H \neq 0$) the edge trajectories for
$\Db_\nea  =\Db_\swa$ are shorter (if $e>0$ is assumed) than those
for $ \Db_\sea= \Db_\nwa$, and therefore $\Db_\nea - \Db_\sea=\Db_\swa - \Db_\nwa>0$.

Since $\Db_\al \propto \frac{1}{a^3} \frac{1}{D_0 a^{-2}}$
and the difference $\Db_\nea - \Db_\sea$ is linear in $\om_H \tau_0$,
one can estimate

\beq
    \Db_\nea - \Db_\sea \propto \frac{1}{a^3} \frac{\om_H \tau_0}{ D_0 a^{-2}} \propto
        e^2 \nu \frac{\rho_{xy}^\text{gr}}{a},
\label{eq:dDest}
\eeq
where
\[
    \rho_{xy}^\text{gr}=\frac{\sig_{xy}^\text{gr}}{(\sig_{xx}^\text{gr})^2}=
    \frac{\om_H \tau_0}{\sig_{xx}^\text{gr}} = \frac{H}{nec}
\]
is the specific HR of the grain material expressed in terms of the
carrier density $n$ in the grains
[Einstein relation $\sig_{xx}^\text{gr}=2e^2 \nu D_0$ was used in Eq.~(\ref{eq:dDest})].
We see that $\Db_\nea - \Db_\sea$ does not depend on the intragrain
disorder, described by the scattering time $\tau_0$.
The proportionality coefficient in Eq.~(\ref{eq:dDest}) is
determined by the shape of the grains only.
Thus, for HC [Eq.~(\ref{eq:sigxy0D})
]
of the system we get
\[
    \sig_{xy}^{(0)} \propto a^{2-d}  G_T^2 \frac{\rho_{xy}^\text{gr}}{a}
\]
and for HR [see also Eq.~(\ref{eq:sigxx0})]
\[
    \rho_{xy}^{(0)} = \frac{\sig_{xy}^{(0)}}{(\sig_{xx}^{(0)})^2}
        \propto a^{d-3}  \rho_{xy}^\text{gr}= a^{d-3}  \frac{H}{nec}
\]
We come to an important conclusion.
The Hall resistivity of the granular system is {\em independent} of the intragrain disorder
and tunneling conductance. It is expressed solely
via the carrier density $n$ of the grain material up to a numerical coefficient
determined by the shape of the grains and the type of granular lattice.

\subsection{\label{sec:classics} Classical picture}
Let us now prove that
Eq.~(\ref{eq:sigxy0D}) for the Hall conductivity 
indeed reproduces the solution of the classical electrodynamics problem,
provided one  treats the tunnel contact as a surface resistor with the conductance $G_T$.

The classical HC of the granular medium in the limit $G_T \ll G_0$
[Eq.~(\ref{eq:granular2})]
can be easily presented in the form of Eq.~(\ref{eq:sigxy0}) (see Fig.~\ref{fig:system}).
The current $I_y=G_T V_y$ running through
the grain in the $y$ direction causes the Hall voltage drop
$V_H=R_H I_y$ between its opposite banks in the $x$ direction,
where $R_H$ is the Hall resistance of the grain.
Since for calculating $\sig_{xy}$ the total voltage drop per lattice period
in the $x$ direction is assumed zero, the same
voltage $V_H$ (but with the opposite sign)
is applied to the contacts in  the $x$ direction.
Thus, the Hall current equals $I_x=G_T V_H = G_T^2 R_H V_y$
which leads to the expression (\ref{eq:sigxy0}) for HC.

The Hall resistance $R_H$
of the grain is
defined via the difference (Hall voltage) of
the electric potential $\varphi(\rb)$ between the opposite banks of the grain
in the $x$ direction,
\beq
    V_H=\varphi(\sbf_r)-\varphi(\sbf_l)=R_H I_y,
\label{eq:VH}
\eeq
when the current $I_y=I$ passes through the grain in the $y$ direction.
The current density
\beq
    \jb(\rb)=- \hat{\sig}_0 \nabla_\rb \varphi(\rb) \equiv
- \begin{pmatrix}
            \sig_{xx}^\text{gr} &  \sig_{xy}^\text{gr} & 0\\
            -\sig_{xy}^\text{gr}  & \sig_{xx}^\text{gr} & 0\\
            0 & 0& \sig_{xx}^\text{gr} \\
\end{pmatrix}
\nabla_\rb \varphi(\rb)
\label{eq:j}
\eeq
($\hat{\sig}_0$ is the conductivity tensor) satisfies the continuity equation
\beq
    \text{div} \jb=q(\rb)
\label{eq:jceq}
\eeq
and the boundary condition
\beq
    (\nb, \jb)|_S=0.
\label{eq:jbc}
\eeq
The charge source function $q(\rb)$ is nonzero on the contacts surface only,
$\int d \s_d q(\s_d)=I$ corresponding to the current $I$ flowing
into the grain through the contact $\s_d$ and
$\int d \s_u q(\s_u)=-I$ corresponding to the current flowing
out of the grain through the contact $\s_u$.
The stationary form of Eq.~(\ref{eq:jceq})
is valid up to the frequencies $\om \sim \sig_{xx}^\text{gr}$,
even if $I=I(t)$ is time-dependent,
compare with discussion of Eq.~(\ref{eq:v}) in Sec.~\ref{sec:Coulomb}.

Inserting Eq.~(\ref{eq:j}) into Eqs.~(\ref{eq:jceq}) and (\ref{eq:jbc}),
we find that  $\varphi(\rb)$ is a solution of the following boundary value problem:
\beq
    -\nabla_\rb^2 \varphi = q(\rb)/\sig_{xx}^\text{gr},\mbox{ }
    (\nb, \nabla \varphi)|_S= \om_H  \tau_0 (\tb, \nabla \varphi)|_S.
\label{eq:phi}
\eeq
Comparing Eq.~(\ref{eq:phi}) with Eqs. (\ref{eq:Db}),
we see that $ \bar{D}(\rb,\rb')$
is a Green function for the problem~(\ref{eq:phi}).
Thus the solution to Eq.~(\ref{eq:phi}) can be written as
\[
    \varphi(\rb)=\frac{1}{2 e^2 \nu} \frac{I}{S_0}
        \lt( \int d\sbf_d \bar{D}(\rb,\sbf_d)-\int d\sbf_u \bar{D}(\rb,\sbf_u)\rt)
\]
(Einstein relation $\sig_{xx}^\text{gr}=2e^2 \nu D_0$ was used).
Inserting $\varphi(\rb)$ in such form into Eq.~(\ref{eq:VH}),
we obtain for the Hall resistance of the grain:
\beq
    R_H=\frac{1}{2e^2 \nu} (\Db_\nea -\Db_\sea + \Db_\swa - \Db_\nwa),
\label{eq:RHexpr}
\eeq
Comparing Eq.~(\ref{eq:sigxy0D}) with Eqs.~(\ref{eq:sigxy0}) and (\ref{eq:RHexpr})
we see that Eq.~(\ref{eq:sigxy0D}) indeed reproduces the classical result.

This establishes the correspondence between  our diagrammatic approach
of considering nonzero diffusion modes and the solution of the
classical electrodynamics  problem for the granular system.

Luckily, for simple geometries of the grain (cubic, spherical)
the Hall resistance $R_H$ can be obtained from symmetry arguments
without solving the problem Eq.~(\ref{eq:phi}).
Suppose the grain has reflectional symmetry in all three
dimensions. Then  it is clear that (1) the largest cross section of
the grain lies in the plane of reflection, (2) the current density
$\jb(\rb)$ is {\em perpendicular} to the plain of reflection at
each point $\rb$ of the cross section, (3) the absolute value
 of  $\jb(\rb)$ is constant on the cross section
and therefore equal to $|\jb(\rb)|=I/S$, where $S$
is the area of the cross section.
So, the Hall voltage Eq.~(\ref{eq:VH})
equals $V_H = \rho_{xy}^\text{gr} |\jb(\rb)| a = a I/S$.
Therefore, the Hall resistance is
\beq
    R_H=\rho_{xy}^\text{gr}a/S
\label{eq:RH}
\eeq
and the Hall resistivity of the granular medium can be expressed in the form
\beq
    \rho^{(0)}_{xy} =
    \frac{\sig^{(0)}_{xy}}{(\sig^{(0)}_{xx})^2}=
    R_H  a^{d-2}= \frac{H}{n^* e c},
\label{eq:rhoxy0text}
\eeq
where
\[
    n^*=a^{3-d} A n,\mbox{ } A=S/a^2 \leq 1.
\]
The quantity $n^*$ defines the effective carrier
density of the granular system. For a 3D sample (many grain monolayers),
$n^*=A n$ differs from the actual carrier
density $n$ of the grain material only by a numerical factor $A$
determined by the shape of the grains and type of the granular lattice.
For a 2D sample $n^*= a A n$ for a single grain monolayer or
$n^*= d_z A n$ in case of several monolayers, where $d_z$ is the thickness
of the sample ($d_z/a$ is the number of monolayers).


We remind the reader, that Eq.~(\ref{eq:rhoxy0text})
was obtain under the following assumptions:
(a) diffusive limit inside the grains, $l \ll a$;
(b) the mean free path $l$ is the same for all grains;
(c) the tunneling conductance $G_T$ is the same for all contacts.
Having established the correspondence between our diagrammatic approach
and the classical solution of the problem,
we can now show that the result (\ref{eq:rhoxy0text}) is actually valid in a much more general case,
when (i) the intragrain disorder is ballistic, $l\gtrsim a$;
(ii) the mean free path $l_\ib$ varies from grain to grain
(iii) the tunneling conductance $G_{T \ib+\ab,\ib}$
varies from contact to contact.
The statement (i) follows from the fact that
the above classical consideration leading to Eq.~(\ref{eq:RH})
involve only the symmetry properties of the current distribution $\jb(\rb)$
and therefore hold for the grains with ballistic disorder ($l\gtrsim a$) as well.
The statement (ii) is true, since the Hall resistance $R_H$ [Eq.~(\ref{eq:RH})]
of the grain is independent of the mean free path $l_\ib$, and, hence, is the
same for all grains $\ib$ provided only their shape is the same.
Finally, using the standard Ohm and Kirchhoff laws, we obtain that
the Hall voltage between the opposite banks of the sample in the $x$ direction
depends on the {\em total} Ohmic current flowing through the sample in the $y$ direction.
Hence, the Hall voltage is essentially independent of the distribution
of the tunneling conductances $G_{T \ib+\ab,\ib}$, and the HR of the system
is still given by Eq.~(\ref{eq:rhoxy0text}), which proves the statement (iii).

Therefore, the result (\ref{eq:rhoxy0text}) for Hall resistivity
is applicable to real granular systems, in which fluctuations of
the intragrain mean free path $l$ and, most importantly,
the intergrain tunneling conductance $G_T$
are always present.

To conclude this section,
we obtain that the classical Hall resistivity [Eq.~(\ref{eq:rhoxy0text})]
of a granular system in the metallic regime,
being independent of parameters that describe Ohmic dissipation,
possesses a great deal of universality, reminiscent of that in
ordinary disordered metals [Eq.~(\ref{eq:rxy})].

Being classical, however, Eq.~(\ref{eq:rhoxy0text}) describes
the behavior of the Hall resistivity at high enough temperatures,
when quantum effects can be neglected.
At sufficiently low temperatures quantum effects of Coulomb interaction
and weak localization set in and can significantly affect electron transport.
In the next section we study quantum corrections to the  obtained results
(\ref{eq:sigxy0D}) and (\ref{eq:rhoxy0text}) due to the Coulomb interaction
between the electrons.

\section{\label{sec:QCoulomb}Quantum effects of Coulomb interaction}
Diagrammatic approach
enables us to incorporate
quantum effects of Coulomb interaction
on the Hall conductivity into the developed scheme.
%
%
We perform calculations to the first order in the screened
Coulomb interaction with the expansion parameter $1/g_T$.
We assume the diffusive limit for the intragrain dynamics
and neglect the diagrams that are small in $\tau_0 \ETh \propto (l/a)^2 \ll 1$.
The ballistic limit can be treated similarly,
although in this case one has to take such diagrams into account.

Technically, one considers the diagrams for the ``bare'' conductivity $\sig_{xy}^{(0)}$
shown in Fig.~\ref{fig:Hall} and connects different electron lines
by the interaction lines corresponding to the screened
Coulomb interaction $V(\Om,\rb_\ib,\rb_\jb')$ [Eq.~(\ref{eq:V})].
It is important that for the quantum interaction
corrections
the zero-mode part $V(\Om,\ib,\jb)$ [Eq.~(\ref{eq:Vq})]
of the interaction $V(\Om,\rb_\ib,\rb_\jb')$  does not drop out and gives
a contribution larger
than the nonzero intragrain modes $\de_{\ib\jb} v(\Om,\rb_\ib,\rb'_\jb)$
[Eq.~(\ref{eq:v})]
(we provide an estimate below).
Therefore for the interaction lines that describe the quantum corrections
to the classical result  we can use
the zero-mode part $V(\Om,\ib,\jb)$ of the interaction.
Further, depending on the sign structure of energies of the Green functions involved,
some interaction vertices are renormalized by the diffusons
and some are not.

%

Two types of diagrams can be identified: (i) the interaction
$V(\Om,\ib,\jb)$ connects different electron loops of the diagrams
like in Figs.~\ref{fig:corrections}(a),(b); (ii) the interaction
$V(\Om,\ib,\jb)$ connects points on the same electron loop, like
in Figs.~\ref{fig:corrections}(c),(d). It is straightforward to
show that the former possibility (i) always gives zero: in each
case contributing diagrams cancel each other identically , an
example is shown in Fig.~\ref{fig:corrections}(a),(b). So, we come
to an important simplification: electron loops in
Fig.~\ref{fig:Hall} are {\em renormalized by the interaction
independently}.

\begin{figure}
\includegraphics{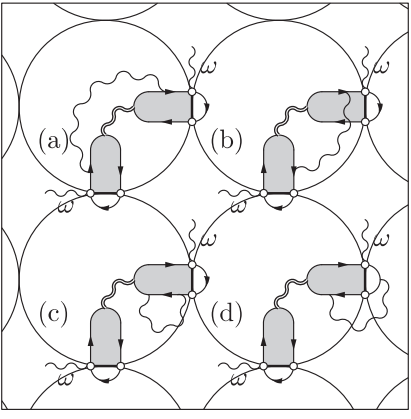}
\caption{Quantum corrections to Hall conductivity of a granular metal
due to Coulomb interaction.
Important point: electron loops in diagrams of Fig.~\ref{fig:Hall}
are renormalized independently: the diagrams
with interaction line connecting different
electron loops cancel each other,
as is the case, e.g., for the diagrams (a) and (b).
Nonvanishing contributions come from the independent
renormalization of electron loops by the interaction, like in diagrams (c) and (d).
}
\label{fig:corrections}
\end{figure}

The temperature $T$, being irrelevant for the single-particle
classical transport [Eq.~(\ref{eq:sigxy0}) for $\sig_{xy}^{(0)}$],
becomes important for quantum effects of Coulomb interaction.
An important energy scale here is the tunneling escape rate $\Ga$.
For $T \gg \Ga$ the thermal length $L^*_T=\sqrt{D_0^*/T} \ll a$ for the intergrain motion
($D_0^*=\Ga a^2$ is the effective diffusion coefficient)
does not exceed the grain size and
only the contributions coming from spatial scales of the order
of the grain size $a$ are significant.
At $T \lesssim \Ga$ the contributions from the scales
$L_T^* \gtrsim a$ exceeding the grain size also become important.

We start with the former regime $T\gg \Ga$ in the following subsection.
Note that the corrections that will be discussed in Sec.~\ref{sec:QCoulombHighT}
are specific to granular systems and absent in HDMs. At the same time
they, as we find,  govern the $T$-dependence of HC and HR
in a wide range of temperatures.
The corrections analogous to those in HDMs (``Altshuler-Aronov'' corrections)
arise from large spatial scales ($\gg a$), are relevant at $T \ll \Ga$
and will be addressed afterwards in Sec.~\ref{sec:QCoulombLowT},
where we consider the case $T \lesssim \Ga$.

\subsection{``High'' temperatures $T \gg \Ga$ \label{sec:QCoulombHighT}}
First,
we consider the range of temperatures
$T \gg \Ga$
 greater than the escape rate $\Ga$.
In this regime each tunneling process brings a small factor $\Ga/T
\ll 1$. Therefore the main contribution comes from the diagrams
which contain minimal number of hops between the grains as
compared to the diagrams for the bare HC $\sig_{xy}^{(0)}$. This
means that
the screened zero-mode interaction can be taken in the form [see Eq.~(\ref{eq:Vq})]:
\beq
    V(\Om,\qb)=   \frac{E_c(\qb)}{1+2 E_c(\qb) \Ga_\qb/(\de|\Om|) }
\label{eq:VqhighOm}
\eeq
and for the diffusons renormalizing the interaction vertex
we can neglect tunneling completely,
i.e., take them as
\[
    \Dc(\Om,\rb_\ib,\rb'_\jb)=\de_{\ib\jb} D(\Om,\rb_\ib,\rb'_\jb).
\]
%
%
%
It is very important that,
since the interaction $V(\Om,\ib,\jb)$ is coordinate-independent
within each grain, the intragrain diffusons renormalizing
the interaction vertex contain only the zero mode $1/(|\Om|\Vc)$,
whereas the nonzero modes {\em drop out automatically} due to the orthogonality
condition (\ref{eq:ortho}) (we do not simply neglect them):
\[
    \int d\rb'_\ib D(\Om,\rb_\ib,\rb'_\ib) V(\Om,\ib,\jb)=\frac{1}{|\Om|} V(\Om,\ib,\jb).
\]

Assuming $T \gg \Ga$,
we {\em do not} assume the temperature $T$ or the frequencies $\Om$ much smaller than the Thouless
energy $\ETh$ in this section.
As we will see, the only way to clearly identify the physics
of the contributions
 and obtain a correct upper cut-off
for the logarithmically divergent quantities
is to include the range $T,\Om \sim \ETh$ into consideration.

%




As explained above, we may renormalize electron loops shown in Fig.~\ref{fig:Hall}
independently of each other.
There are three different types of electron loops in Fig.~\ref{fig:Hall}:
1) the (tunneling current)-(tunneling current)
correlator $\lan I I \ran $ of the diagram in Fig.~\ref{fig:Hall}(a);
2) the (tunneling current)-density correlators $\lan I n\ran$
in Fig.~\ref{fig:Hall}(b) connected by the  screened interaction $v(\om,\rb,\rb')$;
3) the density-density correlators $\lan n n \ran$
[i.e. the intragrain polarization operator $P(\om,\rb,\rb')$, Eq.~(\ref{eq:P})],
in Fig.~\ref{fig:Hall}(c), which are the insertions into the bare Coulomb interaction line.

Since the geometry factor $f_n$ [Eqs.~(\ref{eq:fn}) and (\ref{eq:fnal})]
is the same for all diagrams
and the properties of the granular array are assumed the same
in the $x$ and $y$ directions,
let us draw the diagrams in the ``longitudinal geometry''
(see Figs.~\ref{fig:dP},\ref{fig:dPi01},\ref{fig:dPi02}).
For our purpose it is only important now that there is a
``central'' grain  with  a non-zero-mode diffuson
and there are two ``adjacent'' grains. External
tunneling vertices
are attached to the contacts between the central and adjacent grains.

%
%

For each diagram one has to take 4 possibilities
(2 for each contact)  of attaching
 external tunneling vertices into account,
as in Fig.~\ref{fig:Hall1}.
Only one such possibility is shown in
Figs.~\ref{fig:dP},\ref{fig:dPi01},\ref{fig:dPi02}.
Further, for each diagram one has to
consider (i)  the up/down reversal, if the diagram does not transfer to itself,
(ii) the left/right reversal, if the diagram does not transfer to itself.
We introduce ``up/down reversal'' and
``left/right reversal'' multiplication factors
$\al^\text{u/d}$ and  $\al^\text{l/r}$ correspondingly:
$\al^\text{u/d},\al^\text{l/r}=1$, if the reversal is not possible,
$\al^\text{u/d},\al^\text{l/r}=2$, if the reversal is possible.

The summation region over the fermionic frequency $\eps$ of the electron loop and
bosonic frequency $\Om$ carried by the interaction line is determined
by the analytical properties of the Green functions involved.
%
%
After the integration over the Green functions momenta
the expressions become independent of $\eps$
and the summation over $\eps$ can be performed. This
always results in the sum
\begin{eqnarray}
    2\pi T  \sum_{\substack{0<\Om \le \om,\\ -\Om<\eps<0}} F(\om,\Om) +
    2\pi T \sum_{\substack{\om<\Om,\\ -\om<\eps<0}} F(\om,\Om) =
\nonumber
\\
    =\sum_{0<\Om \le \om} \Om  F(\om,\Om)+ \om \sum_{\Om>\om} F(\om,\Om)
    =\sum_{\Om >0} \theta_{\om,\Om} F(\om,\Om)
\label{eq:sum}
\end{eqnarray}
standard for the first-order interaction corrections calculations \cite{AA}.
We have introduced the function
\[
\theta_{\om,\Om}=\lt\{ \begin{array}{ll} \Om, \mbox{ }&  0<\Om \le \om, \\
                \om, \mbox{ }  & \Om>\om. \end{array}
        \rt.
\]
for compactness. The diagrams considered here may contain either
one or two summations (\ref{eq:sum}).
We introduce the ``sum'' multiplication factor $\al^\text{s}$ correspondingly:
$\al^\text{s}=1$ or $\al^\text{s}=2$.
So, each ``topologically unique''
diagram comes with an overall multiplication factor
\beq
    \al= \al^\text{s}\, \al^\text{u/d} \al^\text{l/r} .
\label{eq:al}
\eeq

We start by considering the corrections to the intragrain polarization operator.

\subsubsection{Corrections to the intragrain polarization operator $P(\om,\rb,\rb')$}

\begin{figure*}
\includegraphics{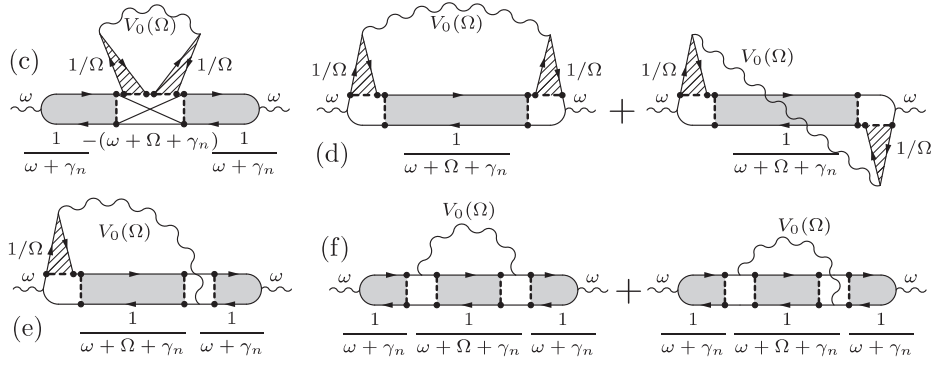}
\caption{Diagrams for the interaction corrections to the intragrain polarization operator
$P_n(\om)=2[1-\om/(\om+\ga_n)]$  [Eq.~(\ref{eq:P})].
The corresponding expressions are given in Table \ref{tab:dP}.
Gray blocks depict nonzero diffusion modes,
rendered with lines blocks  depict zero-mode diffusons $1/\Om$
renormalizing interaction vertices,
dashed lines stand for the impurity correlation function (\ref{eq:UU}).
The crossed block in diagram (c) is the Hikami box (see Fig.~\ref{fig:HB}).
}
\label{fig:dP}
\end{figure*}

\begin{figure}
\includegraphics{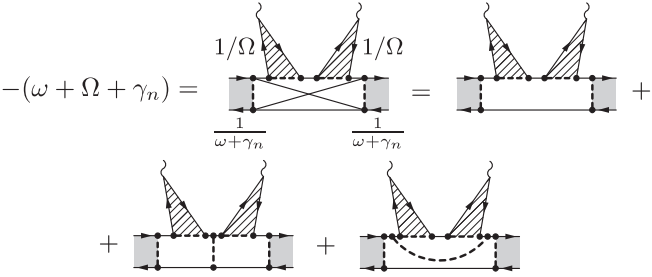}
\caption{Hikami box. Analytical expression for the Hikami box
in case of coordinate-independent interaction potential is $-(\om+\Om+\ga_n)$.}
\label{fig:HB}
\end{figure}

The set of diagrams for the first-order interaction corrections to the intragrain
polarization shown Fig.~\ref{fig:dP} is the same as the one for a bulk system~\cite{PO}.
The crossed region in diagram (c) is a Hikami box \cite{HB}
shown in Fig. \ref{fig:HB}, which analytical expression for
the case of coordinate-independent interaction is $-(\om+\Om+\ga_n)$.

We present the correction to the $n$-th mode
$P_n(\om)=2[1-\om/(\om+\ga_n)]$
of the polarization operator $P(\om,\rb,\rb')$ [Eq.~(\ref{eq:P})] in the form
\beq
    \de P^{(i)}_n(\om)=  2 T  \sum_{\Om>0} \theta_{\om,\Om} \, \al_i \la_n^{(i)}(\om,\Om),
\mbox{ } i=\mbox{c,d,e,f}, \label{eq:dP} \eeq where the
expressions for $\la_n^{(i)}(\om,\Om)$ and the multiplication
factors
$\al_i= \al_i^\text{s} \,\al_i^\text{u/d} \al_i^\text{l/r}$ [Eq.~(\ref{eq:al})] are given in
Table~\ref{tab:dP} (the factor 2 stands for spin degeneracy, the
diagrams are labelled in correspondence with the diagrams for
$\de\Pi_{xy}^{(0,1)}(\om)$ and $\de\Pi_{xy}^{(0,2)}(\om)$ below).
In Table~\ref{tab:dP} and Fig.~\ref{fig:dP}, the quantity
$V_0(\Om)=V(\Om,\ib,\ib)$ is the zero-mode interaction in a given
grain. Summing the contributions (\ref{eq:dP}), we obtain that the
total correction to the intragrain polarization operator
$P_n(\om)$ due to the zero-mode interaction $V_0(\Om)$ vanishes:
\beq
    \de P^\text{c}_n(\om)+  \de P^{\text{d}}_n(\om)   +
     \de P^{\text{e}}_n(\om)    +\de P^{\text{f}}_n(\om)=0
\label{eq:dPeq0}
\eeq
%
This is an expected result, since due to the gauge invariance
the constant interaction potential cannot affect physical quantities \cite{PO,ZNA},
expressed in this case by the density-density correlation function.
\begin{table}
\begin{tabular}{|c|c|l|l|l|}
    \hline
    $i$  & Expression for $\la^{(i)}_n(\om,\Om)$ &
                      $\al_i^\text{s}$ &
                      $\al_i^\text{u/d}$ &
                      $\al_i^\text{l/r}$ \\
    \hline
    c & $ \displaystyle     -(\om+\Om+\ga_n)\frac{1}{(\om+\ga_n)^2}
            \frac{1}{\Om^2} V_0(\Om) $  &  1 & 2 & 1 \\
    d & $ \displaystyle
            \frac{1}{\om+\Om+\ga_n}
            \frac{1}{\Om^2} V_0(\Om)$        & 1 & 2 & 1 \\
    e & $ \displaystyle
            \frac{1}{\om+\ga_n}
                \frac{1}{\om+\Om+\ga_n}
                \frac{1}{\Om} V_0(\Om)$
                             & 1 & 2 & 2 \\
    f & $ \displaystyle
            \frac{1}{(\om+\ga_n)^2}
            \frac{1}{\om+\Om+\ga_n}
            V_0(\Om)    $       &  1 & 2 & 1 \\
    \hline
\end{tabular}
\caption{ Corrections to the intragrain polarization operator
$P_n=2[1-\om/(\om+\ga_n)]$ [Eq.~(\ref{eq:P})].}
\label{tab:dP}
\end{table}

As a result, we obtain that the screened non-zero-mode Coulomb
interaction $v(\om,\rb,\rb')$ [Eq.~(\ref{eq:v}), double wavy line in
Fig.~\ref{fig:Hall}d] {\em does not} acquire any correction.
Therefore we should only renormalize the electron loops $\lan II
\ran$ and $\lan I n \ran$ shown in Figs.~\ref{fig:Hall}a,d
{\em explicitly}.

%

\subsubsection{Interaction corrections to $\Pi_{xy}^{(0,1)}(\om)$ and  $\Pi_{xy}^{(0,2)}(\om)$.}
Now we renormalize the electron loop $\lan I I \ran$
of $\Pi_{xy}^{(0,1)}(\om)$ [Eq.~(\ref{eq:Pixy01}), Fig.~\ref{fig:Hall}a)] and two loops
$ \lan I n \ran $
of $\Pi_{xy}^{(0,2)}(\om)$ [Eq.~(\ref{eq:Pixy02}), Fig.~\ref{fig:Hall}d].
The nonzero modes $v_n(\om)$ of the screened intragrain interaction
(double wavy line in Fig.~\ref{fig:Hall}d) are not renormalized according to the result
of the previous subsection.

All corrections to $\Pi_{xy}^{(0,1)}(\om)$ and  $\Pi_{xy}^{(0,2)}(\om)$
may be presented in the form:
\beq
    \de\Pi^{(i)}_{xy}(\om)=   \frac{g_T^2}{\nu}
          \sum_{n>0} f_n      \,T \sum_{\Om>0} \theta_{\om,\Om} \, \al_i \la_n^{(i)}(\om,\Om).
\label{eq:dPixyi}
\eeq
%
%
The geometry factor $f_n$ [Eqs.~(\ref{eq:fn}) and (\ref{eq:fnal})] arises,
when corrections to all diagrams in Fig.~\ref{fig:Hall1}
from the four closest contacts are taken into account.
The sets of diagrams giving corrections to $\Pi_{xy}^{(0,1)}(\om)$
and
$\Pi_{xy}^{(0,2)}(\om)$ are  shown in Figs.~\ref{fig:dPi01} and \ref{fig:dPi02},
and the corresponding expressions for $\la_n^{(i)}(\om,\Om)$ and
$\al_i=\al_i^\text{s} \al_i^\text{u/d} \al_i^\text{l/r}$
are given in Tables \ref{tab:dPi01} and \ref{tab:dPi02}, respectively.
For each diagram in Figs.~\ref{fig:dPi01} and \ref{fig:dPi02}
one must take four possibilities
[two for each contact according to Eq.~(\ref{eq:I})
and as shown in Fig.~\ref{fig:Hall1}]
of attaching external tunneling vertices
into account.
%
In the expressions,
\[
    V_0(\Om)=V(\Om,\ib,\ib)=
    \sum_\qb V(\Om, \qb)
\]
is the ``on-cite'' interaction of the grain,
\[
    V_1(\Om)=V(\Om,\ib+\eb_i,\ib)=
    \sum_\qb \cos (q_i a) V(\Om, \qb), \mbox{ } i=x,y,
\]
is the interaction between the neighboring grains, and

\[
    V_2(\Om) =V(\Om,\ib+\eb_x+\eb_y,\ib)=
    \sum_\qb \cos(q_x a) \cos(q_y a) V(\Om, \qb)
\]
is the interaction between the next-to-the-nearest-neighboring grains.
The interaction $V(\Om, \qb)$ is given by Eq.~(\ref{eq:VqhighOm}).

\begin{figure*}
\includegraphics{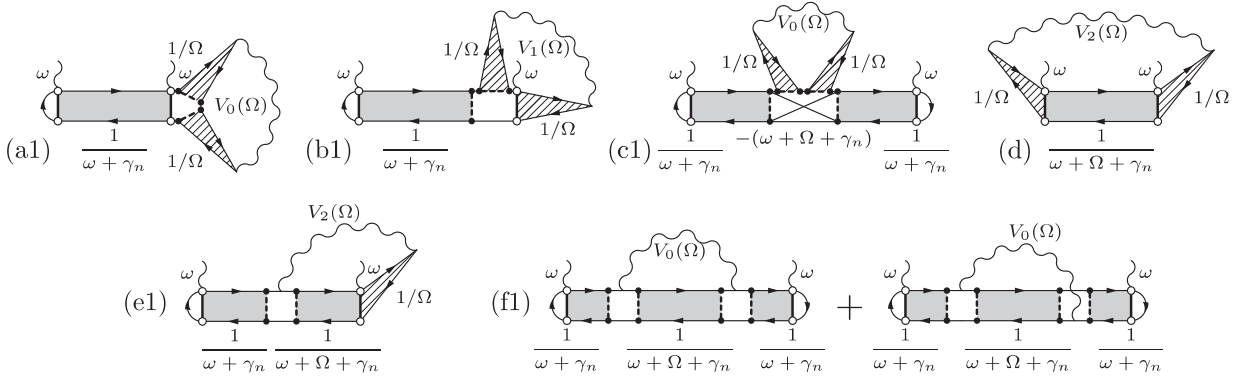}
\caption{Diagrams for the Coulomb interaction corrections to $\Pi_{xy}^{(0,1)}(\om)$
[Eq.~(\ref{eq:Pixy01}), Fig.~\ref{fig:Hall}(a)] describing renormalization
of the (tunneling current)-(tunneling current) correlator $\lan I I \ran$
of Fig.~\ref{fig:Hall}(a). Open circles denote tunneling vertices placed at the contacts.
Other elements are explained in caption to Fig.~\ref{fig:dP}.}
\label{fig:dPi01}
\end{figure*}

\begin{figure*}
\includegraphics{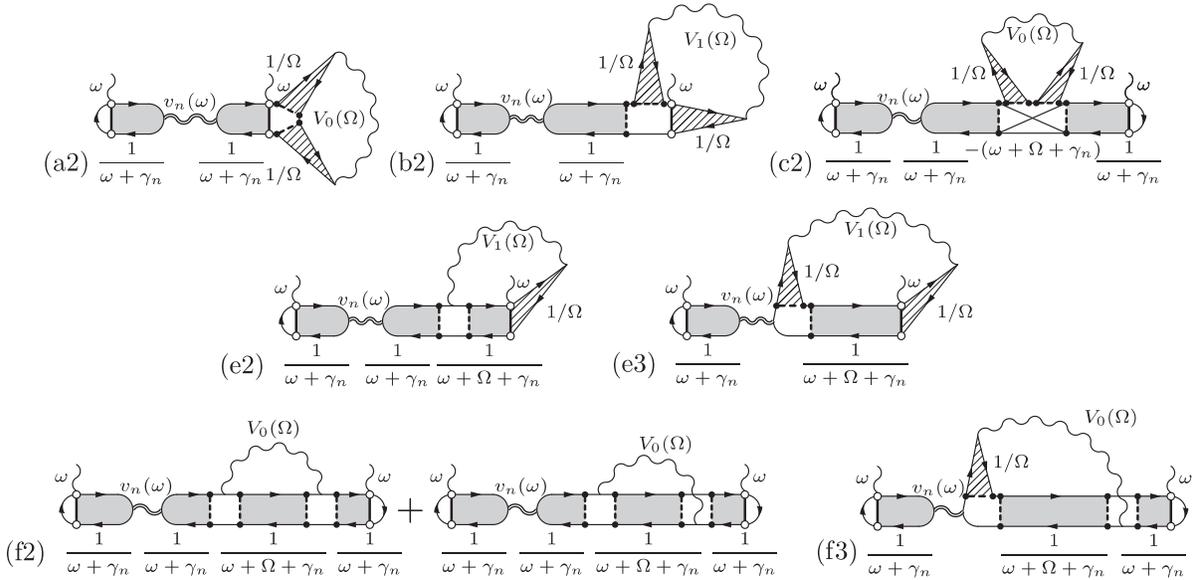}
\caption{
Diagrams for the Coulomb interaction corrections to $\Pi_{xy}^{(0,2)}(\om)$
[Eq.~(\ref{eq:Pixy02}), Fig.~\ref{fig:Hall}(d)], describing renormalization
of the (tunneling current)-density correlators $\lan I n \ran$
of Fig.~\ref{fig:Hall}(d).}

\label{fig:dPi02}
\end{figure*}

\begin{table}
\begin{tabular}{|c|c|l|l|l|}
    \hline
    $i$  & Expression for $\la^{(i)}_n(\om,\Om)$ &  $\al_i^\text{s}$ &
                      $\al_i^\text{u/d}$ &
                      $\al_i^\text{l/r}$
 \\
    \hline
    a1 & $ \displaystyle
        -\frac{1}{\om+\ga_n}\frac{1}{\Om^2} V_0(\Om) $ & 2 & 1 & 2 \\
    b1 & $ \displaystyle
            \frac{1}{\om+\ga_n}\frac{1}{\Om^2} V_1(\Om)  $ & 1 & 2 & 2 \\

    c1 & $ \displaystyle    -(\om+\Om+\ga_n)\frac{1}{(\om+\ga_n)^2}
            \frac{1}{\Om^2} V_0(\Om) $  &  1 & 2 & 1 \\

    d & $ \displaystyle
            \frac{1}{\om+\Om+\ga_n}
            \frac{1}{\Om^2} V_2(\Om)$        & 2 & 1 & 1 \\
    e1 & $ \displaystyle
            \frac{1}{\om+\ga_n}
                \frac{1}{\om+\Om+\ga_n}
                \frac{1}{\Om} V_1(\Om)$
                             & 1 & 2 & 2 \\
   f1 & $ \displaystyle
            \frac{1}{(\om+\ga_n)^2}
            \frac{1}{\om+\Om+\ga_n}
            V_0(\Om)    $       &  1 & 2 & 1 \\

    \hline
\end{tabular}
\caption{ Corrections to $\Pi_{xy}^{(0,1)}(\om)$, [Eq.~(\ref{eq:Pixy01}), Fig.~\ref{fig:Hall}(a)].}
\label{tab:dPi01}
\end{table}
%
%
%
\begin{table}
\begin{tabular}{|c|c|l|l|l|}
    \hline
     $i$  & Expression for $\la^{(i)}_n(\om,\Om)$ &  $\al_i^\text{s}$ &
                      $\al_i^\text{u/d}$ &
                      $\al_i^\text{l/r}$
 \\
    \hline
    a2 & $ \displaystyle -\frac{1}{\om+\ga_n}
                \frac{1}{\Om^2} V_0(\Om) \lt[ \frac{2\om}{\om+\ga_n} v_n \rt]$
        & 2 & 1 & 2 \\
    b2 & $ \displaystyle \frac{1}{\om+\ga_n}
            \frac{1}{\Om^2} V_1(\Om)
            \lt[ \frac{2\om}{\om+\ga_n} v_n  \rt]
                            $ & 1 & 2 & 2 \\

    c2 & $ \displaystyle -\frac{\om+\Om+\ga_n}{(\om+\ga_n)^2 \Om^2} V_0(\Om)
            \lt[ \frac{2\om}{\om+\ga_n} v_n  \rt]$  &  1 & 2 & 2 \\

      e2 & $ \displaystyle
            \frac{1}{\om+\ga_n}
            \frac{1}{\om+\Om+\ga_n}
            \frac{1}{\Om} V_1(\Om)
            \lt[ \frac{2\om}{\om+\ga_n} v_n  \rt]$  & 1 & 2 & 2 \\
    e3 & $ \displaystyle \frac{1}{\om+\Om+\ga_n}
            \frac{1}{\Om^2} V_1(\Om)
            \lt[ \frac{2\om}{\om+\ga_n} v_n  \rt] $  & 1 & 2 & 2 \\

    f2 & $ \displaystyle \frac{1}{(\om+\ga_n)^2}
            \frac{1}{\om+\Om+\ga_n}
                V_0(\Om)
            \lt[ \frac{2\om}{\om+\ga_n} v_n  \rt]$      &  1 & 2 & 2 \\

    f3 & $ \displaystyle \frac{1}{\om+\ga_n}
            \frac{1}{\om+\Om+\ga_n}
            \frac{1}{\Om} V_0(\Om)
            \lt[ \frac{2\om}{\om+\ga_n} v_n  \rt]$      &  1 & 2 & 2 \\

    \hline
\end{tabular}
\caption{Corrections to $\Pi_{xy}^{(0,2)}(\om)$,
        [Eq.~(\ref{eq:Pixy02}), Fig.~\ref{fig:Hall}(d)].}
\label{tab:dPi02}
\end{table}

Note that the diagram (d) in Fig.~\ref{fig:dPi01}, giving a correction to
$\Pi_{xy}^{(0,1)}(\om)$,
does not have an analog for $\Pi_{xy}^{(0,2)}(\om)$,
because this would mean connecting different electron loops
by the interaction line, which gives 0, as discussed before.
The crossed region in diagrams (c1) and (c2) is the Hikami box (Fig. \ref{fig:HB}).
%

Let us perform partial summation of the contributions (\ref{eq:dPixyi}) as follows:
\begin{subequations}
\beq
    \al_\text{a1}\la^\text{a1}_n(\om,\Om)+
    \al_\text{a2}\la^\text{a2}_n(\om,\Om)= - 4 \frac{1}{\Om^2} V_0(\Om) r_n,
\label{eq:contriba}
\eeq
\beq
    \al_\text{b1}\la^\text{b1}_n(\om,\Om)+
    \al_\text{b2}\la^\text{b2}_n(\om,\Om)=  4 \frac{1}{\Om^2} V_1(\Om) r_n,
\label{eq:contribb}
\eeq
\beq
\sum_{\substack{i=\text{c1,c2,}\\\text{f1,f2,f3}}} \al_i\la^{(i)}_n(\om,\Om)
        = 2 \frac{1}{\om+\Om+\ga_n} \frac{1}{\Om^2} V_0(\Om)
                -4 \frac{1}{\Om^2} V_0(\Om) r_n,
\label{eq:contribe}
\eeq
\beq
    \al_\text{d}\la^\text{d}_n(\om,\Om)= 2\frac{1}{\om+\Om+\ga_n}
            \frac{1}{\Om^2} V_2(\Om),
\label{eq:contribd}
\eeq
\beq
\sum_{\substack{i=\text{e1,}\\\text{e2,e3}}} \al_i\la^{(i)}_n(\om,\Om)
            =4 \lt[
        -\frac{1}{\om+\Om+\ga_n}
            \frac{1}{\Om^2} V_1(\Om)
                +  \frac{1}{\Om^2} V_1(\Om)  r_n \rt],
\label{eq:contribc}
\eeq
\end{subequations}
where $r_n=1/\ga_n$ is the ``resistance'' mode arising as a sum of the series
shown in Fig.~\ref{fig:Hall} and given by Eq.~(\ref{eq:rn}).
%

We see that two functionally  different forms arise. One contains
the frequency-independent resistance modes $r_n=1/\ga_n$, the
other contains nonzero diffuson modes $1/(\om +\Om+\ga_n)$.
Summing these two independently, we present the total first-order
interaction correction $\de\Pi_{xy}(\om)$ to the classical
result~(\ref{eq:Pixy0}) for the current-current  correlation
function $\Pi_{xy}^{(0)}(\om)$ as:
\beq
    \de\Pi_{xy}(\om)=\de\Pi^{TA}_{xy}(\om)+\de\Pi^{VD}_{xy}(\om),
\label{eq:dPixy}
\eeq
where
\beq
    \de\Pi^{TA}_{xy}(\om)= - 8\frac{ g_T^2}{\nu} \sum_{n>0} f_n
          T  \sum_{\Om>0} \theta_{\om,\Om} \, r_n
                \frac{1}{\Om^2} [V_0(\Om)- V_1(\Om)]
\label{eq:dPiTA}
\eeq
and
\begin{eqnarray}
    \de\Pi^{VD}_{xy}(\om) &=& 2\frac{ g_T^2}{\nu} \sum_{n>0} f_n    T  \sum_{\Om>0} \theta_{\om,\Om} \,
    \frac{1}{\om+\Om+\ga_n} \nonumber\\
    & &\times \frac{1}{\Om^2} [V_2(\Om)+V_0(\Om)-2 V_1(\Om)].
\label{eq:dPiVD}
\end{eqnarray}


The corresponding correction
$\de\sig_{xy}(\om)=\de\sig^{TA}_{xy}(\om)+\de\sig^{VD}_{xy}(\om)$
to Hall conductivity
\beq
    \sig_{xy}(\om)=\sig^{(0)}_{xy}+\de\sig_{xy}(\om)
,
\label{eq:sigxy2}
\eeq
is  obtained from Eq.~(\ref{eq:dPixy}) according to Eq.~(\ref{eq:sig}):
\beq
    \de\sig_{xy}(\om)=2e^2 a^{2-d} \frac{1}{\om} \de\Pi_{xy}(\om)
\label{eq:dsigxy2}
\eeq

Let us discuss the obtained results
(\ref{eq:dPixy}),(\ref{eq:dPiTA}),(\ref{eq:dPiVD}).
First of all, note that the corrections $\de\Pi^{TA}_{xy}(\om)$ and
$\de\Pi^{VD}_{xy}(\om)$
cannot be attributed to certain separate sets of diagrams:
the diagrams (a1),(a2),(b1),(b2) contribute to $\de\Pi^{TA}_{xy}(\om)$ only
[Eqs.~(\ref{eq:contriba}) and (\ref{eq:contribb})],
the diagram (d) contributes to $\de\Pi^{VD}_{xy}(\om)$ only [Eq.~(\ref{eq:contribd})],
but the rest of the diagrams contain parts corresponding both
to $\de\Pi^{TA}_{xy}(\om)$ and $\de\Pi^{VD}_{xy}(\om)$,
as seen from Eqs.~(\ref{eq:contribe}) and (\ref{eq:contribc})

Note also that the corrections $\de\Pi^{TA}_{xy}(\om)$ and
$\de\Pi^{VD}_{xy}(\om)$ vanish separately in the case of constant
interaction potential, i.e., when $V_0(\Om)=V_1(\Om)=V_2(\Om)$.
This property is not accidental and is enforced by  the gauge
invariance (see, e.g., Refs.~\onlinecite{PO,ZNA}): constant
interaction potential results in a shift of the  chemical
potential of the whole electron system and therefore does not
affect physical quantities expressed diagrammatically as chains of
closed electron loops (see Fig.~\ref{fig:Hall}). This fact serves
as an important check of our results
(\ref{eq:dPixy}),(\ref{eq:dPiTA}), and (\ref{eq:dPiVD}). We remind
the reader that Eq.~(\ref{eq:dPeq0}) is also a consequence of the
gauge invariance.

The physical processes leading to the corrections
$\de\Pi^{TA}_{xy}(\om)$ and
$\de\Pi^{VD}_{xy}(\om)$ are most clearly identified from the diagrams
that contribute to {\em either} one
of the quantities, i.e., from  the diagrams (a1),(a2),(b1),(b2)
for $\de\Pi^{TA}_{xy}(\om)$ and from the diagram (d) for $\de\Pi^{VD}_{xy}(\om)$.

\subsubsection{``Tunneling anomaly'' contribution $\de\Pi^{TA}_{xy}(\om)$}
First, consider the correction $\de\Pi^{TA}_{xy}(\om)$ [Eq.~(\ref{eq:dPiTA})].
Clearly, the diagrams (a1),(a2),(b1),(b2), contributing solely to
$\de\Pi^{TA}_{xy}(\om)$, describe
the effect of Coulomb interaction on the
process of electron tunneling through the contact.
%
Therefore, the correction $\de\Pi^{TA}_{xy}(\om)$
should be attributed to the {\em tunneling anomaly}\cite{AA,TA1,TA2} (TA) effect in a granular metal.
The correction $\de\Pi^{TA}_{xy}(\om)$  corresponds to the independent renormalization of  the tunneling conductances $G_T$
of the individual contacts in formula (\ref{eq:sigxy0})
 for the bare classical HC
$\sig_{xy}^{(0)}\propto G_T^2 R_H$, whereas
the Hall resistance $R_H$ of the grain remains unaffected.

%
%
Indeed, the $\Om$-independent resistance modes $r_n$ [Eq.~(\ref{eq:rn})]
can be taken out of the sum over $\Om$ in Eq.~(\ref{eq:dPiTA}).
After that the summed over $\Om$ expression becomes independent of the mode index $n$
and we can reduce the relative correction to the current-current
correlation function and HC to the form:
%
\beq
    \frac{\de\sig^{TA}_{xy}(\om)}{\sig^{(0)}_{xy}}=
    \frac{\de\Pi^{TA}_{xy}(\om)}{\Pi^{(0)}_{xy}(\om)}= 2 \frac{\de g_T(\om)}{g_T},
\label{eq:dPiTArel}
\eeq
where we have introduced the quantity
\beq
    \frac{\de g_T(\om)}{g_T} =- \frac{1}{\om} 4
 T  \sum_{\Om>0} \theta_{\om,\Om} \frac{1}{\Om^2} [V_0(\Om)- V_1(\Om)].
\label{eq:dgT}
\eeq
The expression in the RHS
of Eq.~(\ref{eq:dgT}) should be treated  as a relative correction to
the tunneling conductance $g_T$ of the  individual contact due to the tunneling anomaly.
[The factor 2 in Eq.~(\ref{eq:dPiTArel}) stands for two contacts
according to the square $G_T^2$ in Eq.~(\ref{eq:sigxy0}) for $\sig_{xy}^{(0)}$.]
Such physical interpretation
most clearly arises from the calculations
of interaction corrections to LC.
The bare LC $\sig_{xx}^{(0)} \propto g_T$ [Eq.~(\ref{eq:sigxx0})]
(in the lowest in $g_T/g_0 \ll 1$ order)
is expressed solely via the tunneling conductance $g_T$  and
the interaction correction to $\sig_{xx}^{(0)}$
corresponds to the renormalization of $g_T$,
since no other physical parameters are available. 
The interaction corrections to LC $\sig_{xx}^{(0)}$
were studied in Refs.~\onlinecite{ET,BELV} and
at $T \gg \Ga$ the correction $\de\sig_{xx}^{TA}(\om)$
to LC
\beq
    \sig_{xx}(\om)=\sig_{xx}^{(0)}+\de\sig_{xx}^{TA}(\om)
\label{eq:sigxx2}
\eeq
was obtained:
\beq
    \frac{\de\sig^{TA}_{xx}(\om)}{\sig^{(0)}_{xx}}=
    \frac{\de g_T(\om)}{g_T}.
\label{eq:dsigxx2}
\eeq

Therefore, according to the form of Eq.~(\ref{eq:dPiTArel}),
the correction $\de\Pi^{TA}_{xy}(\om)$
indeed corresponds to the renormalization of
the tunnelling conductances $G_T$ in Eq.~(\ref{eq:sigxy0})
and does not affect the Hall resistance $R_H$ of the grain.



\subsubsection{Virtual diffusion contribution $\de\Pi^{VD}_{xy}(\om)$}
Now let us discuss the correction $\de\Pi^{VD}_{xy}(\om)$ [Eq.~(\ref{eq:dPiVD})].
The diagram (d), which contributes solely to $\de\Pi^{VD}_{xy}(\om)$,
describes electron diffusion through the central grain.
The corresponding diffuson $D(\om+\Om,\rb,\rb')$ enters
Eq.~(\ref{eq:dPiVD}) as the combination
\beqar
    \sum_{n>0}\frac{f_n}{\om+\Om +\ga_n} &= &D_\nea(\om+\Om) -D_\sea(\om+\Om)\\
                        & & + D_\swa(\om+\Om) - D_\nwa(\om+\Om)
\eeqar
[see Eqs.~(\ref{eq:Daldef}) and (\ref{eq:fn})]
and therefore contains nonzero modes $n>0$ only.


As discussed above,
one cannot ``construct'' the propagator $\Db(\rb,\rb')$
[see Eqs.~(\ref{eq:Db}),(\ref{eq:rn})],
which describes the {\em real} propagation of electron density in metals
(basically, the classical conduction process),
from the diffuson $D(\om+\Om,\rb,\rb')$
by inserting interaction lines into the central diffuson in the diagram (d):
the corresponding diagrams simply cancel each other.
%
%
This  emphasizes {\em virtual} character of the diffusion:
real diffusion is not possible in metallic samples,
since nonequilibrium electron density created
in the course
 of diffusion is screened by Coulomb interaction.

Thus, the correction  $\de\Pi^{VD}_{xy}(\om)$
describes the  process of {\em``virtual diffusion''}
(VD) of electrons through the grain.

\subsubsection{Why $\Om,T \sim \ETh$ are necessary}
We emphasize that it has become possible to identify
two physically different contributions
$\de\Pi^{TA}_{xy}(\om)$ [Eq.~(\ref{eq:dPiTA})]  and $\de\Pi^{VD}_{xy}(\om)$  [Eq.~(\ref{eq:dPiVD})]
to the total correction $\de\Pi_{xy}(\om)$ [Eq.~(\ref{eq:dPixy})]
only because we have included ``high'' frequency range
$\Om \sim \ETh$ in our calculations.
%
%
%
At frequencies $\Om \ll \ETh$
the
expressions
 in Eqs.~(\ref{eq:dPiTA}) and (\ref{eq:dPiVD})
acquire the same functional form,
since $1/(\om+\Om+\ga_n)\approx 1/\ga_n=r_n$
and one cannot distinguish between them.

Including the frequencies $\Om \sim \ETh$
has complicated
the calculations significantly. Indeed,
for $\Om \ll \ETh$ one may consider
the diagrams (a1),(b1),(c1),(d) only.
The rest of the diagrams are smaller
than these ones in $\Om / \ETh \ll 1$ or
$\om/\ETh \ll 1$ and become comparable to them
only at $\Om,\om \sim \ETh $.
Considering the diagrams (a1),(b1),(c1),(d) only
and neglecting the frequencies $\om,\Om$ compared to $\ga_n \gtrsim \ETh$
in the diffusons $1/(\om+\ga_n)$ and  $1/(\om+\Om+\ga_n)$,
we would get the total correction $\de\Pi_{xy}(\om)$ [Eq.~(\ref{eq:dPixy})]
in the form
\begin{eqnarray}
    \de\Pi_{xy}(\om) &=& -2 \frac{g_T^2}{\nu} \sum_{n>0} f_n
        T  \sum_{\Om>0} \theta_{\om,\Om} \, \frac{1}{\ga_n}
            \frac{1}{\Om^2} \times \nonumber \\
            &  \times &  [3V_0(\Om)- 2V_1(\Om)-V_2(\Om)],
\label{eq:dPixy2}
\end{eqnarray}
which is just the sum of $\de\Pi^{TA}_{xy}(\om)$ and $\de\Pi^{VD}_{xy}(\om)$
provided one neglects $\om+\Om$ in $1/(\om+\Om+\ga_n)$.
Clearly, the above physical analysis would not be possible
based on the form of Eq.~(\ref{eq:dPixy2}).

Another related drawback of considering small frequencies $\Om \ll \ETh$ only
will be clear in the next subsection,
where we turn to the temperature dependence of the obtained corrections.



\subsubsection{Temperature dependence of the corrections}
Now let us discuss what consequences
the corrections (\ref{eq:dPixy}),(\ref{eq:dPiTA}),(\ref{eq:dPiVD})
have on the Hall conductivity $\sig_{xy}(\om)$
and, more importantly, on  the Hall resistivity
\[
    \rho_{xy}(\om) =\frac{\sig_{xy}(\om)}{\sig_{xx}^2(\om)}
=\rho_{xy}^{(0)}+\de \rho_{xy}(\om),
\]
which is directly measurable experimentally.
The conductivities $\sig_{xy}(\om)$ and  $\sig_{xx}(\om)$
are given by Eqs.~(\ref{eq:sigxy2}) and (\ref{eq:sigxx2}), respectively,
and the ``bare'' HR $\rho_{xy}^{(0)}$ by Eq.~(\ref{eq:rhoxy0text}).

First of all, since $\rho_{xy}^{(0)}$ [Eq.~(\ref{eq:rhoxy0text})]
simply does not contain $G_T$,
the tunneling anomaly effect
{\em does not} influence the Hall resistivity.
Indeed, according
to Eqs.~(\ref{eq:dPiTArel}) and (\ref{eq:dsigxx2}),
we obtain that the correction to HR due to TA vanishes:
\[
    \frac{\de \rho_{xy}^{TA}(\om)}{\rho_{xy}^{(0)}}=
    \frac{\de\sig^{TA}_{xy}(\om)}{\sig^{(0)}_{xy}}
    -2\frac{\de\sig^{TA}_{xx}(\om)}{\sig^{(0)}_{xx}} \equiv 0.
\label{eq:drhoxyTAhighT0}
\]
Next, since the analog of VD correction (\ref{eq:dPiVD})
is absent for LC $\sig_{xx}(\om)$ [see Eq.~(\ref{eq:sigxx2})]
in the leading in $g_T/g_0 \ll 1$ order\label{VDpage}\footnote{
In fact,
the correction $\de\sig_{xx}^{VD}$ analogous to $\de\sig_{xy}^{VD}$,
also exists for LC.
However, it is a correction to the term $-G_T^2/G_0$ in Eq.~(\ref{eq:sigxxexp})
but not to $G_T$. Therefore it contains an additional smallness $g_T/g_0 \ll 1$
compared to $\de\sig_{xx}^{TA}$ and can be neglected:
$   \de\sig_{xx}^{VD}/\de\sig_{xx}^{TA} \propto g_T/g_0 \ll 1.$} ,
the correction to Hall resistivity due to virtual diffusion process is
\[
    \frac{\de \rho_{xy}^{VD}(\om)}{\rho_{xy}^{(0)}}=
    \frac{\de\sig_{xy}^{VD}(\om)}{\sig_{xy}^{(0)}}=
    \frac{\de\Pi^{VD}_{xy}(\om)}{\Pi^{(0)}_{xy}(\om)}.
\]
Therefore, we obtain that the total correction
\[
    \de \rho_{xy}(\om)=\de \rho_{xy}^{TA}(\om)+\de \rho_{xy}^{VD}(\om)
\]
to HR at temperatures $T \gg \Ga$ is due to VD effect only:
\[
    \frac{\de \rho_{xy}(\om)}{\rho_{xy}^{(0)}}=
    \frac{\de\sig_{xy}^{VD}(\om)}{\sig_{xy}^{(0)}}.
\]

Now, let us discuss
 the temperature dependence of the obtained corrections.
The Coulomb potential (\ref{eq:VqhighOm}) is  completely screened
and equal to the inverse polarization operator,
\[
    V(\Om,\qb)=\frac{\de}{\Pc_0(\Om,\qb)}=\frac{\de \, \Om}{2\Ga_\qb},
\]
for frequencies $\Om \lesssim g_T E_c$.
In the limit of high frequencies $\Om \gtrsim g_T E_c$
the Coulomb potential (\ref{eq:VqhighOm}) remains unscreened
and $V(\Om,\qb) \sim E_c$.
The expression
\beq  
    \frac{1}{\Om^2} V(\Om,\qb) =
    \frac{1}{\Om} \frac{\de}{2\Ga_\qb}
\label{eq:integrand}
\eeq
in Eqs.~(\ref{eq:dPiTA}) and (\ref{eq:dPiVD})  is thus
proportional to $1/\Om$ for $\Om \lesssim g_T E_c$.
Therefore, the sum of over $\Om$ is logarithmically divergent
and we have to determine the low and high frequency cut-offs.

In the d.c. limit $\om \ll T$, the low frequency cut-off
for the sum over $\Om$
is set by the temperature $T$ (for $T \gg \Ga$).
This can be obtained by the analytical continuation
of Eqs.~(\ref{eq:dPiTA}) and (\ref{eq:dPiVD}) to real frequencies $\om$
and taking the limit $\om \ll  T$ according to
\beqar
    T\sum_{\Om_m>0} \theta_{\om_n,\Om_m} F(i\Om_m) |_{i\om_n \rtarr \om+i0, \om \ll T} =\\
    = -\frac{\om}{4\pi} \int_{-\infty}^{\infty} d \eps \frac{d}{d\eps} \lt( \eps \coth\frac{\eps}{2T} \rt) F(\eps),
\eeqar
we do not repeat this standard procedure here
(see, e.g., Refs.~\onlinecite{AAjetp,ZNA},
the integer indices of the Matsubara frequencies
$\om_n$ and
$\Om_m$ were written for clarity).

The high-frequency cut-off is different for TA
contribution $\de\Pi_{xy}^{TA}$ and VD contribution $\de\Pi_{xy}^{VD}$,
which is a direct consequence of their different physical origin.
According to Eq.~(\ref{eq:dgT}) the upper cut-off for $\de\Pi_{xy}^{TA}$
is $g_T E_c$. At $\Om \gtrsim g_T E_c$  we have
$V(\Om)/\Om^2 \sim E_c/\Om^2$ and the sum converges.
Therefore, the correction (\ref{eq:dgT}) to the tunneling conductance $g_T$
takes the form\cite{ET,BELV}

\beq
    \frac{\de g_T(T)}{g_T} =-\frac{1}{2\pi g_T d} \ln \frac{g_T E_c}{T}
\label{eq:dgT2}
\eeq
and for the TA correction to HC from Eqs.~(\ref{eq:dPiTA}) and (\ref{eq:dPiTArel})
we get:
\beq
    \frac{\de \sig_{xy}^{TA}(T)}{\sig_{xy}^{(0)}} =
    -\frac{1}{\pi g_T d} \ln \frac{g_T E_c}{T}.
\label{eq:dsigxyTAhighT}
\eeq
The lattice-specific factor $1/d$ arises as an integral
\beq
    \frac{1}{d}=\int \frac{a^d d^d\qb}{(2\pi)^d} \frac{1-\cos q_x a}
                {\sum_\be (1-\cos q_\be a)}.
\label{eq:1d}
\eeq

For the VD contribution $\de\Pi_{xy}^{VD}$
the summed over $\Om$ expression
contains additional $\Om$ in the denominator
coming from the intragrain nonzero diffusion mode $1/(\om+\Om+\ga_n)$.
Therefore, the expression is proportional to $1/\Om$
provided not only $\Om \lesssim g_T E_c$, but also $\Om \lesssim \ETh$.
Thus, the upper cut-off is the minimum
of the two quantities, i.e., $\min(\ETh,g_T E_c)$.
Calculating the sum  in Eq.~(\ref{eq:dPiVD})
and extracting $\sig^{(0)}_{xy}$ with the help of Eq.~(\ref{eq:sigxy0D}), we get
\beq
    \frac{\de\sig^{VD}_{xy}(T)}{\sig^{(0)}_{xy}} =\frac{c_d}{4 \pi g_T }
        \ln \lt[\frac{\min(g_T E_c,\ETh)}{T}\rt],
\label{eq:dsigxyVDhighT}
\eeq
where
\beq
    c_d=\int \frac{a^d d^d\qb}{(2\pi)^d} \frac{(1-\cos q_x a)(1- \cos q_y a)}
            {\sum_\be (1-\cos q_\be a)}
\label{eq:cd}
\eeq
is the lattice form-factor.
Appearance of the Thouless energy $\ETh$
as an additional cut-off in Eq.~(\ref{eq:dsigxyVDhighT})
reflects the diffusive nature of the contribution.
Virtual diffusion process is suppressed
for $T \gtrsim \ETh$, since in this case
the intragrain thermal length $L_T=\sqrt{D_0/T} \lesssim a$ becomes smaller
than the size $a$ of the grain.

We conclude that,
the total correction to HR at temperatures $T\gg\Ga$
is due to virtual diffusion process only and equals
\beq
     \frac{\de\rho_{xy}(T)}{\rho^{(0)}_{xy}} =
    \frac{\de\sig^{VD}_{xy}(T)}{\sig^{(0)}_{xy}} =
    \frac{c_d}{4 \pi g_T }
        \ln \lt[\frac{\min(g_T E_c,\ETh)}{T}\rt].
\label{eq:drhoxyhighT}
\eeq




We see that due to the logarithmic divergence
of the corrections (\ref{eq:dsigxyTAhighT}) and (\ref{eq:dsigxyVDhighT})
one is forced to go to the frequencies $\Om \sim \ETh$
in order to get a correct upper cut-off,
even if one considers the temperatures $T \ll \ETh$.
This has direct consequences
on physical quantities.
The upper cut-off also
determines the upper
bound for the temperature range, in which
the corrections have the $\ln T $-dependence
of Eqs.~(\ref{eq:dsigxyTAhighT}) and (\ref{eq:dsigxyVDhighT}).
So, HR $\rho_{xy}(T)=\rho_{xy}^{(0)}+\de \rho_{xy}(T)$
is $\ln T $-dependent according to Eq.~(\ref{eq:drhoxyhighT})
for $\Ga \lesssim T \lesssim \min(g_T E_c,\ETh)$,
whereas LR $\rho_{xx}(T)=1/\sig_{xx}(T)$ is
$\ln T $-dependent according to Eq.~(\ref{eq:dgT2})
for $\Ga \lesssim T \lesssim g_T E_c$.
For $T \gtrsim \min(g_T E_c,\ETh)$ the relative correction
to HR $ \de\rho_{xy}(T)/\rho^{(0)}_{xy} \lesssim 1/g_T$ becomes insignificant.
The ratio of  $g_T E_c$ and $\ETh$
can be arbitrary in a real system.
In case $\ETh \ll g_T E_c$ HR $\rho_{xy}(T)$
is $\ln T$-dependent in a narrower
range $\Ga \lesssim T \lesssim \ETh$ than LR $\rho_{xx}(T)$.
We emphasize, that perturbative approach used by us
is applicable as long as the relative correction (\ref{eq:drhoxyhighT})
is small.

Note that the integrals over $\qb$ in Eqs.~(\ref{eq:1d}) and (\ref{eq:cd})
do not diverge at small $q a \ll 1$,
although $\Ga_\qb=2\Ga \sum_\be (1-\cos q_\be a) \rtarr 0$ as $q a \rtarr 0$.
This is a consequence of the gauge-invariance mentioned previously.
At the same time, this is not the case for the density of states\cite{ET}.

The lattice factors $1/d$ and $c_d$ in
Eqs.~(\ref{eq:dsigxyTAhighT}),(\ref{eq:dsigxyVDhighT}),(\ref{eq:drhoxyhighT})
are not universal and are specific for quadratic/cubic lattice we considered.
However the $\ln T$-dependence itself
is a result of the screened form of the Coulomb interaction
in granular metals
and is robust to the lattice structure.
For a different type of lattice
the logarithmic dependence of Eqs.~(\ref{eq:dsigxyTAhighT}),(\ref{eq:dsigxyVDhighT}),(\ref{eq:drhoxyhighT})
remains the same, although numerical prefactors may be different.

Moreover, we expect the logarithmic form of Eqs.~(\ref{eq:dsigxyTAhighT}),(\ref{eq:dsigxyVDhighT}),(\ref{eq:drhoxyhighT})
to persist even if one allow for fluctuations (at least, moderate) of tunneling conductances $G_T$
from contact to contact.
The reason is that the logarithmic contribution arises from the
{\em integration over frequency}: $\int^{g_T E_c}_{T} d\Om/\Om =\ln (g_T E_c/T)$,
which is  {\em``decoupled''} from the {\em integration over quasimomentum} $\qb$
in Eqs.~(\ref{eq:dPiTA}),(\ref{eq:dPiVD}) [see Eq.~(\ref{eq:integrand})]
in the considered range of $\Om$.
The tunneling conductance $g_T$ enters as an upper cutoff, which
precise value with logarithmic accuracy is not important.
So, for a realistic array with randomly distributed tunneling condcutances
the form of Eqs.~(\ref{eq:dsigxyTAhighT}),(\ref{eq:dsigxyVDhighT}),(\ref{eq:drhoxyhighT})
should remain, although the lattice structure factors $1/d$ and $c_d$
should be replaced by some other distribution-dependent factors of the order of unity
and $g_T$ by some distribution-averaged value.

We mention in this respect that
the  logarithmic renormalization of individual conductances
for an array with randomly distributed tunneling conductances
was studied by Feigelman,  Ioselevich, and Skvortsov
in Ref.~\onlinecite{FIS} and, indeed, it was shown that
the $\ln T$-dependence of Eq.~(\ref{eq:dgT2})
for the {\em effective} tunneling conductance persists.


\subsubsection{Estimate for the contribution from nonzero interaction modes.}
As we claimed above,
coordinate-dependent intragrain interaction modes
$\de_{\ib\jb} v(\Om,\rb_\ib,\rb'_\jb)$ [Eq.~(\ref{eq:v})] of
the screened potential $V(\Om,\rb_\ib,\rb'_\jb)$
 [Eq.~(\ref{eq:V})]
give a smaller contribution
than the zero-mode part $V(\Om,\ib,\jb)$ [Eq.~(\ref{eq:Vq})]
in the relevant range $\Om,T \lesssim \ETh$. We provide an estimate here.

%

Let us revise, for example, the diagrams (a1),(a2)
in Figs.~\ref{fig:dPi01} and \ref{fig:dPi02}
taking now nonzero modes of the screened interaction into account.
For the block in the right grain, we have
\beqar
    T \sum_{\Om>0} \int d\rb\, d\rb' D(\Om,\s,\rb) V(\Om,\rb,\rb') D(\Om,\rb',\s)=
\\ =T \sum_{\Om>0}  \lt[ \frac{1}{\Om^2} V_0(\Om)+
    \sum_{n>0} \frac{|\phi_n(\s)|^2}{(\Om+\ga_n)^2}
        \frac{1}{\nu} v_n(\Om) \rt]
\eeqar The first term in the RHS is the contribution from the
zero-mode interaction $V_0(\Om)$ we had before, whereas the second
one is the contribution from nonzero modes $v_n(\Om)$, which we
want to estimate now ($\s$ is a point on the contact).
Since $V_0(\Om) \sim \de \Om/\Ga$ for $\Om \lesssim g_T E_c$,
the first term gives the logarithmic contribution (\ref{eq:dgT2}) to the conductivity:
\beq
    T \sum_{\Om>0} \frac{1}{\Om^2} V_0(\Om) \sim \frac{1}{g_T} \ln \frac{g_T E_c}{T}
\label{eq:zeromode}
\eeq
As $v_n(\Om) \sim (\Om+\ga_n)/\ga_n$
for $\Om \lesssim \min(1/\tau_0,\sig^\text{gr}_{xx})=1/\tau_0$,
the contribution from nonzero modes is estimated as
\[  T \sum_{\Om>0}  \de \sum_{n>0} \frac{1}{\Om+\ga_n} \frac{1}{\ga_n}
    \sim\frac{1}{\nu D_0} \int_{T}^{\infty} d\Om \int_{1/a}^{\infty} \frac{dq}{\Om+D_0 q^2} \sim
\]
\beq
    \sim \frac{1}{(p_F l)^2}[ 1- \al \sqrt{\tau_0 \max(T,\ETh)}]
\label{eq:nonzeromode}
\eeq
(here $\al\sim 1$ is  a cutoff-dependent number).
The obtained contribution
from nonzero modes is nothing else
but the tunneling anomaly in 3D case \cite{AAssc1,AAjetp,AAssc2}
(since the grains are three-dimensional), the square root
describing well-known singularity.
We see, that the contribution (\ref{eq:nonzeromode}) from nonzero modes
is smaller than the contribution (\ref{eq:zeromode}) from zero modes
if the condition
\beq
    g_0/g_T \gg a/l
\label{eq:non0smaller0}
\eeq
is met. Since $g_0 \gg g_T$ [Eq.~(\ref{eq:granular2})],
for ballistic grains ($l\sim a$) this is always the case.
More important, however, is that in the relevant temperature range $T \lesssim \ETh$
the contribution (\ref{eq:nonzeromode}) is $T$-independent.
Therefore, even if the condition (\ref{eq:non0smaller0})
is not well met, nonzero modes give an inessential $T$-independent
renormalization of the bare quantities for $T\lesssim \ETh$
not affecting the overall $T$-dependence of the HC and HR in that range.

\subsection{``Low'' temperatures $T \lesssim \Ga$ \label{sec:QCoulombLowT}}

Now we want to include
the region of temperatures $T\lesssim\Ga$
of the order or smaller than the escape rate $\Ga$
into our considerations.
In this regime the thermal length for the intergrain motion
$L_T^* =\sqrt{ \Ga a^2/T } \gtrsim a$ is greater than the size of the grain $a$
and quantum phenomena can occur not only
inside the grains,
but also at spatial scales much exceeding the grain size.

The technical complication in considering $T \lesssim \Ga$
instead of $T \gg \Ga$ is that we have to take tunneling
fully into account.
At the same time we still have a small parameter $\Ga/\ETh \ll 1$
[Eq.~(\ref{eq:granular1})], which allows us to neglect nonzero intragrain modes
compared to the zero modes
{\em as long as} we consider the frequencies $\Om \ll \ETh$.


As we have discussed in the previous section,
even for temperatures $T \ll \ETh$
considering $\Om \sim \ETh$
in the expression
\[
    \de\Pi_{xy}(\om) = \sum_{\Om>0} \theta_{\om,\Om} F(\Om)
\]
for the correction to HC
is necessary in order to obtain a correct upper cut-off
for logarithmically the diverging
quantities.
At the same time for $\Om \sim \ETh$ the results
must match with those of the previous
section, since for $\Om \gg \Ga$ one can
neglect tunneling in the expressions.

Therefore, to simplify our calculations we do assume $\Om \ll \ETh$
in this section. Having obtained the results
limited by $\Om \ll \ETh$, we determine the upper cut-off
by matching them with the results of the previous section
in the range $\Ga \ll \Om \ll \ETh$, where both
results are applicable.

The diagrams for $T \lesssim \Ga$ can be obtained from
the diagrams in Figs.~\ref{fig:dPi01} and \ref{fig:dPi02}
for $T \gg \Ga$ by including higher orders in tunneling.
As we restrict ourselves to low frequencies $\Om \ll \ETh$, only
four diagrams (a1),(b1),(c1),(d) should be considered and
the rest of the diagrams are smaller in $\Om/\ETh \ll 1$.
Taking tunneling into account we make sure
that in each diagram {\em only one grain contains nonzero
diffusion modes} necessary to have a nonvanishing
contribution to the Hall current, whereas in the rest
of the grains only the zero mode is retained.
The zero-mode interaction potential $V(\Om,\ib,\jb)$
is now taken in the form (\ref{eq:Vq}).
Considering the range $\om,\Om \ll \ETh$, we also neglect
the frequency dependence of nonzero modes now:
$1/(\om +\Om+\ga_n)=1/\ga_n$ and 
$1/(\om+\ga_n) 
=1/\ga_n$.

%


\subsubsection{Short-scale contribution \label{sec:shortscale}}

Including higher orders in tunneling in diagrams (a1), (b1), and (c1)
in Fig.~\ref{fig:dPi01} is straightforward. One has to substitute
the zero-mode diffusons
$\de_{\ib\jb} /\Om$
renormalizing the interaction vertices
by their form $\Dc_0(\Om, \ib,\jb)$ [Eq.~(\ref{eq:Dc0})],
which takes tunneling into account:
\[
    \de_{\ib\jb} \frac{1}{\Om} \rtarr \Dc_0(\Om,\ib,\jb)
,\mbox{ } \Dc_0(\Om,\qb)=\frac{1}{\Om+\Ga_\qb}.
\]
This results in the replacement
of $V(\Om,\ib,\jb)/\Om^2$ in the expressions
for $\la^{(i)}_n(\om,\Om)$, $i=\text{a1},\text{b1},\text{c1}$,
(see Table \ref{tab:dPi01}) by
\[
    \Vt (\Om,\ib,\jb)=\sum_{\kb,\lb} \Dc_0(\Om,\ib,\kb) V(\Om,\kb,\lb)
            \Dc_0(\Om,\lb,\jb)  ,
\]
\[
    \Vt (\Om,\qb) = \Dc_0^2(\Om,\qb) V(\Om,\qb).
\]
Consequently, instead of  Eqs.~(\ref{eq:contriba}), (\ref{eq:contribb}), and (\ref{eq:contribe})
 we get
[Fig.~\ref{fig:lowT}(a),(b),(c)]
\begin{subequations}
\beq
    \al_\text{a}\la^\text{a}_n(\om,\Om)= - 4 \Vt_0(\Om) \frac{1}{\ga_n},
\label{eq:contrib2a}
\eeq
\beq
    \al_\text{b}\la^\text{b}_n(\om,\Om)=
    4 \Vt_1(\Om) \frac{1}{\ga_n},
\label{eq:contrib2b}
\eeq
\beq
    \al_\text{c}\la^\text{c}_n(\om,\Om)= - 2 \Vt_0(\Om) \frac{1}{\ga_n},
\label{eq:contrib2c}
\eeq
where $\Vt_0(\Om)=\Vt(\Om,\ib,\ib)$ and $\Vt_1(\Om)=\Vt(\Om,\ib+\eb_x,\ib)=\Vt(\Om,\ib+\eb_y,\ib)$.

The diagram (d) in Fig.~\ref{fig:dPi01} has to be considered carefully.
It contains three diffusons:
one ``central'' diffuson  $1/(\om+\Om+\ga_n)$ describing
diffusion through the single grain and
two ``adjacent'' diffusons $1/\Om$
renormalizing the interaction vertices.
%
%
In the general case of arbitrary compared to $\Ga$ temperatures $T$,
the diagram (d) for the current-current correlation function $\Pi_{\ab\bb}(\om,\ib,\jb)$
corresponds to the process of virtual diffusion,
when an electron, ``created'' at the contact
 $(\jb+\bb,\jb)$
by the applied bias,
gets diffusively, without additional applied bias,
to the contact $(\ib+\ab,\ib)$, thus contributing
to the current.
Only in the limit $T \gg \Ga$ the main contribution comes from
the closest contacts of a single grain.
This is a virtual process, since for a real electron
its charge would be screened.


Accounting for tunneling in the diagram (d) in Fig.~\ref{fig:dPi01},
one should, in principle,
replace each one of  the three diffusons
by the ``exact'' diffuson $\Dc$ [Eq.~(\ref{eq:Dcdef})].
However, in the lowest nonvanishing in $g_T/g_0 \propto \Ga/\ETh \ll 1$ order
it is sufficient to leave the non-zero-mode diffuson $\Db(0,\rb,\rb')$
[Eq.~(\ref{eq:Dbexpr})]
in only {\em one} of the grains and retain only the zero modes $1/(|\Om| \Vc)$
in the rest of the grains.



\begin{figure}
\includegraphics{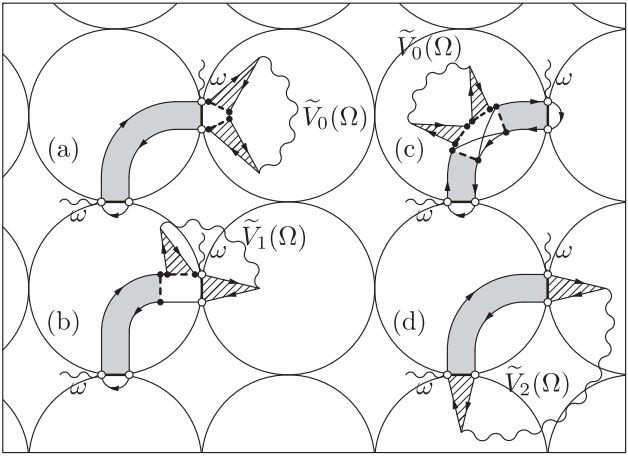}
\caption{
Diagrams for the interaction corrections to Hall conductivity
arising from spatial scales of the order of the grain size (``short-scale'' contributions)
at arbitrary compared to $\Ga$ temperatures $T$.
Gray blocks denote nonzero-mode intragrain diffusons
$\Db(0,\rb,\rb')$ [Eq.~(\ref{eq:Dbexpr})],
whereas rendered with lines blocks denote
zero-mode diffusons $\Dc_0(\Om, \ib,\jb)$ [Eq.~(\ref{eq:Dc0})] of the whole system.}

\label{fig:lowT}
\end{figure}

One possibility is to leave the ``central'' diffuson
as it is in Fig.~\ref{fig:dPi01}, i.e.,  as a non-zero-mode diffuson
$\Db(\om,\rb,\rb')$ of a single grain, and to
take ``adjacent'' diffusons as zero-mode diffusons
$\Dc_0(\Om,\ib,\jb)$ [Fig.~\ref{fig:lowT}(d)].
%
This gives the contribution analogous to Eq.~(\ref{eq:contribd}):
\beq
    \al_\text{d}\la^\text{d}_n(\om,\Om)=  2 \Vt_2 (\Om) \frac{1}{\ga_n},
\label{eq:contrib2d}
\eeq
\end{subequations}
where $\Vt_2(\Om)=\Vt (\Om,\ib+\eb_x,\ib+\eb_y)$.
The rest of the diagrams arising from the diagram (d)
in Fig.~\ref{fig:dPi01} are discussed in the next section.



Let us sum the contributions
(\ref{eq:contrib2a}), (\ref{eq:contrib2b}),
(\ref{eq:contrib2c}), and (\ref{eq:contrib2d}).
For the corresponding correction
\[
    \de\Pi_{xy}(\om)= \frac{g_T^2}{\nu} \sum_{n>0} f_n
          T  \sum_{\Om>0} \theta_{\om,\Om}
             \sum_{i=\text{a},\text{b},\text{c},\text{d}}
            \al_i \la^{(i)}_n(\om,\Om)
\]
to the bare current-current correlation function $\Pi^{(0)}_{xy}(\om)$ [Eq.~(\ref{eq:Pixy0})] we obtain
\begin{eqnarray}
    \de\Pi_{xy}(\om) &=& -2 \frac{g_T^2}{\nu} \sum_{n>0} f_n
          T  \sum_{\Om>0} \theta_{\om,\Om} \times \nonumber \\
        & \times &
                 \frac{1}{\ga_n} [3\Vt_0(\Om)- 2\Vt_1(\Om)-\Vt_2(\Om)].
\label{eq:dPixylowT}
\end{eqnarray}
Compare the correction (\ref{eq:dPixylowT}) with
the correction (\ref{eq:dPixy2}) obtained in the regime
$T \gg \Ga$, in which $\Om,T \ll \ETh$ was also assumed.
Equation~(\ref{eq:dPixylowT}) differs from Eq.~(\ref{eq:dPixy2})
only by having $\Dc_0(\Om,\qb)=1/(\Om+\Ga_\qb)$ instead of $1/\Om$
for the diffusons renormalizing the interaction vertices.
This difference is relevant at $T,\Om \lesssim \Ga$,
when tunneling comes into play.
Nevertheless, the correction (\ref{eq:dPixylowT})
still arises from the spatial scales of the order of the grain size $a$,
even if $T \ll \Ga$.
At $T,\Om \gg \Ga$, when one can neglect tunneling
in Eq.~(\ref{eq:dPixylowT}) [$\Dc_0(\Om,\qb) \rtarr 1/\Om$],
the corrections (\ref{eq:dPixylowT}) and  (\ref{eq:dPixy2}) coincide.

To get a temperature dependence of the correction  (\ref{eq:dPixylowT})
we note that the summed over $\Om$ expression
is still proportional to $1/\Om$ in the wide range
$\Ga \lesssim \Om \lesssim g_T E_c$, giving a logarithmic contribution.
The question about the lower cut-off is easily resolved:
it equals $T$ at $T \gg \Ga$ and $\Ga$ at $T \ll \Ga$.
Thus, the lower cut-off is $\max(T,\Ga)$.
To get a correct upper cut-off we
separate Eq.~(\ref{eq:dPixylowT}) {\em artificially} into two parts
according to the form of Eqs.~(\ref{eq:dPiTA}),(\ref{eq:dPiVD}):
\[
    \de\Pi_{xy}(\om)=\de\Pi^{TA}_{xy}(\om)+\de\Pi^{VD}_{xy}(\om),
\]
where
\beq
    \de\Pi^{TA}_{xy}(\om)= -8 \frac{g_T^2}{\nu} \sum_{n>0} f_n
          T  \sum_{\Om>0} \theta_{\om,\Om}  \frac{1}{\ga_n}
                \frac{1}{\Om^2} \lt[\Vt_0(\Om)- \Vt_1(\Om)\rt]
\label{eq:dPiTAlowT}
\eeq
and
\begin{eqnarray}
    \de\Pi^{VD}_{xy}(\om) &= & 2 \frac{g_T^2}{\nu} \sum_{n>0} f_n    T  \sum_{\Om>0} \theta_{\om,\Om} \,
    \frac{1}{\ga_n} \times \nonumber \\
            &\times& \frac{1}{\Om^2} \lt[ \Vt_2(\Om)+\Vt_0(\Om)-2 \Vt_1(\Om) \rt].
\label{eq:dPiVDlowT}
\end{eqnarray}
Keeping in mind that Eqs.~(\ref{eq:dPiTAlowT}) and
(\ref{eq:dPiVDlowT}) must match with Eqs.~(\ref{eq:dPiTA}) and
(\ref{eq:dPiVD}) at $\Om \sim \ETh$, we have to attribute a
cut-off $g_T E_c$ to  $\de\Pi^{TA}_{xy}(\om)$ and
$\min(g_T E_c,\ETh)$ to $\de\Pi^{VD}_{xy}(\om)$. Doing so, for the
corresponding corrections to HC we obtain
\beq
    \frac{\de\sig^{TA}_{xy}(T)}{\sig^{(0)}_{xy}} =-\frac{1}{ \pi g_T d}
        \ln \lt[\frac{g_T E_c}{\max(T,\Ga)}\rt]  \mbox{ for } T \lesssim g_T E_c
\label{eq:dsigxyTAlowT}
\eeq
and
\beq
    \frac{\de\sig^{VD}_{xy}(T)}{\sig^{(0)}_{xy}} =\frac{c_d}{4 \pi g_T }
        \ln \lt[\frac{\min(g_T E_c,\ETh)}{\max(T,\Ga)}\rt]
\label{eq:dsigxyVDlowT}
\eeq
for $T \lesssim \min(g_T E_c, \ETh)$.

Let us write the total correction $\de\sig_{xy}$
to HC
\[
    \sig_{xy}=\sig_{xy}^{(0)}+\de\sig_{xy}
\]
as
\beq
    \de\sig_{xy}=\de\sig^{TA}_{xy}+\de\sig^{VD}_{xy}+\de\sig^{AA}_{xy},
\label{eq:dsigxylowT}
\eeq
where  $\de\sig^{TA}_{xy}$ and $\de\sig^{VD}_{xy}$ are given by
Eqs.~(\ref{eq:dsigxyTAlowT}) and (\ref{eq:dsigxyVDlowT}) and
$\de\sig^{AA}_{xy}$ comes from the rest of the diagrams,
arising from the diagram (d) in Fig.~\ref{fig:dPi01} when
we take tunneling into account. We consider the latter ``large-scale''
contribution now.

\subsubsection{ Large-scale  contribution \label{sec:largescale}}


\begin{figure*}
\includegraphics{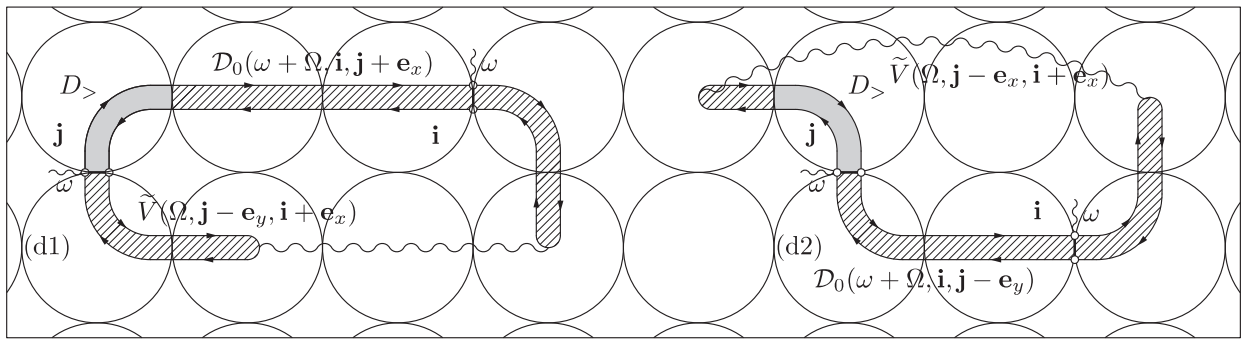}
\caption{
Diagrams for the interaction corrections to Hall conductivity
arising from large spatial scales at arbitrary  compared to $\Ga$ temperatures $T$.
The diagrams (d1) and (d2) cancel each other
leading to the vanishing ``Altshuler-Aronov''
correction to the  Hall conductivity: $\de\sig_{xy}^{AA}=0$.}

\label{fig:lowTlargescale}
\end{figure*}


%
%


As explained above, in order to generalize the diagram (d) in Fig.~\ref{fig:dPi01}
to  include the temperatures $T \lesssim \Ga$
one has to replace all three diffusons
in it by the zero-mode diffusons $\Dc_0$,
and then insert a non-zero-mode part $\Db$ [Eq.~(\ref{eq:Dbexpr})] of the diffuson
$D$ [Eq.~(\ref{eq:Dexpr})] into one of the grains.
Having considered the diagram (d) in Fig.~\ref{fig:lowT},
we are left with two following possibilities.

(i) One can insert the non-zero-mode part $\Db$ [Eq.~(\ref{eq:Dbexpr})]
somewhere into the middle of one of the three diffusons
$\Dc_0$. 
One can show straightforwardly that all the contributions from
such diagrams, when summed up, cancel each other exactly.

(ii) The less trivial possibility is
to consider nonzero modes $\Db$ in the grains
directly adjacent to the
contacts $(\ib+\eb_x,\ib)$ and $(\jb+\eb_y,\jb)$,
corresponding to external current vertices
for the correlation function $\Pi_{xy}(\om,\ib,\jb)$,
i.e., in the grains $\ib+\eb_x$ or $\ib$ for
contact $(\ib+\eb_x,\ib)$ and
in the grains $\jb+\eb_y$ or $\jb$ for
contact $(\jb+\eb_y,\jb)$.
However, the total contribution from such diagrams
also vanishes: for each diagram of this type there exists another diagram,
the contribution of which is exactly opposite and, consequently, their sum is zero.
One of such pairs are the diagrams (d1) and (d2) shown Fig.~\ref{fig:lowTlargescale},
their contributions to the current-current correlation function being
\beqar
    \de\Pi^\text{d1}_{xy}(\om) &= &   \frac{g_T^2}{\nu}
         2   T  \sum_{\Om>0} \theta_{\om,\Om} D_> \times \\
    & \times &  \sum_\jb \Ga\Dc_0(\om+\Om,\ib,\jb+\eb_x)
 \Vt(\Om,\jb-\eb_y,\ib+\eb_x),
\eeqar
\beqar
    \de\Pi^\text{d2}_{xy}(\om) &=&  - \frac{g_T^2}{\nu}
         2  T  \sum_{\Om>0} \theta_{\om,\Om} D_>  \times \\
    &  \times & \sum_\jb \Ga \Dc_0(\om+\Om,\ib,\jb-\eb_y)
 \Vt(\Om,\jb-\eb_x,\ib+\eb_x).
\eeqar
Here $D_>$ is the  non-zero-mode intragrain diffuson connecting contacts
in the counterclockwise direction, i.e., in the direction of the edge drift.
Using the translational invariance and symmetry in each component
of the zero-mode diffuson $\Dc_0$ and the screened Coulomb interaction $\Vt$,
e.g., that $\Dc_0(\Om,\ib-\jb)=\Dc_0(\Om,i_x-j_x,i_y-j_y)=\Dc_0(\Om,j_x-i_x,i_y-j_y)$,
we obtain that the sum over $\jb$ in $\de\Pi^\text{d2}_{xy}(\om)$
is identical to the one in $\de\Pi^\text{d1}_{xy}(\om)$,
and thus the contributions cancel each other:
\[ \de\Pi^\text{d1}_{xy}(\om)+    \de\Pi^\text{d2}_{xy}(\om)=0. \]


Therefore, we obtain that the total contribution from the diagrams
of types (i) and (ii) {\em vanishes identically}:

\beq
    \de\sig_{xy}^{AA}=0.
\label{eq:dsigxyAAlowT}
\eeq


As a result, all nonvanishing contributions to HC
at arbitrary  compared to $\Ga$ temperatures $T$
are given by the diagrams (a),(b),(c),(d) in Fig.~\ref{fig:lowT},
which lead to the corrections (\ref{eq:dsigxyTAlowT}) and (\ref{eq:dsigxyVDlowT}).
Note that these contributions
arise from the spatial scales of the order of the grain size $a$,
even for temperatures $T \ll \Ga$.
On the contrary, the eventually vanishing contributions of
the diagrams of types (i) and (ii) [as  (d1) and (d2) in Fig.~\ref{fig:lowTlargescale}]
arise from the spatial scales exceeding $a$.
The reason is that in these diagrams the contacts with external tunneling
vertices are connected by the diffuson $\Dc_0(\Om,\qb)$,
which ``size'' in the real space is determined by the thermal length $L_T^*=\sqrt{\Ga a^2/T}$.
In case of low temperatures $T \ll \Ga$, the thermal length $L_T^* \gg a$
can exceed the grain size $a$ significantly.

It is always instructive to compare the
results for a granular metal with those for an
ordinary homogeneously disordered metal (HDM).
For quantities arising from large spatial scales
(i.e., much greater than the grain size $a$ for granular metals
and the mean free path for diffusive HDMs)
one expects correspondence between the two,
since at such scales the microscopic structure
of the system becomes irrelevant.
%
The first-order Coulomb interaction correction $\de\sig_{xy}$
to Hall conductivity of HDM
was first studied in Ref.~\onlinecite{AKLL}
and the correction was found to vanish, $\de\sig_{xy}=0$.
So, indeed, our result (\ref{eq:dsigxyAAlowT})
for the ``large-scale'' contribution $\de\sig_{xy}^{AA}$
agrees with that for HDMs.

Note that even our approach of calculating
$\de\sig_{xy}^{AA}$ is quite similar to
that of Ref.~\onlinecite{AKLL}.
The authors of  Ref.~\onlinecite{AKLL}
calculated $\de\sig_{xy}$ perturbatively in magnetic
field $H$ (assuming $\om_H \tau_0 \ll 1$)
by inserting ``magnetic vertex'' $-\frac{e}{mc} \Ab \hat{\pb}$
in all possible ways
into the diagrams for zero magnetic field.
(i) Insertions of magnetic vertex into the diffusons
were found to cancel and
(ii) insertions of magnetic vertex into the block of
Green functions at the current vertices (Fig.~6 in Ref.~\onlinecite{AKLL})
were found to cancel.
These two steps resemble those (i) and (ii) of our approach,
insertion of magnetic vertex corresponding to insertion
of non-zero-mode intragrain diffuson $\Db$,
which contains all information about magnetic field,
into the zero-mode ``large-scale'' diffusons $\Dc_0$.

We want to stress that the ``large-scale'' AA contribution $\de\sig_{xy}^{AA}$
is not any different physically from the ``short-scale'' VD contribution $\de\sig_{xy}^{VD}$.
They both correspond to the process, when
electron gets diffusively from the contact
 $(\jb+\eb_y,\jb)$
to the contact $(\ib+\eb_x,\ib)$.
It just happens, that the contribution $\de\sig_{xy}^{VD}$
from the diffusion processes through a single grain does contribute to HC,
whereas the total contribution $\de\sig_{xy}^{AA}$
to HC  from the diffusion through
 more than one grain vanishes.


As a result of Secs. \ref{sec:shortscale} and \ref{sec:largescale},
we obtain that the {\em total} correction $\de\sig_{xy}$ [Eq.~(\ref{eq:dsigxylowT})]
to HC
at arbitrary temperatures $T$
is given by two short-scale contributions
$\de\sig^{TA}_{xy}$ [Eq.~(\ref{eq:dsigxyTAlowT})] and
$\de\sig^{VD}_{xy}$ [Eq.~(\ref{eq:dsigxyVDlowT})],
whereas the large-scale  contribution  $\de\sig^{AA}_{xy}=0$
[(Eq.~(\ref{eq:dsigxyAAlowT})] vanishes in agreement
with the theory of HDMs.
For the discussion of the corresponding corrections

\[
    \de\rho_{xy}=\de\rho^{TA}_{xy} +\de\rho^{VD}_{xy} +\de\rho^{AA}_{xy}
\]
to the Hall resistivity
\[
    \rho_{xy}=\frac{\sig_{xy}}{\sig_{xx}^2} =\rho_{xy}^{(0)}+\de\rho_{xy}.
\]
we refer the reader to the Results section \ref{sec:results} starting from Eq.~(\ref{eq:rhoxy}).


\section{Conclusion\label{sec:conclusion}}

In conclusion, we presented a theory of the Hall effect in granular metals.
In spite of its importance this question
has not been addressed before. It turned out that considering only
zero intragrain spatial harmonics that was very successful in
describing the longitudinal conductivity \cite{ET,BELVreview} is
not sufficient for calculating Hall conductivity and we
were forced to take nonzero harmonics into account.
Proceeding in this way we have shown that at
high enough temperatures the Hall resistivity is given by the
classical expression, from which one can extract the effective
carrier density of the system. At lower temperatures, quantum
effects come into play,
their most significant effect being
logarithmic temperature-dependent contributions to the Hall conductivity and resistivity
due to Coulomb interaction, which are absent in ordinary disordered metals.

We emphasize, however, that Eqs.~(\ref{eq:dsigxyTA}),(\ref{eq:dsigxyVD})
give the first-order interaction corrections
and the result is valid as long as these correction are small.
%
%
One could try to account for higher orders
using a renormalization group analysis
(since the contributions are logarithmic),
for example, like the one used for the longitudinal conductivity
in Ref.~\onlinecite{ET}.
In order to write down proper
renormalization group equations, methods
more sophisticated than the present
diagrammatic approach are needed.

%
%
%


Concerning the experimental situation
related to the theory developed here,
the logarithmic dependence $\rho _{xx}(T)=R_1-R_2\ln T$ of the longitudinal resistivity
in good conducting ($g_T \gtrsim 1$) granular materials,
corresponding to the logarithmic renormalization\cite{ET,BELV}
of the integrain tunneling conductance $g_T$ [Eq.~(\ref{eq:dgT2})],
has been observed experimentally (see Refs.~\onlinecite{expRxx}).
Clearly, measurements of the Hall resistivity $\rho _{xy}$ of such
granular samples could be also performed and our theory could be
thus tested. Unfortunately, the known to us experimental papers
(Refs.~\onlinecite{expRxy2}) on convetional Hall effect in
granular materials mostly deal with the systems in the regime of
low tunneling conductance $g_T \ll 1$, opposite to the metallic
regime $g_T \gg 1$ studied by us, which does not allow us to make
a detailed comparison now.
%
%
We mention that our theory may be also applied to indium tin oxide (ITO) materials
(see, e.g., Refs.~\cite{ITO}). Another related effect is the anomalous
Hall effect in ferromagnetic granular materials \cite{Mitra}.

We hope that more experiments on this  subject will be
done in the nearest future and that Hall measurements
will evolve into an important method of
characterization of granular materials.

We thank Igor S. Beloborodov, Yuli V. Nazarov and Anatoly F. Volkov for illuminating
discussions. Part of this work was done at the MPI-PKS Dresden in
the framework of the workshop "Dynamics and Relaxation in Complex
Quantum and Classical Systems and Nanostructures". The work was
financially supported by Degussa AG (Germany), SFB Transregio 12,
the state of North-Rhine Westfalia, and the European Union.


\appendix
\section{\label{app:bc}Boundary condition}

\begin{figure}
\includegraphics{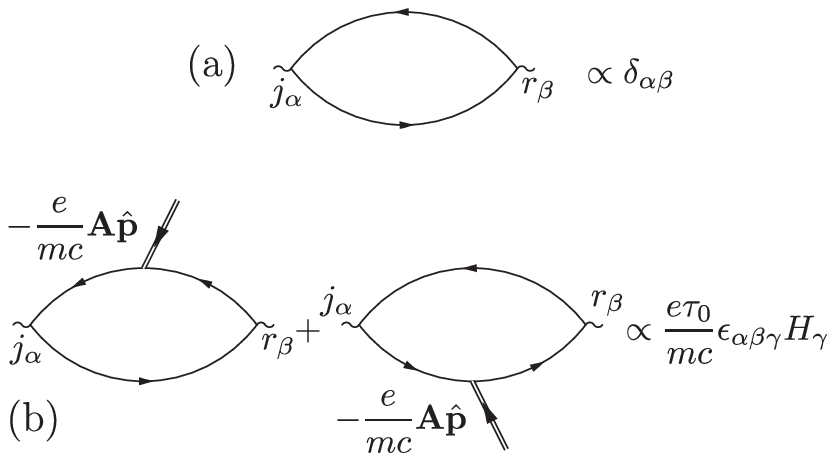}
\caption{
Diagrams for the current-(radius vector) correlator $\lan j_\al r_\be \ran$
[Eq.~(\ref{eq:j3})], if one directly expands Green functions
to the linear order in vector potential $\Ab$.
Fermionic lines denote the Green function
$[G(\eps,\pb)]^{-1}=i \eps - \xi(\pb) + \frac{i}{2 \tau_0} \sgn \eps$ of a bulk metal with $H=0$.
(a) Magnetic-field-independent part of  $\lan j_\al r_\be \ran$
giving the LHS of the boundary condition (\ref{eq:Dbc}).
(b) Linear in magnetic field part of $\lan j_\al r_\be \ran$
obtained by inserting ``magnetic vertex'' $-\frac{e}{mc} \Ab \hat{\pb}$
in all possible ways into the diagram (a) and giving the RHS of Eq.~(\ref{eq:Dbc}).
}
\label{fig:jr}
\end{figure}

To obtain the boundary condition for the diffuson $D(\om,\rb,\rb')$
[Eq.~(\ref{eq:Ddef})], we recall
 its physical meaning:
In the real-time representation, the quantity
\[
    D(t,\rb,\rb')=\int_{-\infty}^{+\infty} d \tilde{\om} e^{-i\tilde{\om} t}
    [ D(\om_n,\rb,\rb')]|_{\substack{i\om_n \rtarr \tilde{\om}+i 0,\\ \om_n>0}}
\]
gives the probability density to find an electron at point $\rb$
at time $t$ provided it was at point $\rb'$ at time $t=0$.
Therefore, according to the formal definition (\ref{eq:Ddef}) of
the diffuson, we can write down the probability current
corresponding to the diffusion process as
\beq
    \jb(\om,\rb,\rb')=
    \frac{1}{2 \pi \nu} \lan \hat{\jb}_\rb \lt[\Gc(\eps+\om,\rb,\rb')
            \Gc(\eps,\rb',\rb) \rt] \ran_U,
    (\eps+\om)\eps < 0,
\label{eq:jdef}
\eeq
where
\begin{widetext}
\begin{eqnarray}
    \hat{\jb}_\rb [ \Gc(\eps+\om,\rb,\rb') \Gc(\eps,\rb',\rb)]  &=&
    \frac{1}{2m} [ \Gc(\eps,\rb',\rb)
 (-i\nabla_\rb) \Gc(\eps+\om,\rb,\rb') +\nonumber\\
%
%
    &+& \Gc(\eps+\om,\rb,\rb') (i\nabla_\rb) \Gc(\eps,\rb',\rb)]
    - \frac{e}{mc} \Ab(\rb) \Gc(\eps+\om,\rb,\rb')
 \Gc(\eps,\rb',\rb)
\label{eq:joper}
\end{eqnarray}
\end{widetext}
is the current operator acting
on the product of two Green functions and $\Ab(\rb)$
is a vector potential corresponding to magnetic field $\Hb$.
Since electron cannot escape from an isolated grain, the normal component
of the current $\jb=\jb(\om,\rb,\rb')$ must vanish at the grain boundary:
\beq
    (\nb,\jb)|_{\rb\in S}=0.
\label{eq:jbcapp}
\eeq
Here, $\nb$ is a unit vector normal to the grain boundary
and pointing outside the grain.
Equation (\ref{eq:jbcapp}) together with Eqs.~(\ref{eq:jdef}) and (\ref{eq:joper})
gives a general form of the boundary condition
for  the diffusion propagator $D(\om,\rb,\rb')$.
Further simplifications depend on the model
used. In the case of white-noise disorder,
using the integral equation (\ref{eq:Dint}) for the diffuson,
we obtain
\begin{eqnarray}
    \jb(\om,\rb,\rb') &= &\hat{\jb}_\rb [D_0(\om,\rb,\rb')]  \nonumber \\
        &  + &\frac{1}{\tau_0} \int d\xb\, \hat{\jb}_\rb [D_0(\om,\rb,\xb)] D(\om,\xb,\rb')
\label{eq:jint}
\end{eqnarray}
where $\hat{\jb}_\rb$ acts on the Green functions of
$D_0(\om,\rb,\rb')=G(\eps+\om,\rb,\rb') G(\eps,\rb',\rb)$
as in Eq.~(\ref{eq:joper}).


We now exploit the diffusive limit.
Since $D_0(\om,\rb,\rb')$ varies on the spatial scale $l \ll a$,
we can
(i) neglect the first term in the RHS
 of Eq.~(\ref{eq:jint});
(ii) expand $D(\om,\xb,\rb')$ writing it as
\[
    D(\om,\xb,\rb') \approx D(\om,\rb,\rb')+(\xb-\rb)_\be \nabla_{\rb\be} D(\om,\rb,\rb')
\]
to get
\beq
    j_\al(\om,\rb,\rb') =\frac{1}{\tau_0} \lan j_\al r_\be \ran
                \nabla_{\rb\be} D(\om,\rb,\rb')
\label{eq:j2},
\eeq
where
\beq
    \lan j_\al r_\be \ran= \int d\xb \hat{j}_{\rb\al} [D_0(\om,\rb,\xb)]
        (\xb-\rb)_\be
\label{eq:j3}
\eeq
is the current-(radius vector) correlation function.
(iii)~In the diffusive limit, details of the boundary scattering
become unimportant as soon as we move away from the boundary into
the bulk of the grain over the distance of the order of  the mean free path $l$.
Since $D(\om, \rb,\rb')$ in Eq.~(\ref{eq:j2}) varies on the scales $a \gg l$,
the condition (\ref{eq:jbcapp}) may be evaluated not
exactly at the boundary, but at some point a few $l$ away from it.
This allows us to use for Green functions $G$
in Eq.~(\ref{eq:j3}) their expressions for the bulk.

The correlation function $\lan j_\al r_\be \ran$ can be calculated
via  diagrammatic technique either by directly expanding Green
functions in vector potential $\Ab(\rb)$ (Fig.~\ref{fig:jr}) or
using an explicitly gauge-invariant approach developed by Khodas
and Finkel'stein in Ref.~\onlinecite{Khodas}. Proceeding in either
way,
in the linear order in $H$
we get

\beq
    \lan j_\al r_\be \ran=C \lt( \de_{\al\be} + \frac{e\tau_0}{m c} \epsilon_{\al\be\ga} H_\ga \rt),
\label{eq:jr}
\eeq
where $ \epsilon_{\al\be\ga} $ is the totally antisymmetric tensor, $\epsilon_{xyz}=1$,
and $C=-(2\pi/3) \nu l^2$ 
is an irrelevant for the boundary condition (\ref{eq:jbcapp}) prefactor.
Inserting Eqs.~(\ref{eq:j2}) and (\ref{eq:jr}) into Eq.~(\ref{eq:jbcapp}),
we obtain the boundary condition for the diffuson $D=D(\om,\rb,\rb')$
in the presence of magnetic field:
\[
    n_\al \lt( \de_{\al\be} + \frac{e\tau_0}{m c} \epsilon_{\al\be\ga} H_\ga \rt)
    \nabla_{\rb\be} D
|_{\rb\in S}=0,
\]
which can be easily expressed in the form of Eq.~(\ref{eq:Dbc}).



\end{document}